\PassOptionsToPackage{table}{xcolor}

\documentclass[sigconf]{acmart}

\usepackage{soul}
\usepackage{enumitem}
\usepackage{tabularx}
\usepackage{multirow}
\usepackage{color}
\usepackage{xcolor}
\usepackage{xspace}
\usepackage{subfig}
\usepackage{graphicx}
\usepackage{listings}
\usepackage{hhline}
\usepackage{acmart-taps}

\makeatletter

\newcommand{\bpstart}[1]{
\noindent{\textbf{#1}}%
}

\newcommand{\app}{\mbox{DataPilot}\xspace}

\newcommand{\paragraphHeadingSpace}{\vspace{4px}}

\newcommand{\pc}[1]{$B_{#1}$}
\newcommand{\pq}[1]{$Q_{#1}$}
\newcommand{\pe}[1]{$D_{#1}$}
\newcommand{\pde}[1]{$E_{#1}$}
\newcommand{\ses}[1]{$S_{#1}$}

\definecolor{DD}{HTML}{00856b} 
\definecolor{QQ}{HTML}{6c5ce7}
\definecolor{BB}{HTML}{e17055}
\definecolor{DE}{HTML}{4834d4}
\definecolor{red}{HTML}{ff0000}
\definecolor{linkcolor}{HTML}{647382}
\definecolor{citecolor}{HTML}{647382} %
\definecolor{urlcolor}{rgb}{0.4,0.2,0.2}
\definecolor{sqlcolor}{HTML}{965d67}
\definecolor{smtcolor}{HTML}{5d968c}
\definecolor{commentcolor}{HTML}{4B0082}
\definecolor{high}{HTML}{2ecc71}
\definecolor{medium}{HTML}{f1c40f}
\definecolor{low}{HTML}{e74c3c}
\definecolor{unchecked}{HTML}{efefef}
\definecolor{cellhighlight}{HTML}{cefaee}

\definecolor{DD}{RGB}{0, 133, 107} 
\definecolor{QQ}{RGB}{108, 92, 231}
\definecolor{BB}{RGB}{225, 112, 85}
\definecolor{DE}{RGB}{72, 52, 212}
\definecolor{red}{RGB}{255, 0, 0}
\definecolor{linkcolor}{RGB}{100, 115, 130}
\definecolor{citecolor}{RGB}{100, 115, 130} %
\definecolor{sqlcolor}{RGB}{150, 93, 103} 
\definecolor{smtcolor}{RGB}{93, 150, 140}
\definecolor{commentcolor}{RGB}{75, 0, 130}
\definecolor{high}{RGB}{46, 204, 113}
\definecolor{medium}{RGB}{241, 196, 15}
\definecolor{low}{RGB}{231, 76, 60}
\definecolor{unchecked}{RGB}{239, 239, 239}
\definecolor{cellhighlight}{RGB}{206, 250, 238}
\definecolor{darkgreen}{HTML}{009432}
\definecolor{darkblue}{HTML}{2980b9}
\definecolor{darkred}{HTML}{c0392b}

\newcommand{\cut}[1]{}
\newcommand{\edit}[1]{#1}
\newcommand{\add}[1]{#1}

\AtBeginDocument{%
  }

\copyrightyear{2023}
\acmYear{2023}
\setcopyright{rightsretained}
\acmConference[CHI '23]{Proceedings of the 2023 CHI Conference on
Human Factors in Computing Systems}{April 23--28, 2023}{Hamburg,
Germany}
\acmBooktitle{Proceedings of the 2023 CHI Conference on Human Factors
in Computing Systems (CHI '23), April 23--28, 2023, Hamburg,
Germany}
\acmDOI{10.1145/3544548.3581509}
\acmISBN{978-1-4503-9421-5/23/04}

\begin{document}

\title[DataPilot: Utilizing Quality and Usage Information for Subset Selection during Visual Data Preparation]{DataPilot: Utilizing Quality and Usage Information\\ for Subset Selection during Visual Data Preparation}

\author{Arpit Narechania}
\affiliation{%
  \institution{Georgia Institute of Technology}
  \city{Atlanta}
  \country{USA}
}
\email{arpitnarechania@gatech.edu}

\author{Fan Du}
\affiliation{%
  \institution{Adobe Research}
  \city{San Jose}
  \country{USA}}
\email{dufan2013@gmail.com}

\author{Atanu R Sinha}
\affiliation{%
  \institution{Adobe Research}
  \city{Bengaluru}
  \country{India}}
\email{atr@adobe.com}

\author{Ryan A. Rossi}
\affiliation{%
  \institution{Adobe Research}
  \city{San Jose}
  \country{USA}}
\email{ryrossi@adobe.com}

\author{Jane Hoffswell}
\affiliation{%
  \institution{Adobe Research}
  \city{Seattle}
  \country{USA}}
\email{jhoffs@adobe.com}

\author{Shunan Guo}
\affiliation{%
  \institution{Adobe Research}
  \city{San Jose}
  \country{USA}}
\email{sguo@adobe.com}

\author{Eunyee Koh}
\affiliation{%
  \institution{Adobe Research}
  \city{San Jose}
  \country{USA}}
\email{eunyee@adobe.com}

\author{Shamkant B. Navathe}
\affiliation{%
  \institution{Georgia Institute of Technology}
  \city{Atlanta}
  \country{USA}
}
\email{sham@cc.gatech.edu}

\author{Alex Endert}
\affiliation{%
  \institution{Georgia Institute of Technology}
  \city{Atlanta}
  \country{USA}
}
\email{endert@gatech.edu}

\renewcommand{\shortauthors}{Narechania et al.}


\begin{abstract}
Selecting relevant data subsets from large, unfamiliar datasets can be difficult. We address this challenge by modeling and visualizing two kinds of auxiliary information: (1) \emph{quality} -- the validity and appropriateness of data required to perform certain analytical tasks; and (2) \emph{usage} -- the historical utilization characteristics of data across multiple users. Through a design study with 14 data workers, we integrate this information into a visual data preparation and analysis tool, \app. \app presents visual cues about ``the good, the bad, and the ugly'' aspects of data and provides graphical user interface controls as interaction affordances, guiding users to perform subset selection. Through a study with 36 participants, we investigate how \app \edit{helps} users\cut{to} navigate a large, unfamiliar tabular dataset, prepare a relevant subset, and build a visualization dashboard. We find that users selected smaller, effective subsets with higher quality and usage, and with greater success and confidence.
%
\end{abstract}



\maketitle

\section{Introduction}


Data are never truly raw~\cite{gitelman2013raw} but still require processing through cleaning, integration, transformation, and selection before they can be utilized for their intended purposes~\cite{pyle1999data}. 
Modern organizations often ingest all incoming data in their native form with the intent of performing analytics later~\cite{farid2016clams}.
The inherent information overload due to this ``load-first'' philosophy poses several challenges in data navigation and knowledge discovery~\cite{deng2017data,fernandez2018aurum,gotz2016adaptive}. 
For example, consider a user task, \emph{``analyze a large e-commerce dataset and build a dashboard visualizing recent geographic trends~\add{for predicting future sales}.''} To perform this task, users must first identify relevant data attributes pertaining to customers' locations (e.g., ``ZipCode'') and then select the desired data records by applying a temporal filter (e.g., \edit{monthly}\cut{retrieve transactions only for the current month}).
\edit{Unfortunately, new users unfamiliar with the data may adopt ``trial and error'' inspection strategies~\add{\cite{national2000people}} resulting in the selection of irrelevant, inferior attributes while missing out on important attributes, undermining the outcome of the subsequent analysis.}
\cut{Unfortunately, new users who are unfamiliar with the data may adopt ``trial and error'' inspection strategies~\add{\cite{national2000people}} that may result in them selecting irrelevant, inferior attributes (e.g., ``Email'' attribute with mostly missing values) while potentially also missing out on important, high-quality attributes (e.g., ``State'' attribute with an analysis-worthy distribution of values), undermining the outcome of the subsequent analysis.}
Even experienced users may rely upon their own past usage and not explore new attributes of a new dataset, also putting the analysis outcome into question.
Furthermore, users \edit{may spend} more time finding relevant data than performing the\cut{actual} analytic task at hand~\cite{fernandez2018aurum}.
Thus, we ask, \emph{``How \edit{to}~\add{design user interfaces that} provide guidance to users to analyze large, unfamiliar datasets and select relevant and effective subsets for downstream analytics and visualization tasks such as building dashboards and customer segmentation?''}

We interviewed 14 data workers from a large technology company who\cut{often} select data subsets (extract a smaller set of attributes and records from a larger dataset) for\cut{usage in downstream applications such as} making dashboards (data analysts), training machine learning models (data scientists), and running digital marketing campaigns (marketers).
All\cut{of the} data workers\cut{reinforced our initial hunch about}~\edit{communicated the importance of} the quality of data\cut{(e.g., implications of missing and incorrect values in data)}; some of them, who relied on others for preparing these data subsets as they lacked the necessary skill set, also reflected on the potential of surfacing other data characteristics such as their usage across users\cut{(e.g., popularity of data attributes when utilized during analytics)}.
\edit{This feedback from the data workers call for an interactive, self-service tool that facilitates data preparation with two kinds of auxiliary information: (1) quality and (2) usage. We model this auxiliary information using the data, associated meta-data, and corresponding usage logs and visually present it to users to guide them during subset selection and analysis, a task that they all perform for different purposes.}


\add{Prior art defines data quality from multiple perspectives: consumer~\cite{furber2016semantic}, business~\cite{otto2009identification, herzog2007data, fleckenstein2018modern, mahanti2019data, pipino2002data}, and standards-based~\cite{medic2016new, chang2019nist}. A single definition covering the different contexts is difficult~\cite{furber2016semantic}.} \edit{Contextual to this work, we} define \emph{quality} as \emph{``the validity and appropriateness of data required to perform certain analytical tasks.''} Quality is important because data are often messy, and organizations' ``load-first'' philosophy often results in ``big data graveyards''~\cite{stein2014pwcreport} comprising large volumes of missing, erroneous, and irrelevant information. Ideally, these data deficiencies would trigger corrective measures or even non-use; however, most organizations fail to maintain data quality standards~\cite{nagle2017only} as ``everyone wants to do the [ML] model work, not the data work''~\cite{sambasivan2021everyone}. 
In this work, we model three quality dimensions~\cite{pipino2002data}, deemed important by \edit{the} experts:

\begin{enumerate}[nosep,leftmargin=0.5cm]
    \item \emph{completeness}: frequency of non-missing values in the data.
    \item \emph{correctness}: frequency of \edit{correct} values in the data.
    \item \emph{objectivity}: \edit{extent that values conform to a target distribution}\cut{amount of distortion in the distribution of data}.
\end{enumerate}

We define \emph{usage} as \emph{``the historical utilization characteristics of data across multiple users,''} inspired by the ``data utility''\cut{data flow} descriptor~\cite{schulz2017systematic}.~\cut{In a workplace setting}Users often collaborate at work~\cite{almahmoud2021teams, sambasivan2021everyone, kim2017data, zhang2020data, drosos2020wrex}, \edit{but much more around code than around data}~\cite{koesten2019collaborative}. Understanding how data are created and shared inside an organization is\cut{less well}~\edit{under}explored~\cite{koesten2019collaborative}.
\edit{We believe leveraging usage logs of current and past users, and meta-data can be one way to guide other users.}\cut{For example, frequent usage of the ``Profit'' attribute in past reports from the sales function is potentially indicative of its high importance; presenting this information to future report makers can nudge them to use it in their own reports or at least perform a sanity check before signing-off, all within the self-service tool and without reaching out to a peer.}
Motivated by use cases from the data workers, we derive three dimensions of usage for a subset selection and dashboard building task, where data \edit{refers to} attributes and records of a tabular dataset: 

\begin{enumerate}[nosep,leftmargin=0.5cm]
    \item \emph{in-subsets}: percentage of users that put the data in their subset.
    \item \emph{in-filters}: percentage of users that applied a filter on the data.
    \item \emph{in-visualizations}: percentage of users that visualized the data.
\end{enumerate}

We integrate both quality and usage information into a visual data preparation and analysis tool, \app. \edit{DataPilot} facilitates preparing a subset from a large tabular dataset\cut{(e.g., with hundreds of attributes)} for building a visualization dashboard. 
Specifically, \app computes a standardized score out of 100 for each of the quality and usage dimensions, e.g., \emph{in-subsets} score for the ``Profit'' attribute is 94 out of 100.
\app also presents visual cues to guide users about the ``good'' and ``bad'' aspects of their data, e.g., highlighting missing and incorrect data values by coloring them in red. 
Lastly, \app provides graphical user interface (GUI) controls as interaction affordances to assist users during subset selection, e.g., range sliders to filter out less popular data and sorting widgets to order and group data with similar characteristics together. 
Modern data tools~\cite{tableauprep,powerbi,metaplane,uberdatabook,dremio,trifactawrangler,amazondeequ,ham2013openrefine} provide a myriad of features \edit{such as interactive GUIs} to help prepare data; however, to the best of our knowledge, no tool leverages usage information from the usage logs and associated meta-data to provide interaction affordances that facilitate interactive subset selection and analysis.

We conducted a user study with 36 participants to investigate how the \app user interface guides users (nudging them one way or another) to navigate a large and unfamiliar tabular dataset, prepare a relevant subset, and build a visualization dashboard.
Our findings indicate that quality and usage information together help users to create smaller, effective subsets with greater success and confidence. 
We define an effective subset as one that has a higher percentage of attributes and records with high overall scores on quality and usage.
Importantly, participants expressed caution about excessive reliance on usage behaviors of previous users as it can reduce exploration of quality data (pursuing novelty less), in favor of exploitation (repeating what has worked so far).
Challenging convention, our findings also call for visual data analysis tools to prioritize and integrate data preparation affordances directly into analysis workflows to foster more effective use of data.

The primary contributions of this work include:
\begin{enumerate}[nosep]\leftskip-2pt
    \item A design study with 14 data workers about tasks and challenges during data preparation and analysis that revealed the importance of data quality and the potential of surfacing additional characteristics such as usage to improve these workflows and also improve user collaboration (Section~\ref{sec:interview}).
    \item Modeling of two kinds of auxiliary information: \textbf{quality} and \textbf{usage}, by leveraging the data, associated meta-data, and usage logs of users (Section~\ref{sec:metadata}),
    \item A visual data preparation and analysis tool, \textbf{\app}, integrated with quality and usage information to guide users during subset selection and analysis (Section~\ref{sec:datapilot}),
    \item A user study with 36 participants that revealed how \app helped users to select smaller, effective subsets from large, unfamiliar datasets  with greater success and confidence during a subset selection and dashboard building task (Section~\ref{sec:study}).
\end{enumerate}

Note that judging the true effectiveness of the selected subsets and the created dashboards depends on the end goals and other contextual circumstances, requiring expert assessment; we did not pursue this angle because our participants were not domain experts. Also, while \app focuses on subset selection, additional tools and studies are needed to evaluate other downstream analytics tasks such as ranking and clustering across other applications.


\section{Related Work}

\add{\subsection{Data Preparation}}
Data preparation (or pre-processing) involves analyzing the data to ensure high-quality results through collection, integration, transformation, cleaning, reduction, and discretization~\cite{dataprep2003zhang}.
\subsubsection{\edit{Subset Selection\cut{during Data Preparation}}}
Subset selection (or data reduction) involves reducing the size of the dataset~\cite{fodor2002survey,john1994irrelevant, pudil1998novel}; it can be performed in two ways: feature set reduction (attributes or columns of a tabular dataset) or sample set reduction (records or rows of a tabular dataset).
Feature set reduction is common when training ML models wherein users either drop irrelevant features~\cite{featureselection2015jovic} or reduce them through dimensionality reduction techniques~\cite{fodor2002survey}. 
Sample set reduction is common during market segmentation~\cite{tynan1987market} wherein select groups of consumers are shortlisted to satisfy\cut{certain} segment specific goals.
These techniques have been used to combat selection bias, e.g., by visualizing how a\cut{selected} subset compares \edit{to} the original dataset~\cite{gotz2016adaptive,borland2019selection}.
\edit{In this work, we support subset selection tasks by presenting data quality and usage information to users.}

\subsubsection{\edit{Data Quality Assessments and Tools}}
Real-world datasets are often ``dirty'' and \edit{include} a variety of data quality problems~\cite{kaggle2019survey} that speculatively cost organizations trillions of dollars\cut{each year}~\cite{badDataCostsMillions,haug2011costs}. Data quality is crucial to ensure that systems using the data can perform the intended task in a performant, scalable, accurate, and unbiased \edit{manner}~\cite{halevy2009unreasonable, chu2016data, arbesser2016visplause, amazonHiringAISexist, mehrabi2021survey}. 
\add{Umbrich~et~al.~\cite{umbrich2015quality} point out that even low meta-data quality (missing meta-data) affects both the discovery and the consumption of the datasets.}
However, Kandel~et~al.~\cite{kandel2012enterprise} revealed that practitioners consider data wrangling tedious and time-consuming. Sambasivan~et~al.~\cite{sambasivan2021everyone} provided empirical evidence of ``Data Cascades'' -- compounding events that cause negative, downstream effects from data quality issues\cut{ because ``everyone wants to do the model work, not the data work.''}. A growing body of work, thus, has been focused on understanding and improving data quality to avoid the ``garbage in, garbage out'' problem~\cite{hazelwood2018applied, redman2018if}.

\add{The Data Nutrition Label~\cite{holland2020dataset} framework, like the Nutrition Facts label on food, highlights the ``ingredients'' of a dataset to help determine if the dataset is healthy for a particular statistical use case. \app provides similar at-a-glance information about a dataset's quality.}
\add{Tableau Prep~\cite{tableauprep}, OpenRefine~\cite{ham2013openrefine}, and Wrangler~\cite{kandel2011wrangler} are self-service data preparation tools that provide interactive affordances to explore, clean, structure, and shape the data before analysis.}\cut{In terms of data cleaning, numerous techniques have been studied based on integrity constraints~\cite{chu2013holistic}, qualitative methods~\cite{chu2016qualitative}, statistics~\cite{song2007missing,mayfield2010eracer,peng2021dataprep}, machine learning~\cite{yakout2013don}, and GUI-based interactions~\cite{raman2001potter,kandel2011wrangler,luo2020interactive,luo2020visclean,ham2013openrefine}.}\cut{OpenRefine~\cite{ham2013openrefine} and Wrangler~\cite{kandel2011wrangler} are similar tools for working with messy data: cleaning them; transforming them from one format into another; and extending them with web services and external data.}
Most relevant to \app is Profiler~\cite{kandel2012profiler}, a visual analysis tool for assessing quality issues in tabular data; Profiler applies data mining methods to automatically flag problematic data~\cut{(e.g., missing, erroneous, inconsistent, and extreme values, and key violations)} and also suggests coordinated summary visualizations for assessing the data in context, albeit without \app-like usage information.

Pipino~et~al.~\cite{pipino2002data} first presented sixteen objective and subjective dimensions for assessing data quality, that have since been extended~\cite{otto2009identification, sidi2012data, laranjeiro2015survey, schulz2017systematic, vaziri2019measuring, abedjan2016detecting} as there is ``no one size fits all set of metrics''~\cite{pipino2002data} and also ``no single dominant tool''~\cite{abedjan2016detecting}. 
Based on feedback from our domain expert interviewees, we model three of these dimensions (\emph{completeness}, \emph{free-of-error}, and \emph{objectivity}) to guide users about potential data quality concerns. 
While data cleaning (e.g., imputing missing values)~\cite{chu2013holistic,chu2016qualitative,song2007missing,mayfield2010eracer,peng2021dataprep,yakout2013don,raman2001potter,kandel2011wrangler,luo2020interactive,luo2020visclean,ham2013openrefine} is deferred to future work, \app currently provides novel GUI interaction affordances~\cut{(e.g., range sliders)} to support subset selection\cut{(e.g., filter out all low quality data attributes with a score below 50 out of 100)}.

\subsubsection{\edit{Collaboration among Users}\cut{in Data Preparation and Analysis}}

Prior work has\cut{extensively} examined human collaboration for information sharing and access~\cite{sambasivan2021everyone,zhang2020data, koesten2019collaborative, kim2017data, isenberg2011collaborative, bhardwaj2014datahub}. 
Social translucence theory~\cite{socialtranslucence2000erickson} describes designing digital systems to support 
collaboration \edit{in} large groups by making participants and their activities visible to one another.
These collaborations have direct~\cut{monetary} costs (e.g., employee salaries\cut{benefits in exchange for participants' time}) and indirect costs (e.g., time delays\cut{caused} due to user preferences and \edit{availability})~\cite{rogelberg2012wasted, thom2010you,thom2009s}, motivating efforts to mitigate\cut{associated} inefficiencies.

In the visualization domain, collaborative systems~\cite{isenberg2011collaborative,willett2011commentspace,wang2015docuviz,manyeyes2007viegas} have focused on supporting both synchronous~\cite{viegas2006communication} and asynchronous~\cite{heer2007voyagers,willett2007scented} models.\cut{In the data science domain, Zhang~et~al.~\cite{zhang2020data} found that data science workers extensively collaborate during a data science project, albeit much less around datasets than code, identifying a need to understand how users share and reuse data with limited interaction with the creator~\cite{koesten2019collaborative}.} 
In the data science domain, Auto-suggest recommends data preparation steps in computational notebooks~\cite{yan2020auto}, albeit based on previously written code and not usage logs. Presenting users with readable, reusable code (over manual programming) for data wrangling in computational notebooks has been shown to increase user efficiency, trust, and confidence~\cite{wrex2020drosos}.

In the database domain, providing a rich set of \emph{starter queries} \edit{from} experts has been shown to empower non-experts to use SQL for data analysis with ad hoc databases~\cite{howe2011automatic}. 
\cut{Gebru~et~al.~\cite{gebru2021datasheets} found that data documentation, a crucial aspect of facilitating collaboration~\cite{bhardwaj2014datahub, buneman2001and}\add{,} also suffers from a lack of standards and conventions.}
In the machine learning (ML) and artificial intelligence (AI) community, Almahmoud~et~al.~\cite{almahmoud2021teams} studied how different team members communicate about the quality of ML models; they found a mismatch between user-focused and model-focused notions of performance and a difficulty in understanding concerns beyond one's\cut{own} role. 
Ehsan~et~al.~\cite{ehsan2021expanding} found that social transparency can potentially calibrate trust in AI, improve decision-making, facilitate organizational collective actions, and cultivate holistic explainability. 
This prior work motivated us to model \add{and present} usage characteristics from prior utilization of data\cut{and present them to users} to \add{help} \edit{users} during subset selection.
\app's modeling of usage information is a key novelty in addition to data quality.



\subsection{Data,~\cut{Meta-data,}Analytic Provenance, and Guidance}
With the proliferation of big data, more~\cut{and more}data~\cite{deng2017data} and meta-data~\cite{metadata2022ulrich} (e.g., application logs) are being stored and processed.
In database contexts, \add{``data provenance'' is used to reason about the current state of a data object~\cite{zafar2017trustworthy}}, e.g., \edit{describe} its provenance characteristics\cut{and utility} (``Data Descriptors''~\cite{schulz2017systematic}), \add{study secure provenance schemes and associated issues (Zafar~et~al.~\cite{zafar2017trustworthy}), and} document the purpose, performance, safety, and security\cut{, and provenance information} of data and models (``FactSheets''~\cite{arnold2019factsheets}) \add{and computational workflows (Wings/Pegasus~\cite{kim2008provenance})}.

\add{Provenance information has also been explored for dataset reuse. Koesten~et~al.~\cite{koesten2020dataset} described a case study that determined how dataset provenance information in the form of GitHub-specific engagement metrics (e.g., the number of forks, watchers, stars, and committers) can predict a dataset's likeliness of reuse. Facilitating data navigation and fostering reuse, many open data portals utilize and/or present provenance information~\cite{portal2021cern} to users including ``most-viewed''~\cite{portal2012ogdindia}, ``high-value''~\cite{portal2012ogdindia}, and ``trending''~\cite{portal2016eu, kaggle2022misc} datasets and access to example projects and user discussion boards~\cite{koesten2020dataset, kaggle2019survey}.}

In visualization contexts and \add{most relevant to this work,} ``analytic provenance'' \edit{tracks} users' interactions with a visualization system to provide an overview of their sensemaking process~\cite{north2011analytic}. 
This information is then used for product and user behavior analytics purposes such as generating personalized content~\cite{wedig2006large}, mitigating biased analytic behaviors~\cite{narechania2021lumos, wall2021left}, increasing user trust~\cite{arnold2019factsheets}, recommending alternate design choices~\cite{dumais2014understanding, liu2020data}, and visual data exploration~\cite{willett2007scented}. 
In HCI contexts, traces of prior interactions have been applied in revisiting common regions of a page using scrollbar history~\cite{alexander2009revisiting}, tracking user interactions with documents~\cite{hill1992edit}, facilitating groupware coordination~\cite{gutwin2002traces}, and tracking user focus while browsing a webpage using eye-\cut{tracking}~\cite{nielsen2010eyetracking} and mouse-tracking~\cite{arroyo2006usability}\cut{gear}. 

\edit{Characterizing provenance in visualization and data analysis,} Ragan~et~al.~\cite{ragan2015characterizing} present an organizational framework comprising five types and six purposes of analytic provenance\add{;}\cut{of which,} our work most closely falls in the ``Data'' type (\emph{the history of changes and movement of data, which can include subsetting, data merging, formatting, transformations, or execution of a simulation to ingest or generate new data}) and ``Collaborative Communication'' purpose (\emph{communicating and sharing data, information, and ideas with others who are conducting the same analysis}). 
\cut{Ceneda~et~al.~\cite{ceneda2016characterizing} define guidance in visual analytics as \emph{``a computer-assisted process that aims to actively resolve a knowledge gap encountered by users during an interactive visual analytics session.''}} 
Ceneda~et~al.~\cite{ceneda2016characterizing} characterize guidance in visual analytics along three degrees (orienting, directing, prescribing) that specify the extent to which guidance is required by users and provided by the system. DataPilot provides the least intrusive, ``orienting'' guidance through visual cues hinting at the good and bad aspects of data quality and usage.
\cut{No prior work has used analytic provenance to provide such guidance during data preparation and we are the first to build and evaluate a tool that facilitates it.}



\section{Design Study and Expert Interviews}
\label{sec:interview}


To learn about user tasks and challenges associated with \add{subset selection} as part of data preparation before analysis, we adopted a design study methodology~\cite{designstudy2012sedlmair} comprising semi-structured interviews and brainstorming and feedback sessions centered around iterative prototype design and development.

We interviewed 14 data workers (\textcolor{DE}{\pde{1-14}}; 10 males; 4 females) from a large technology company to learn about their current tasks,  challenges, and requirements. These experts comprised \emph{data engineers (3: \textcolor{DE}{\pde{1-3}})}, \emph{data analysts (3: \textcolor{DE}{\pde{4-6}})}, \emph{data scientists (4: \textcolor{DE}{\pde{7-10}})}, \emph{digital marketers (3: \textcolor{DE}{\pde{11-13}})}, and a \emph{user interface engineer (1: \textcolor{DE}{\pde{14}})}, with relevant working experience ranging 3-34 years (median: 15.5; $\mu$=16.86). We recruited these experts from within our enterprise network using a combination of targeted emailing and snowballing strategies. 
We conducted these interviews remotely using Microsoft Teams~\cut{(a teleconferencing application with video, screensharing, and recording capabilities)} over the course of twelve weeks; the earlier sessions were more frequent and spontaneous than the latter weekly sessions.

\add{In total, there were eight sessions (including three follow-up sessions) that lasted} \edit{about 45 minutes each}, with 1-7 domain experts and 2-3 study administrators participating on the calls (sessions \ses{1}:~\textcolor{DE}{\pde{2}}, \ses{2}:~\textcolor{DE}{\pde{7,8}}, \ses{3}: \textcolor{DE}{\pde{3,9}}, \ses{4}: \textcolor{DE}{\pde{11,12}}, \ses{5}: \textcolor{DE}{\pde{1,4-6,10,13,14}}, \ses{6}: \textcolor{DE}{\pde{2}}, \ses{7}: \textcolor{DE}{\pde{4,5}}, \ses{8}: \textcolor{DE}{\pde{14}}). \add{These experts had an established working relationship and were aware of each others' strengths and expertise, unlike a group meeting with complete strangers. Discussing problems and solutions from such a cohesive group was valuable for us. For instance, the program manager often relied on the engineers' opinion regarding the technical feasibility of an idea; similarly, engineers alluded to the program manager for questions around prioritization and timelines.} 
\edit{During these sessions, one study} administrator shared a PowerPoint presentation and another took notes while facilitating a conversation structured around the following questions: 
\begin{enumerate}[nosep]
    \item \emph{``What (kind of) tasks related to data and analytics do you accomplish on a day-to-day basis?}
    \item \emph{``What (types of) data do you work with? How do you prepare this data? What tools do you use?''}
    \item \emph{``(How) do you collaborate with other people within/outside your organization over your tasks?''}
    \item \emph{``What are some challenges that you face while working on your tasks? How do you overcome them?''}
\vspace*{-6pt}
\end{enumerate}

\subsection{User Tasks and Challenges}
We coded the domain experts' spoken quotes using inductive thematic analysis~\cite{boyatzis1998transforming}, categorized them based on their roles, tasks performed, challenges faced, and opinions about quality and usage.~\cut{To maintain confidentiality, we do not include individual quotes in supplemental material but share selected findings.}
\add{We make these available in supplemental material, albeit anonymized.}

\add{We found that tasks varied quite a bit based on the different user roles.}
\edit{Data analysts \add{select key performance indicators (attributes) and subset relevant records to} design interactive dashboards~\cut{(e.g., to monitor real time sales)} and prepare reports~\cut{(e.g., on monthly sales)} for business executives.
Marketers \add{subset customer behavior data and demographic data by devising} strategic segmentation rules, e.g. a filter criteria to shortlist customers for running targeted digital marketing communications.
Data scientists select \add{a subset of} relevant attributes (features) and records (observations) from existing datasets to build predictive models.~\cut{(e.g., to forecast future sales)}
Data engineers help other users (e.g., marketers) prepare their data for various analytical or operational uses; they also monitor the organization's data repositories to control their storage and cost footprints.
User interface (UI) engineers help design scalable interactive web applications for various user-facing use-cases.}

We also learned many general as well as domain-specific challenges that these experts faced while performing their tasks. 
Everyone~\cut{reinforced our initial hunch}\edit{communicated to us} that data quality is important, e.g., \emph{``Are the data complete? Correct? Unbiased? Having the correct datatypes?''} and demanded that users be made more aware of quality issues \edit{with additional guidance} during data preparation and analysis. 
In particular, \add{\textcolor{DE}{\pde{1}}~(data engineer) acknowledged that skewness is an important problem, \emph{``What if all of the data came from Northern California (in USA)?''}}
\edit{\textcolor{DE}{\pde{11}}~(marketer) acknowledged from a marketing standpoint that segmenting a customer dataset by a skewed (and/or sparse) attribute can result in suboptimal targeting of communications; \textcolor{DE}{\pde{7}}~(data scientist) affirmed this concern from an ML~standpoint.}

\add{With respect to sparseness, \textcolor{DE}{\pde{3}} (data engineer) shared from experience that missing data can be, \emph{``an empty string (``''), ASCII-space only string (``~~''), null string (``null''), missing string (), which is different from [an] explicitly null [value]; depending on how the data comes in, these can [mess] up your segmentation logic.''}}
\add{\textcolor{DE}{\pde{2}} (data engineer) raised the computational cost of ``bad'' filtering strategies (e.g., if the first filter minimizes the search-set by 95\%; then users often undo that operation by running a new query which is expensive). Instead, \textcolor{DE}{\pde{2}} suggested showing (quality) insights upfront as it may \emph{``instill feelings of curiosity and care in the user''} and help catch the ``bad'' filter(s).}
Another challenge commonly faced was selecting \emph{``important/best/effective/relevant''} attributes and records from a large dataset for preparing effective dashboards (\textcolor{DE}{\pde{4}}), training accurate and fair ML models (\textcolor{DE}{\pde{9}}), and defining segmentation rules for running successful marketing communications (\textcolor{DE}{\pde{11}}).
 
Many domain experts noted that collaboration during data preparation could be tedious, and thus advocated for better tools to support this process.
\textcolor{DE}{\pde{11}} (marketer) mentioned that conveying their data-related requirements to the data engineers is often a tedious process, requiring multiple iterations that take time and critical information may also get lost during the exchange, advocating for an interactive self-service tool.
~\textcolor{DE}{\pde{1}} (data engineer) suggested provisioning a visual report-card similar to LinkedIn's~\cite{linkedin} profile completeness, but for datasets as that \emph{``can also give decision-makers outside of users who are executing these tasks an idea of how good their data is, prompting them to enforce data policies.''} 
Some suggested surfacing insights based on the usage of data within the organization.
\textcolor{DE}{\pde{3}}~\add{(data engineer)} first sighed that \emph{``even if the ingested data might be of low quality, users sometimes don't really care; they aren't actually using it. We just have to accept that.''} However, \edit{\textcolor{DE}{\pde{3}} also noted} that, \emph{``the consumption [usage] of data can be really important here as it can provide a different kind of awareness''} and \edit{defined} two types of usage dimensions that may be beneficial from a business standpoint: \emph{``the `boolean logic' [filter criteria] to perform customer segmentation (e.g., `Age' > 18 is a common criteria to target adults) and `projections' [referenced attributes], e.g., the `Name' and `Email' attributes are frequently utilized \add{to target users} during campaigns.''}
\add{\textcolor{DE}{\pde{3}} further noted that, \emph{``If I'm defining segments and referencing fields that might not be good [or] garbage, with more nulls than expected, I want to know them. So I love the usage aspect here.''}}

\textcolor{DE}{\pde{9}}~\add{(data scientist)} suggested that \edit{data provenance (e.g., \emph{``when, where, and by whom the datasets were last used''}) can be used to assist with data housekeeping as \emph{``there are several unused, low-quality datasets just lying around that may be archived.''}}  
\add{\textcolor{DE}{\pde{4}} (data analyst) suggested generalizing this idea by curating \emph{``a KPI (key performance indicator) catalog comprising different metrics, how they are calculated, how often they are used in dashboards and in segments and in journeys, and these scores would sit right along the side.''}}
\textcolor{DE}{\pde{2}}~\add{(data engineer)} concluded that \emph{``it takes hard work to get data into a high quality form so any kind of re-use is a good thing, whether it is the output or the workflow used to obtain that output.''} Inspired by these expert endorsements, we confirmed that guiding users about the quality and usage characteristics of data can be a promising way to help \edit{them} better accomplish their tasks.

\vspace*{-6pt}
\subsection{Design \add{Requirements} Exercises}
After understanding the user tasks and challenges, we conducted two follow-up sessions, one each with \textcolor{DE}{\pde{2}} (data engineer) and \textcolor{DE}{\pde{4,5}} (data analysts), \add{to discuss architecting \app and handling data in terms of access (authentication and ethical considerations), processing (scalable computation strategies), and persistence (optimal storage mechanisms).}
Through a follow-up session with \textcolor{DE}{\pde{14}} (user interface engineer), we conducted design exercises wherein we sketched low-fidelity designs digitally as well as on paper and presented them for feedback. These sketches included visualizations, widgets, layouts, workflows, and interactions in the UI with an intention to catch errors that could surface later. We brainstormed on the pros and cons of each design resulting in multiple changes and refinements. For some designs, we developed rapid software prototypes with a dual purpose of exploring potential technologies (such as software libraries) and evaluating \edit{their feasibility}, which further helped discard less-useful designs, refine the user tasks to be supported, and distill design goals, described next.

\vspace*{-6pt}
\subsection{Design Goals} 
We derived \textbf{six} key design goals from our expert interviews that drove the design and development of \app.

\paragraphHeadingSpace\bpstart{DG1. Facilitate data preparation \edit{and} visual data analysis, in situ.}
Data preparation is a necessary step \emph{before} analysis. However, users must often revisit the data preparation step even \emph{during} analysis. We derived this core design goal to support both aspects within the same tool (in situ), minimizing unnecessary learning of and switching between multiple tools and windows. In particular, \app supports building a visualization dashboard
from a subset selected after navigating a large, unfamiliar tabular dataset.

\paragraphHeadingSpace\bpstart{DG2. Model data quality and usage information as standardized scores.}
\cut{As per our domain experts, non-technical end-users (e.g., business users) who may lack the necessary skillset to execute complex queries often had to rely on data engineers for relevant subsets or associated auxiliary information.}
\edit{Because non-technical marketers (\textcolor{DE}{\pde{11}}) often had to rely on data engineers (\textcolor{DE}{\pde{1}}),} we derived this goal to model each dimension of quality and usage information into a standardized score (out of 100) using smart, heuristically determined rules. This scoring strategy will enable comparisons and aggregations across dimensions, making it a ``self-service'' experience for users, minimizing their (over)reliance on and tedious exchanges with other users. Power-users can still interactively specify the constraints by themselves, gaining some configuration control.

\paragraphHeadingSpace\bpstart{DG3. Provide visual guidance about data quality and usage while balancing user agency and control.}
This goal translates to providing guidance to users about their data's quality and usage characteristics. To balance user agency and control as desired by the domain experts, we provision the least intrusive ``orienting'' guidance~\cite{ceneda2016characterizing}, providing visual hints (e.g., highlight missing values; show the computed scores) without disrupting users' \edit{analysis}.

\paragraphHeadingSpace\bpstart{DG4. Provide interaction and specification affordances for data discovery, subset selection, and visualization dashboard creation.}
A key novelty, this goal \edit{involves} providing ``self-service'' interaction affordances (e.g., sort and filter UI controls) to help users inspect quality and usage information of data attributes and records in the original datasets, as well as the selected subsets; and specification affordances to assign data attributes to visual encodings (e.g., X axis, Y axis) to build a visualization dashboard.

\paragraphHeadingSpace\bpstart{DG5. \edit{Enable control and context through configurability.}}
\textcolor{DE}{\pde{3}} (data engineer) had noted that quality and usage information may not always be available or applicable (e.g., there is no usage information yet for a newly uploaded dataset). Hence, this goal translates to providing options within the \app UI to configure the visibility of different components pertaining to quality and usage (i.e., one, none, or both), enabling multiple levels of control and context across users and applications.

\paragraphHeadingSpace\bpstart{DG6. \edit{Design for scalability and performance.}}
\textcolor{DE}{\pde{14}} (UI engineer) had reiterated the challenges associated with presenting large amounts of data on the UI (e.g., slow load times and sluggish interaction experience). We derived this goal to design a performant frontend application, offloading complex operations to a scalable backend server for an overall fluid user experience~\cite{elmqvist2011fluid}.


\section{Modeling Data Quality and Usage}
\label{sec:metadata}


Based on the design study described in Section~\ref{sec:interview}, we now discuss how we modeled quality and usage information.
Strategies to model quality and usage information depend on the types of data, users, and applications. In this work, we focus on a dashboard application in which users first upload a tabular dataset, prepare a relevant subset (by selecting relevant attributes and filtering out irrelevant records), and use it to create visualizations that constitute a dashboard. User-defined constraints and interactions with GUI elements (e.g., attribute-level selection checkboxes, range sliders for record-level filters) are used to model, quantify, and also interact with quality and usage information.

\subsection{Quality}
\edit{Based on existing challenges from our domain experts around data skewness (\textcolor{DE}{\pde{1}, \pde{7}, \pde{11}}), sparseness (\textcolor{DE}{\pde{1}, \pde{2}, \pde{3}, \pde{11}}), and incorrectness (\textcolor{DE}{\pde{2}}) and prior work~\cite{pipino2002data},} we modeled three dimensions of quality at an attribute-level: \emph{completeness}, \emph{correctness}, \emph{objectivity} and two dimensions at a record-level: \emph{completeness}, \emph{correctness}~\textbf{(DG2)}.

\subsubsection{Attribute-level Quality Dimensions}\hfill
\paragraphHeadingSpace\bpstart{Completeness} is the percentage of \emph{non-missing values} among an attribute's values, e.g., if 10 of 50 attribute values are \emph{nulls} or \emph{empty strings}, its completeness \edit{is} 100*(50-10)/50 = 80\%. Completeness can help users detect sparse attributes that can, for example, alter \edit{how well} ML algorithms \edit{can} make accurate predictions.

\paragraphHeadingSpace\bpstart{Correctness} is the percentage of \emph{correct values} among an attribute's values, e.g., if 5 out of 50 attribute values are incorrect, then its correctness \edit{is} 100*(50-5)/50 = 90\%. To calculate correctness, \edit{businesses can preconfigure}\cut{DataPilot is preconfigured with} SQL-like constraints~\edit{in the \app source code} through relations~($>$,$<$,$=$), range~(BETWEEN), pattern matching~(LIKE), and membership~(IN) operators; e.g., ``\texttt{\textcolor{sqlcolor}{SELECT} \textcolor{sqlcolor}{Count(*)} \textcolor{sqlcolor}{WHERE} email \textcolor{sqlcolor}{NOT LIKE} `\%\_@\_\_\%.\_\_\%'}'' computes the number of records with incorrect email addresses. With correctness, users can assess the accuracy of individual attributes.

\paragraphHeadingSpace\bpstart{Objectivity} is the extent that values conform to a target distribution, e.g., if the \emph{Gender} attribute has 120 males and 45 females, then it is evidently skewed towards males and hence, from a gender equality standpoint, not objective. We utilize Wall~et~al.'s Attribute Distribution (AD) metric~\cite{wall2017warning} for measuring the deviation between the observed and the expected objective distribution (baseline); AD scores range from [0,1] so we standardize them by multiplying by 100. With this dimension, users can detect anomalous phenomena, e.g., 
 if the majority of applicants are of a specific gender, 
against expectations.~\edit{Like \emph{correctness}, businesses can preconfigure \emph{objectivity} constraints in the \app source~code.}

\subsubsection{Record-level Quality Dimensions}\hfill

\paragraphHeadingSpace\bpstart{Completeness} is the percentage of \emph{non-missing} values in each dataset record, e.g., if a record has 50 values (one for each attribute), 20 of which are \emph{nulls} or \emph{empty strings}, then its completeness \edit{is} 100*(50-20)/50 = 60\%. With this dimension, users can, e.g., discard sparse customer profiles (records) for marketing campaigns where success is determined by the~\add{profiles'} richness.

\paragraphHeadingSpace\bpstart{Correctness} is the percentage of correct values in each record, e.g., if a record has 50 \edit{attribute} values, 15 of which are incorrect (based on set constraints), its correctness is 100*(50-15)/50 = 70\%. With this dimension, marketers can discard customer profiles (records) with invalid email addresses and social media handles that are useless for running marketing campaigns.

\paragraphHeadingSpace\bpstart{Objectivity} is inapplicable for record-level dimensions as each record comprises values from different, incomparable attributes.

\subsubsection{Overall Scores: Aggregations and Customizations.}\hfill

\paragraphHeadingSpace\noindent We compute a configurable heuristics-based \emph{overall} score for each attribute and record that defaults to the arithmetic \emph{mean} of the corresponding dimensions. 
Based on work by Vaziri~et~al.~\cite{vaziri2019measuring}, users can specify different weights for different dimensions (e.g., a user might prefer an overall dimension that comprises 75\% completeness and 25\% correctness, \edit{and ignores} objectivity) as well as different attributes and records (e.g., a digital marketer may want to weigh the ``Phone'' attribute more than ``Email Address'' \edit{for} correctness).

\subsection{Usage}
Based on positive feedback from our domain experts, we modeled usage information \textbf{(DG2)} across three dimensions at an attribute-level: \emph{in-subsets}, \emph{in-filters}, and \emph{in-visualizations} and one dimension at a record-level: \emph{in-subsets}.

\subsubsection{Attribute-level Usage Dimensions}\hfill

\paragraphHeadingSpace\bpstart{In-subsets} score of an attribute is the percentage of users who selected that attribute to be in their subset for later use, e.g., if 15 out of 20 users select a feature for training an ML model, then the \emph{in-subsets} score \edit{is} 100*15/20 = 75\%. With this dimension, new users can, e.g., perform quick and efficient analysis by selecting highly used (important?) attributes based on subsets of prior users.

\paragraphHeadingSpace\bpstart{In-filters} score is the percentage of users who~applied a filter on that attribute, e.g., by choosing a multiselect dropdown option (\emph{Gender=``Female''}) or dragging range slider handles (\emph{Age $\in$ [40,50]}). With this dimension, digital marketers can, e.g., determine segmentation rules (filter criteria to pick certain customer profiles) for running marketing campaigns based on previous ones. \add{Note that \emph{in-filters} is not a subset of \emph{in-subsets}; users can filter (or not) by an attribute and (not) select it in their subset and vice versa.}

\paragraphHeadingSpace\bpstart{In-visualizations} score is the percentage of users who assigned that attribute to one or more visual encodings (e.g., X axis) and utilized the resultant visualization in a dashboard. With this dimension, users can refer to popular (important?) attributes from past business reports to assist with the design of present ones.

\subsubsection{Record-level Usage Dimensions}\hfill

\paragraphHeadingSpace\bpstart{In-subsets} score of a record is the percentage of users who selected that record to be in their subset (as a result of filters). With this dimension, users can, e.g., select a subset of popular (important?) records and re-run new marketing campaigns by targeting customer profiles (records) from previous successful campaigns.
This dimension is in essence the same as record-level \emph{in-filters} and \emph{in-visualizations} usage dimensions because \app treats a filtered dataset as the selected subset that is used in the visualization.

\subsubsection{Overall Scores: Aggregations and Customizations}\hfill

\paragraphHeadingSpace\noindent Like overall quality, we computed a heuristics-based \emph{overall} score for each attribute and record, but as the \emph{maximum} of the constituent dimensions. Because attributes are seldom utilized simultaneously in subsets, filters, and visualizations, choosing \emph{mean} would result in low scores that would be ineffective and demotivating for the user; hence, we chose \emph{maximum}. Users can ignore one or more usage dimensions, e.g., \emph{In-filters} usage, if it is irrelevant to their use-case.


\section{\app User Interface}
\label{sec:datapilot}


\label{subsection:user-interface}

\begin{figure*}
    \includegraphics[width=\textwidth]{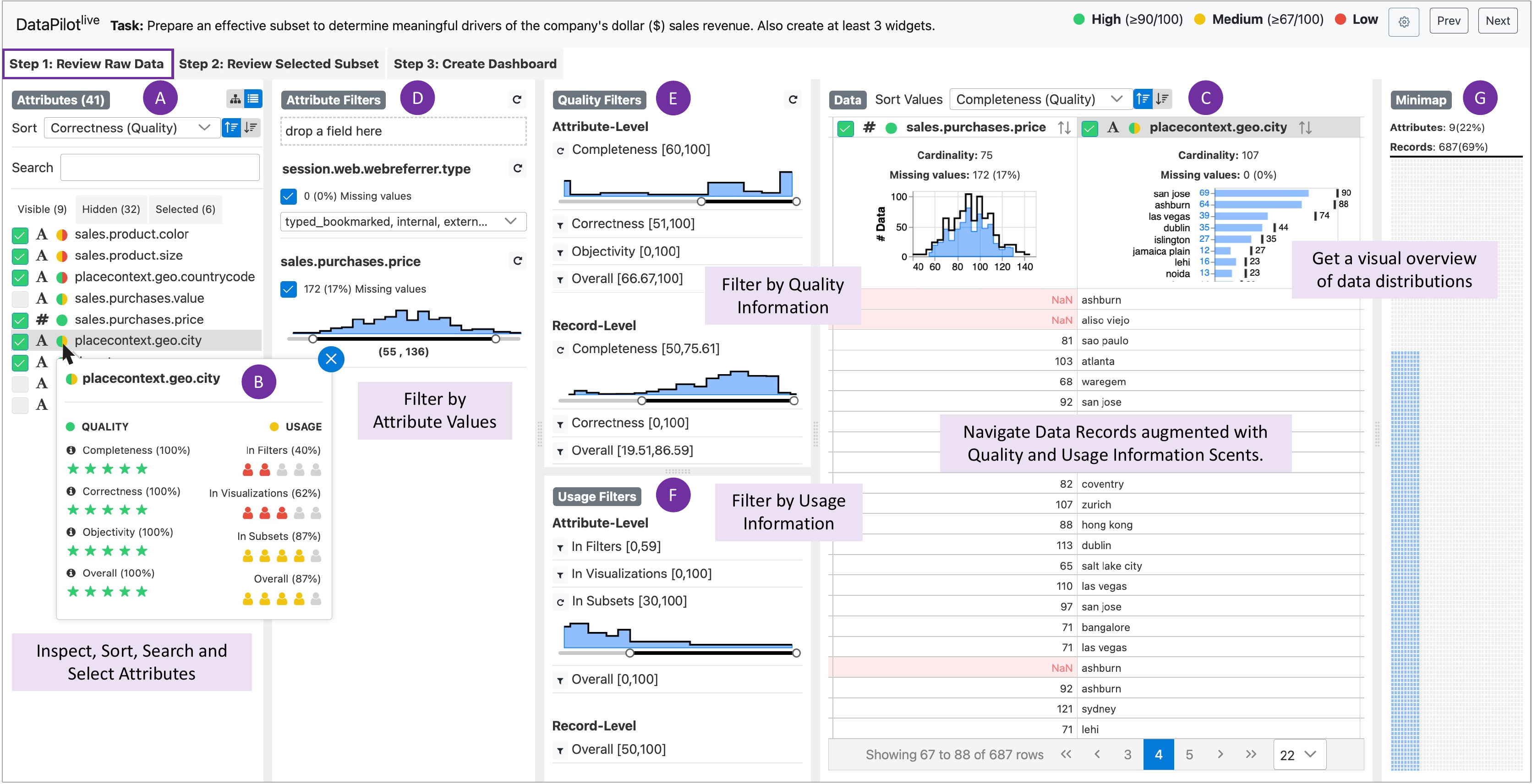}
    \caption{\textbf{\edit{The} \app user interface} showing Step 1 (Review Raw Data) of the three-\edit{step} workflow. Users~\add{can} inspect the list of dataset attributes (\textbf{\edit{A.~Attribute View}}), inspect quality and usage dimension scores for an attribute (\textbf{\edit{B.~Attribute Detail View}}), visualize attribute distributions and navigate dataset records (\textbf{\edit{C.~Data View}}), incrementally filter records by attribute values (\textbf{\edit{D.~Attribute Filter View}}), incrementally filter attributes and records by \add{both} quality (\textbf{\edit{E.~Quality Filters View}}) and usage dimensions (\textbf{\edit{F.~Usage Filters View}}) to reduce the search space, get a visual summary of this filtered dataset (\textbf{\edit{G.~Minimap View}}), and explicitly select attributes (\textbf{\edit{A.~Attribute View}}) and records (automatically selected based on filters) \edit{for} the desired subset.}
    \label{fig:step-1}
    \Description{Screenshot of the landing page of the DataPilot user interface showing Step 1 (Review Raw Data) of the three-stepped visual data preparation and analysis workflow. At the top, there are three tabs: (1) Review Raw Data, (2) Review Selected Data, and (3) Create Dashboard with the first tab selected. The area below these tabs can be viewed as a table with four columns. Reading from left to right, Attribute View (A): where users inspect the list of dataset attributes and their respective quality and usage dimension scores; Attribute Filter View (D): where users incrementally filter data records based on attribute values; Quality Filters View (E) and Usage Filters View (F): where users incrementally filter attributes and records by quality and usage, respectively; Data View (C): where users visualize attribute distributions and navigate dataset records; and Minimap View (G): where users get a visual overview of their filtered dataset. The Quality Filters View (E) and Usage Filters View (F) in the third column are stacked on top of each other and each occupy exactly half the column height.}
\end{figure*}

\begin{figure*}
    \centering
    \includegraphics[width=\textwidth]{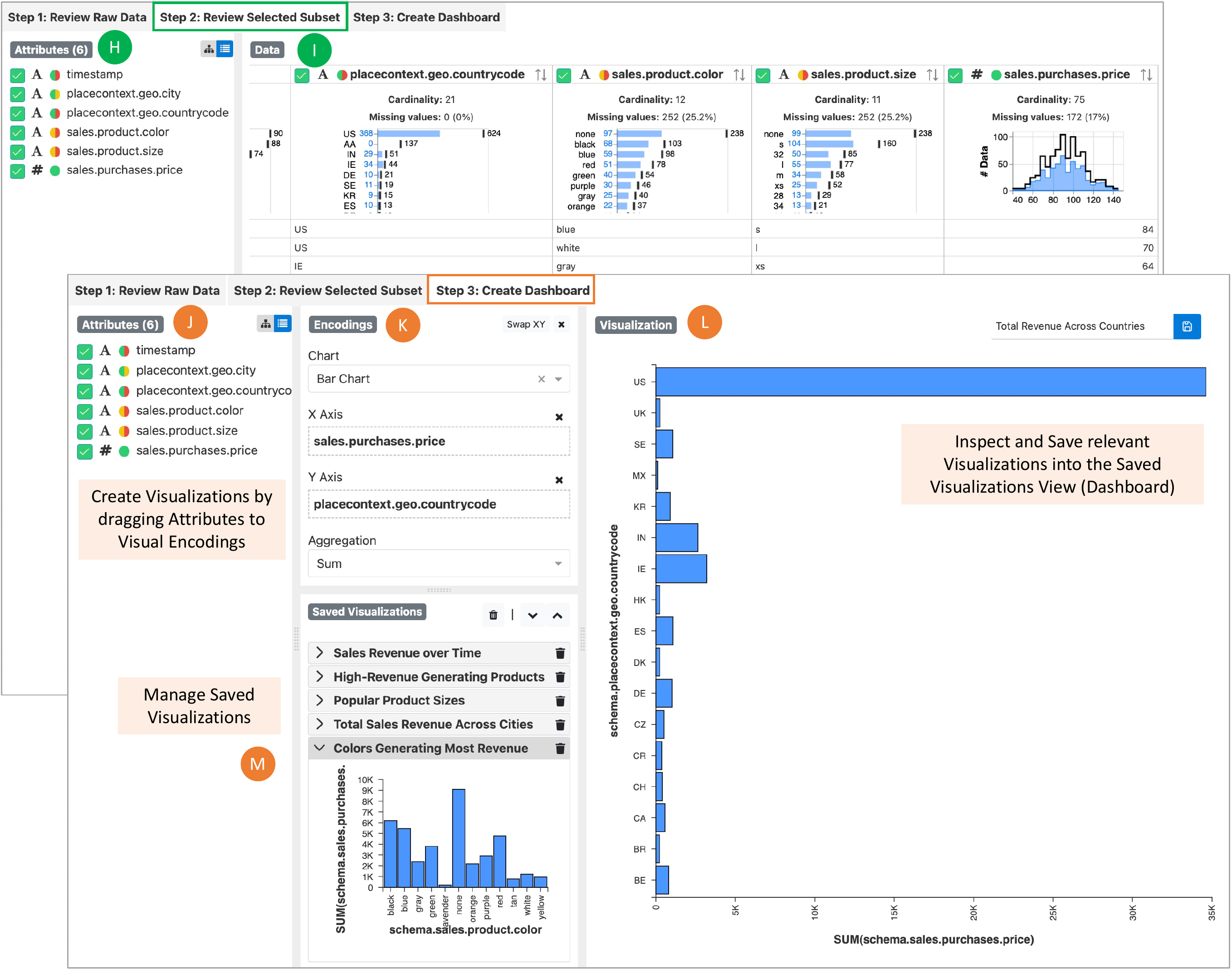}
    \caption{\textbf{\app Step 2 (Review Selected Subset) and Step 3 (Create Dashboard)}. Users review their selected attributes (\textbf{\edit{H.~Attribute View}}) and records (\textbf{\edit{I. Data View}}), assign attributes (\textbf{\edit{J. Attribute View}}) to encodings (\textbf{\edit{K. Encodings View}}), inspect the resulting visualization (\textbf{\edit{L. Visualization Canvas}}) and save it to the dashboard (\textbf{\edit{M. Saved Visualizations}}). Users can freely navigate between the three steps.}
    \label{fig:step-2-3}
    \Description{Step 2 (Review Selected Subset) and Step 3 (Create Dashboard) of DataPilot's three-stepped visual data preparation and analysis workflow. Step 2's screenshot is under Step 3's screenshot such that the former is partially visible and the latter is completely visible. In Step 2, a two-column layout is shown; reading from left to right, Attribute View (H): where users review their selected data attributes; and Data View (I): where users review their selected data records. In Step 3, a three-column layout is shown; reading from light to right, Attribute View (J): where users can see their list of selected attributes; Encodings View (K): where users can assign the selected attributes to visual encodings (e.g., X and Y axes) to create charts (e.g., scatterplot); Visualization Canvas (L): where users see and interact with the resultant visualizations. The second column also includes a Saved Visualizations View (M) below the Encodings View (K): wherein users can see and modify their saved visualizations.}
\end{figure*}

To support subset selection and analysis in the same tool\add~(\textbf{DG1}), we designed the \app UI to have a three-step workflow with each step navigable from \edit{others via} the top left corner (Figures~\ref{fig:step-1} and~\ref{fig:step-2-3}).
We finalized this design based on pilot studies with four users; Section~\ref{sec:alternatives} discusses some of the alternate, discarded designs.


\subsection{Step 1: Review Raw Data}
This step, also the landing page~\add{of \app}, enables users to analyze a dataset and select a relevant subset (Figure~\ref{fig:step-1}). It consists of the following views:

\paragraphHeadingSpace\noindent\textbf{(A) Attribute View} shows all attributes as a flattened list (\includegraphics[height=1.2\fontcharht\font`\B]{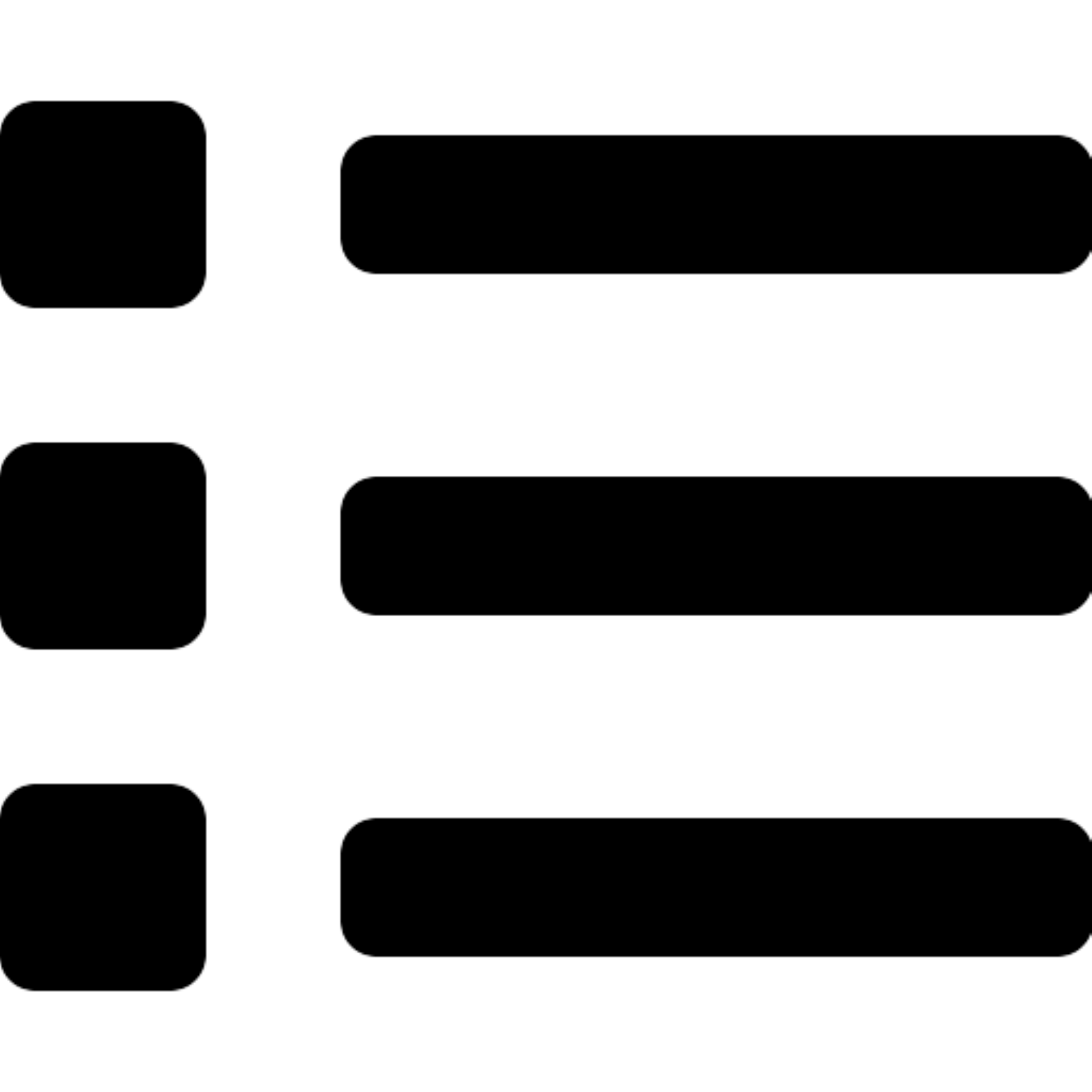}) or as a nested list 
(\includegraphics[height=1.2\fontcharht\font`\B]{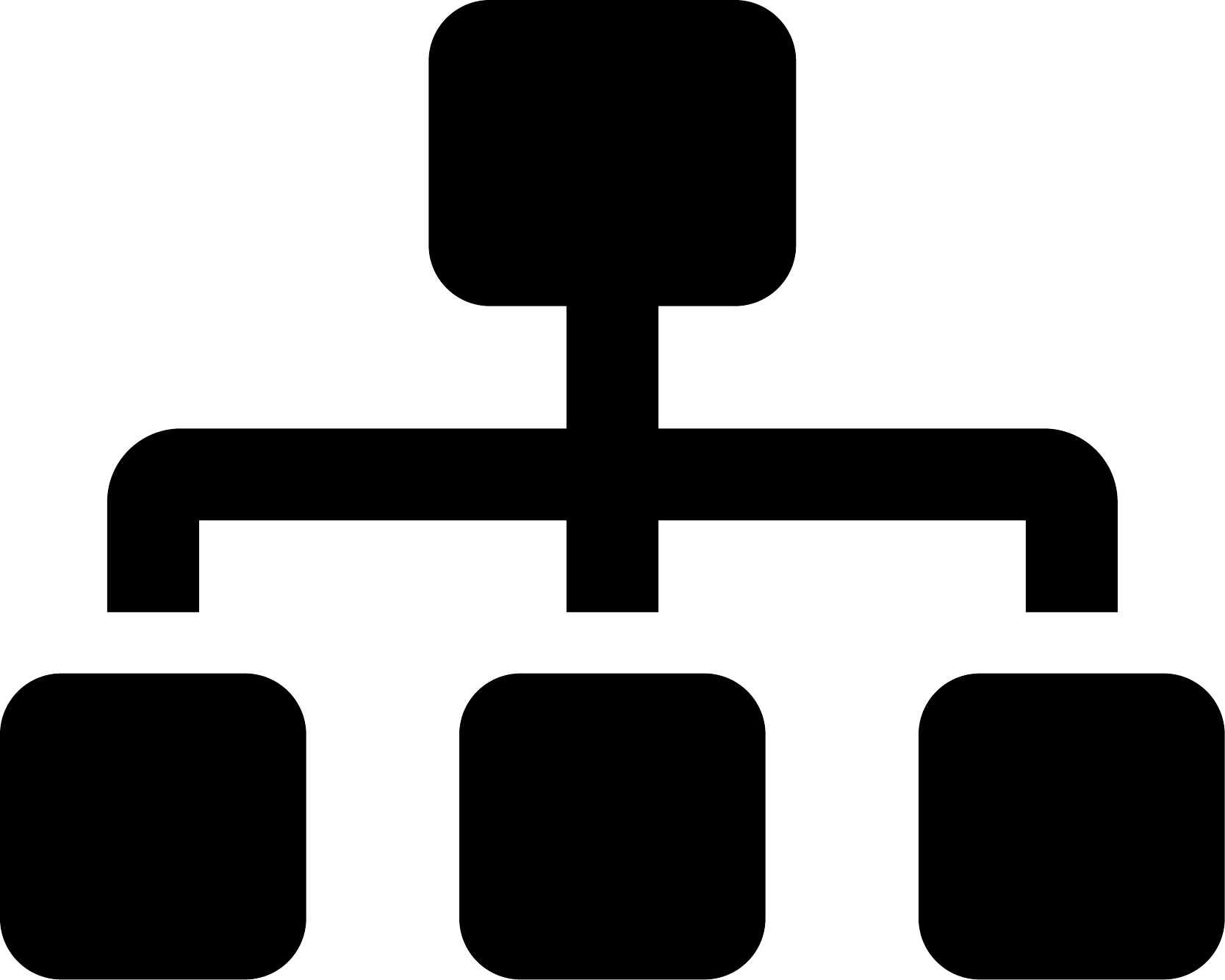}), 
the latter being helpful for hierarchical datasets. To efficiently display~\add{a} large number of attributes, we utilize the virtual scrolling principle preventing unnecessary rendering of objects not visible in the viewport (\textbf{DG6}). A search field allows quick attribute lookup via keyword-based queries. Users can also sort by quality and usage dimensions at the attribute-level. Each list item shows the attribute's name (e.g., ``sales.product.name''), its datatype (e.g., 
\includegraphics[height=1.2\fontcharht\font`\B]{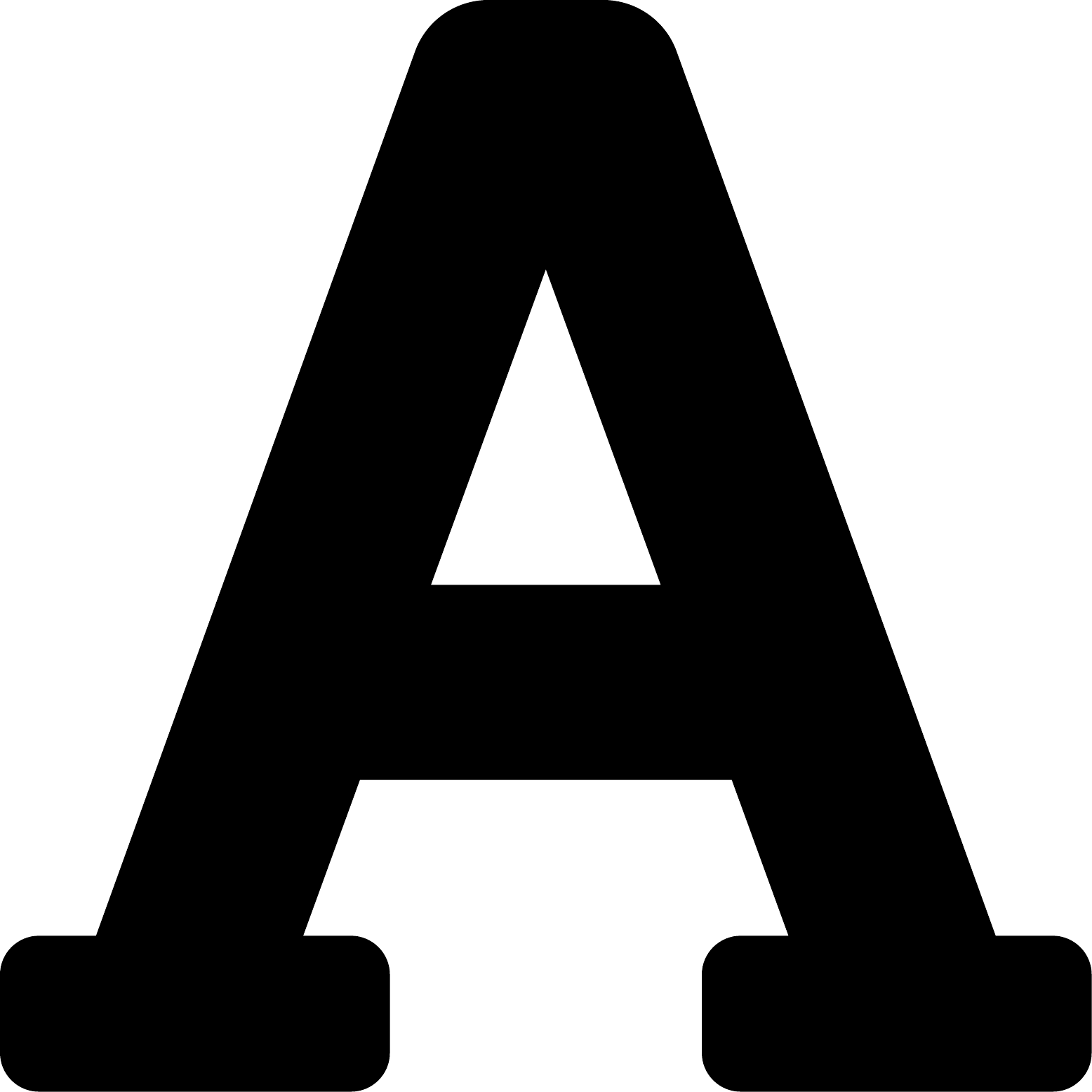}: 
Categorical, 
\includegraphics[height=1.2\fontcharht\font`\B]{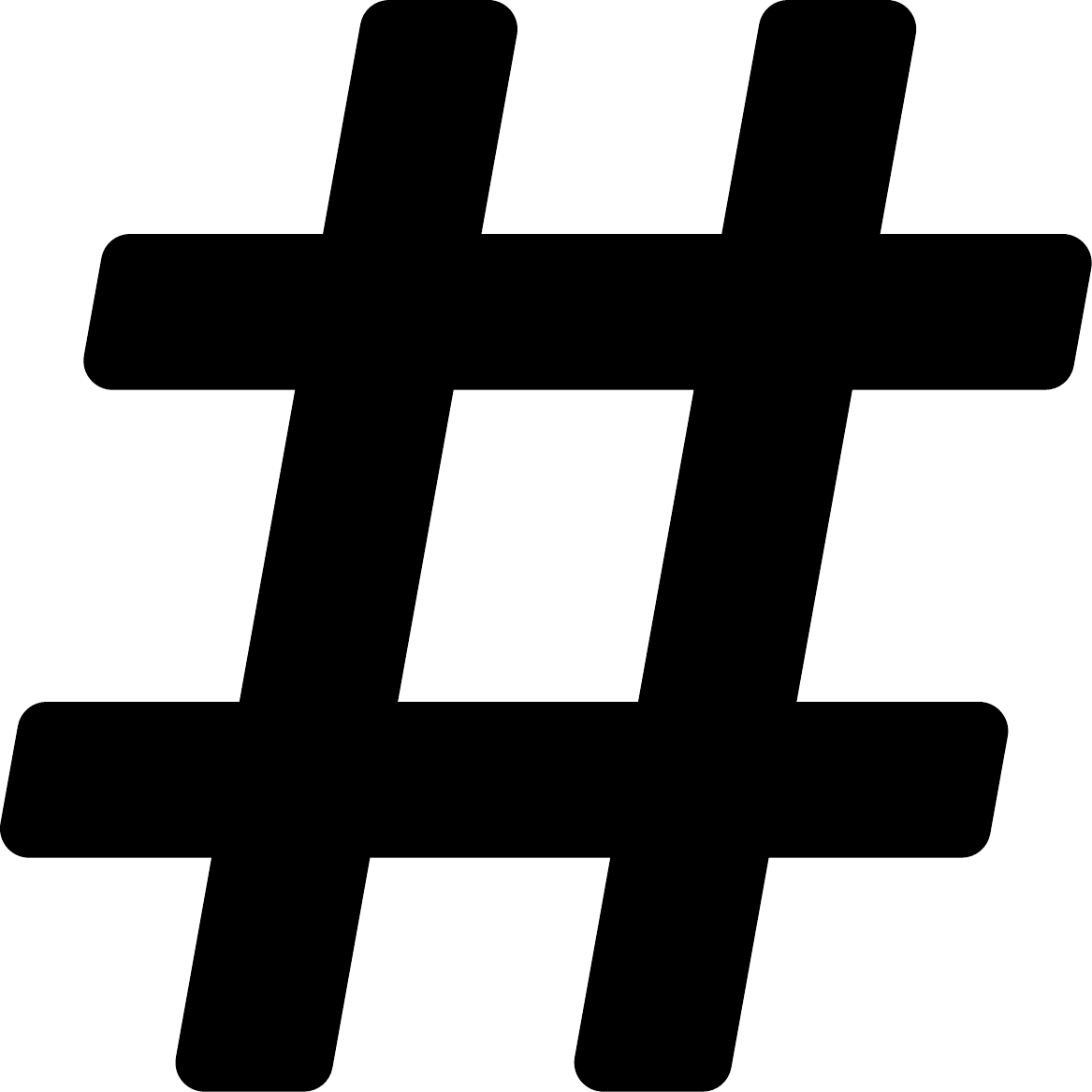}: 
Numerical), a bi-colored circular glyph (\textbf{DG3}), e.g., \includegraphics[height=1.2\fontcharht\font`\B]{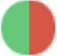} 
(combination of green 
\includegraphics[height=1.2\fontcharht\font`\B]{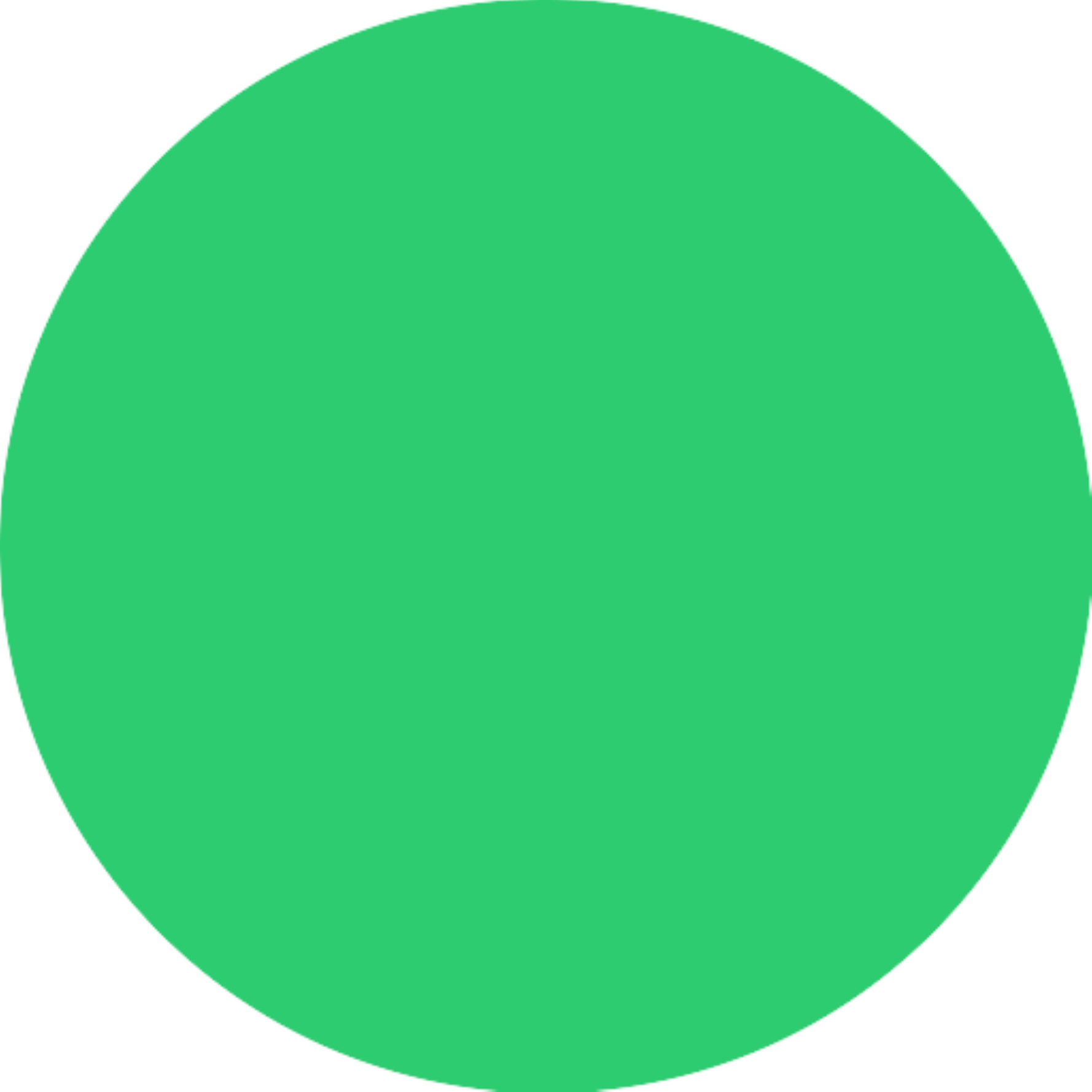}, 
yellow 
\includegraphics[height=1.2\fontcharht\font`\B]{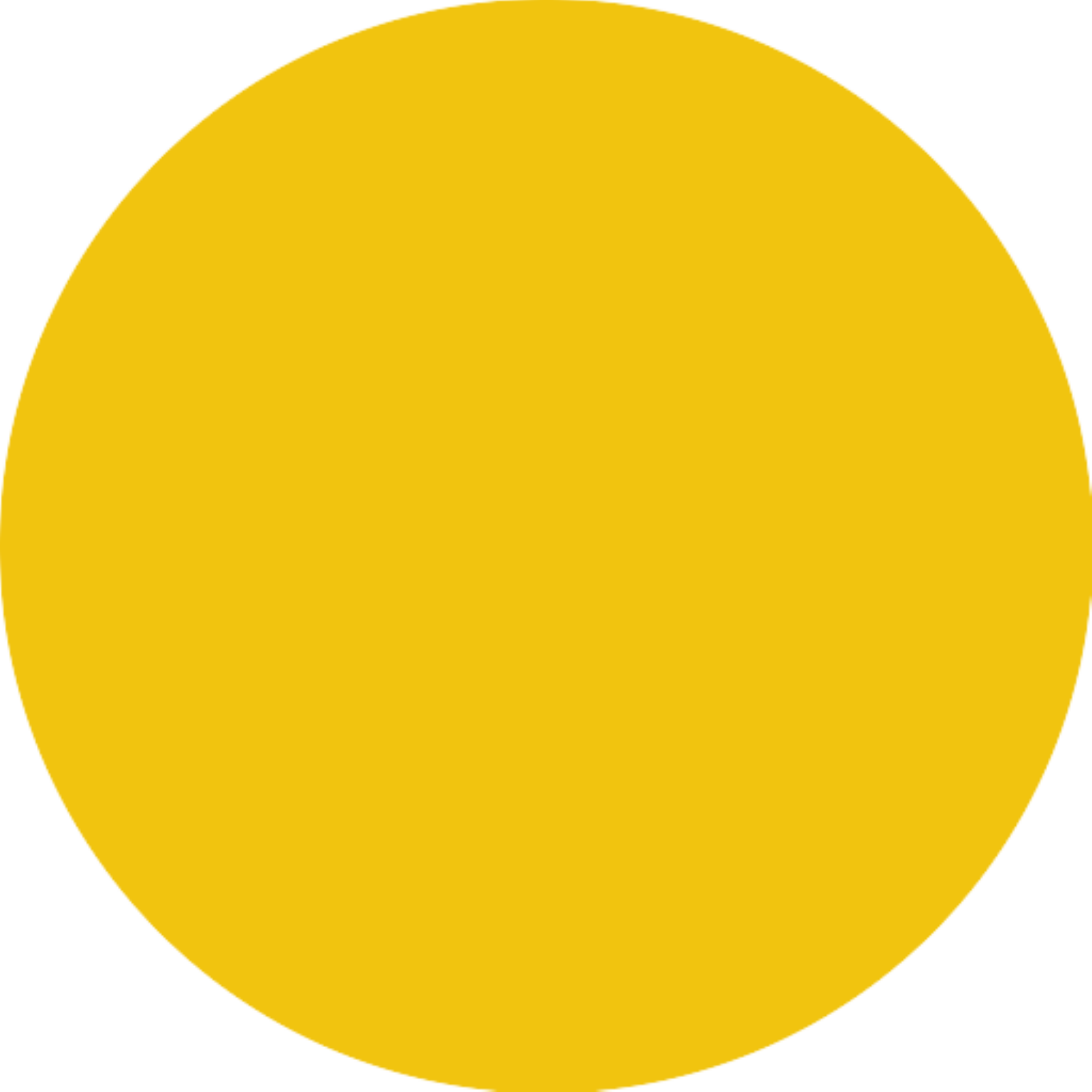}, 
red 
\includegraphics[height=1.2\fontcharht\font`\B]{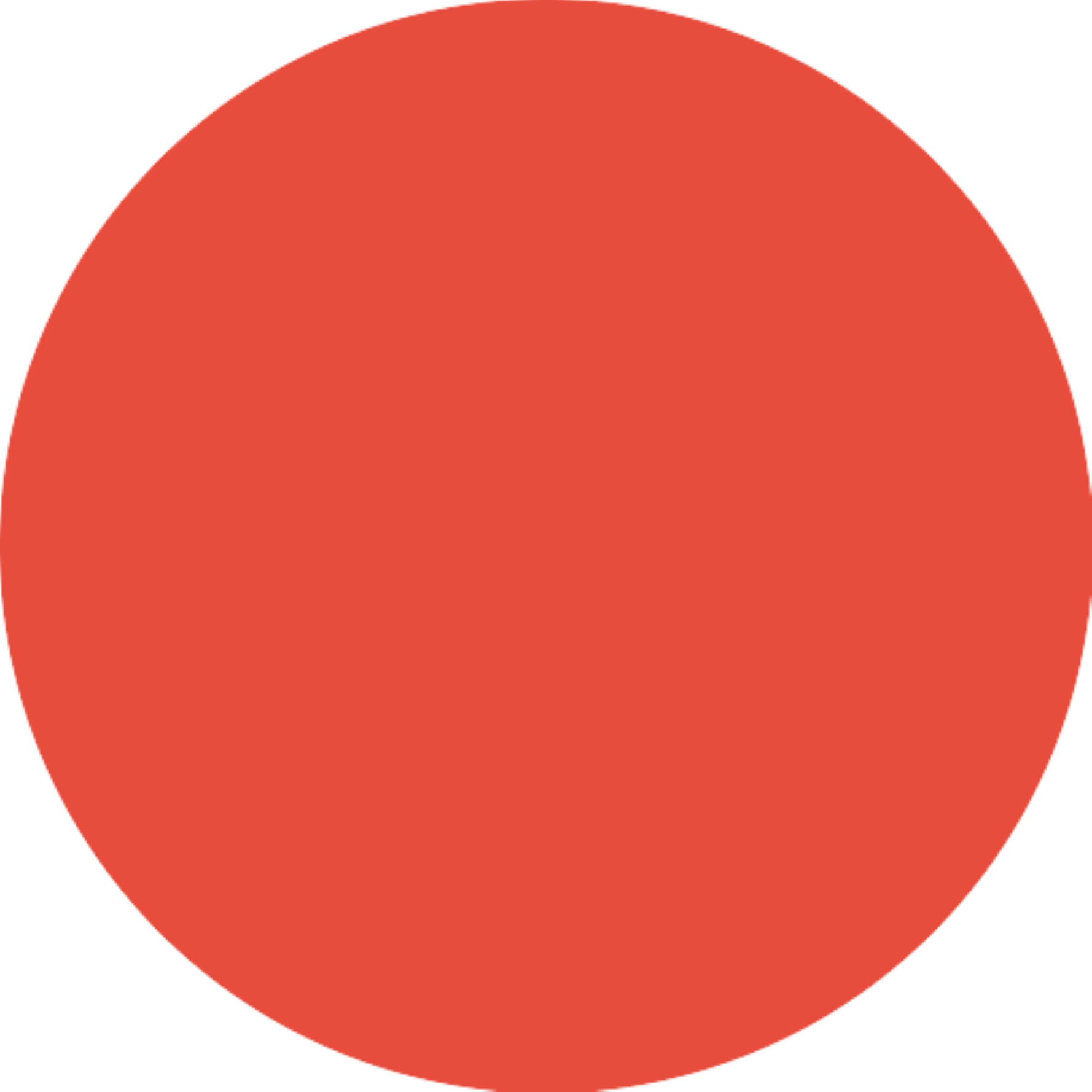}
colors), where the left-half shows the \emph{overall quality} score and the right-half shows the \emph{overall usage} score. 
Note that when the uploaded dataset has only either quality or usage information available, these bi-colored glyphs automatically transform into single-colored glyphs; users can also manually configure~\add{them} from the 
\includegraphics[height=1.2\fontcharht\font`\B]{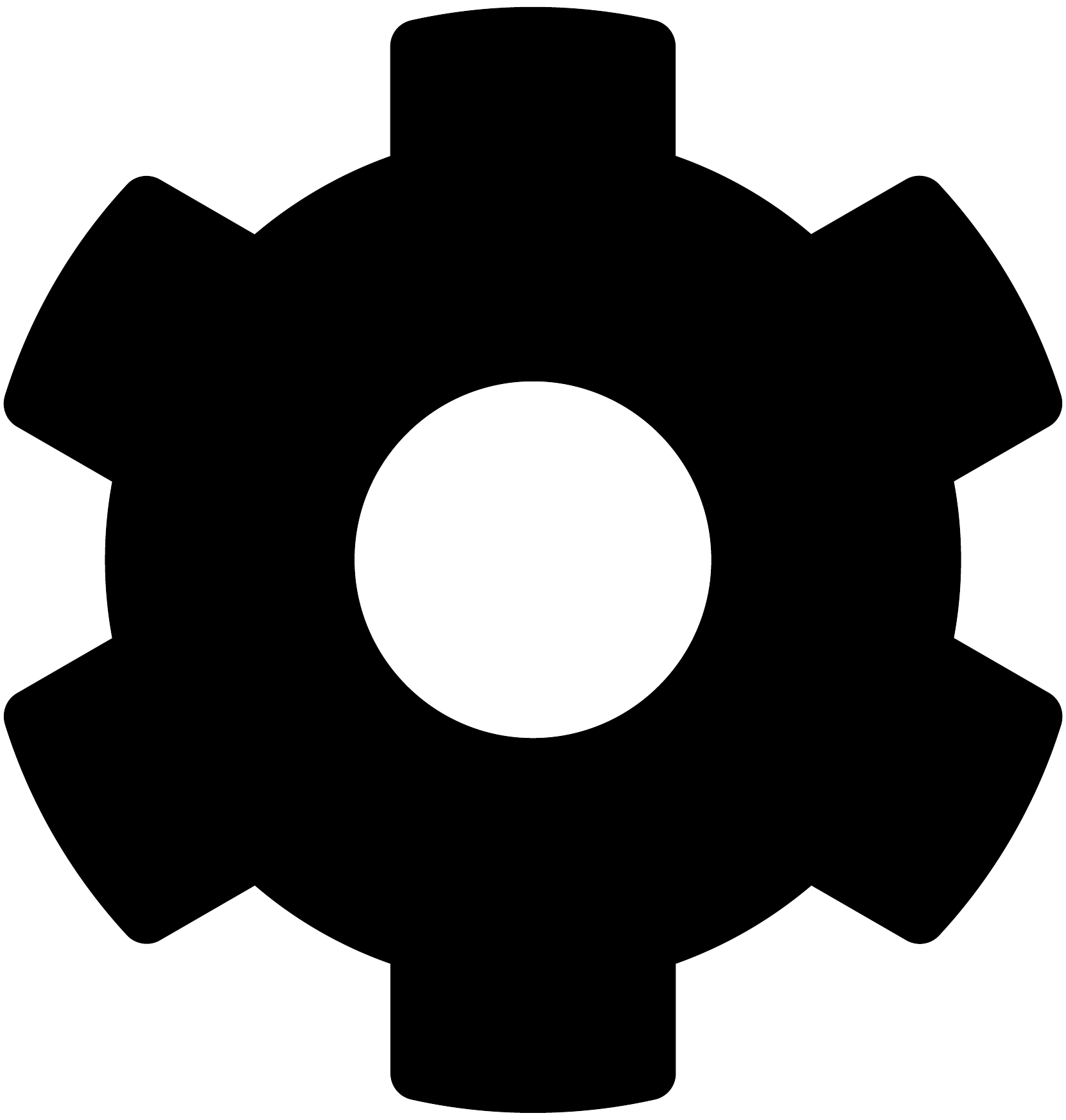}
~settings in the top-right corner (\textbf{DG5}).
The high ($\ge$90), medium ($\ge$67 but $<$90), low cutoffs (that determine the three categories) and the corresponding colors (to accommodate color-related accessibility concerns), can be configured from the legend in the top-right corner. 
Each checkbox allows users to select 
\includegraphics[height=1.2\fontcharht\font`\B]{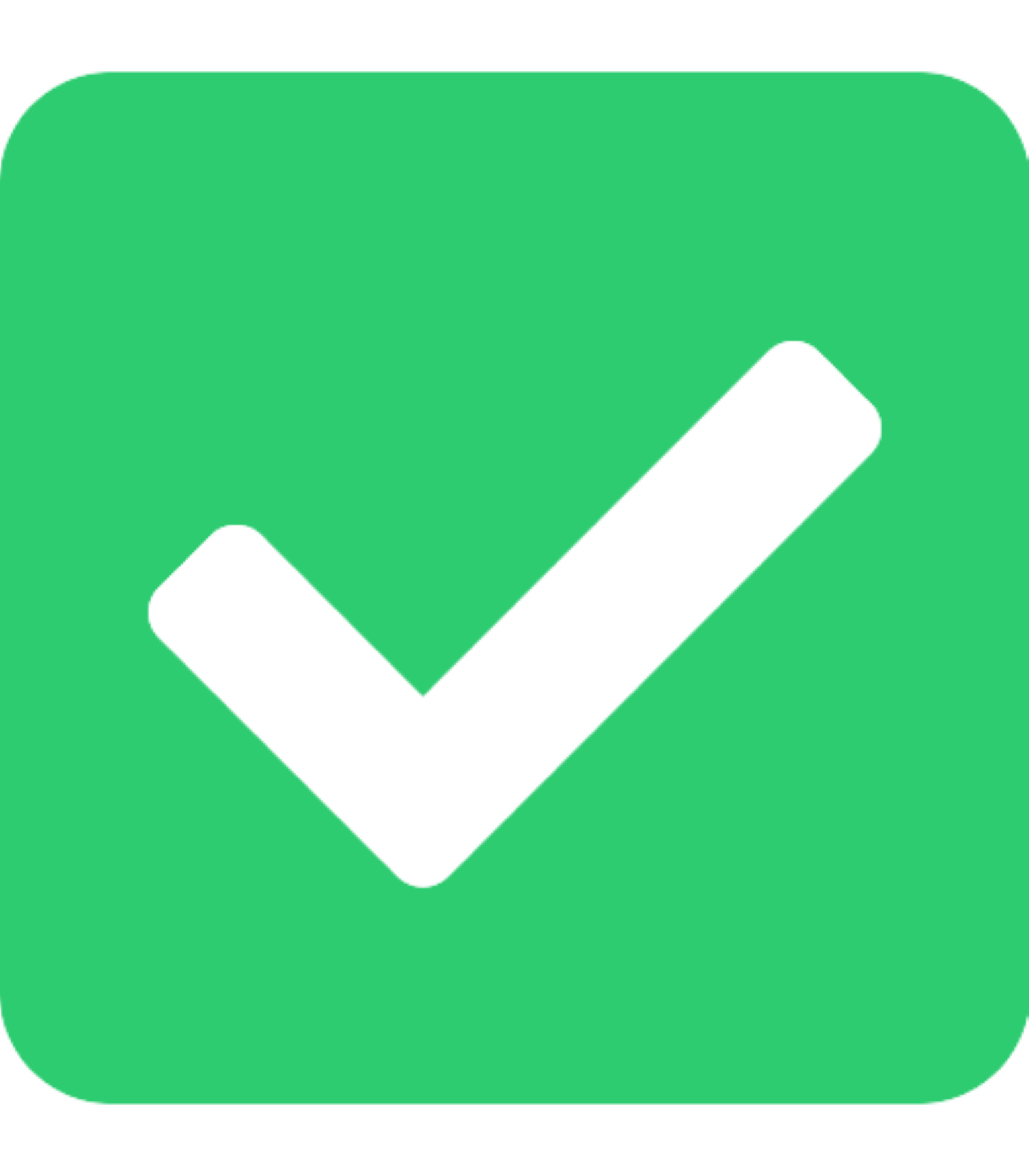}
or deselect 
\includegraphics[height=1.2\fontcharht\font`\B]{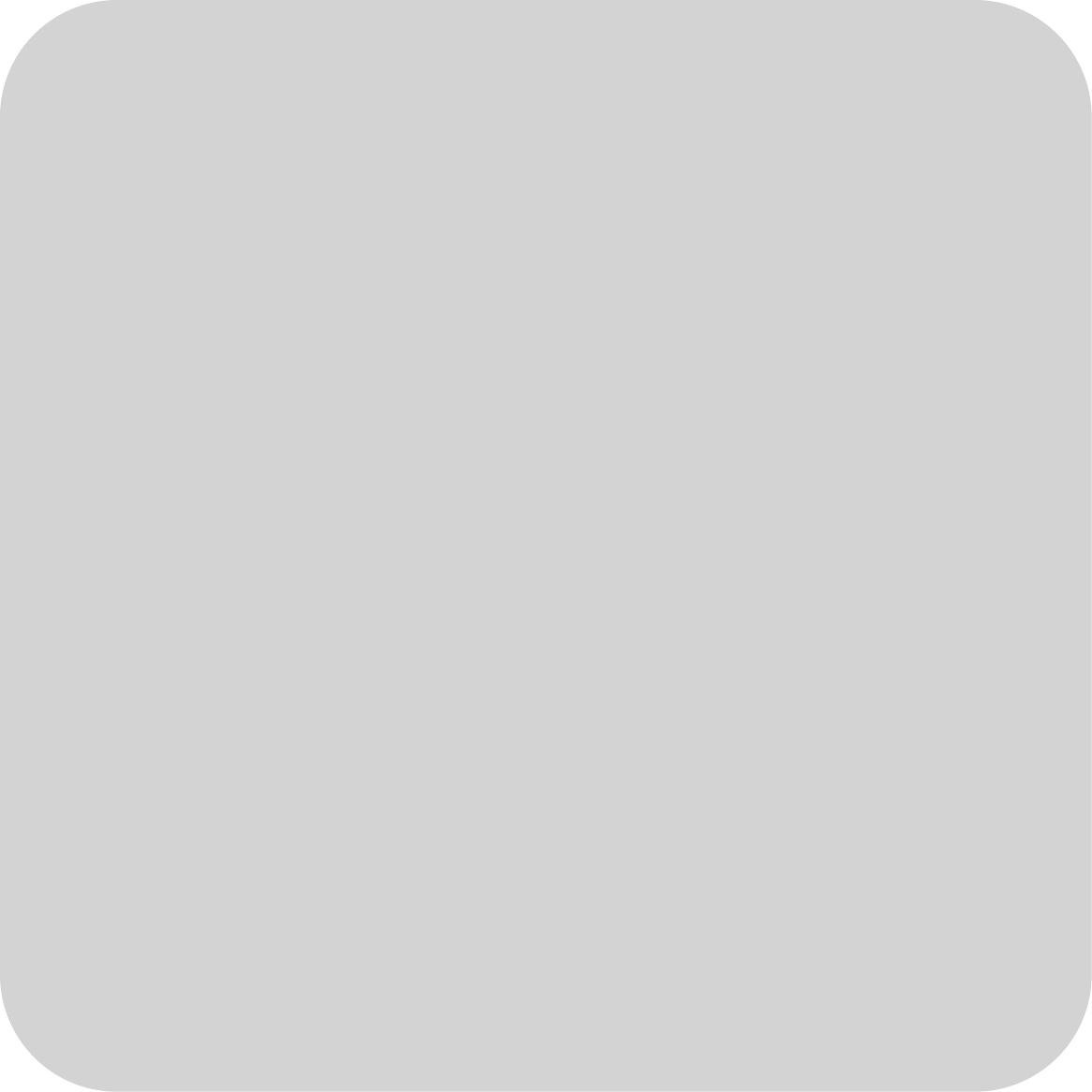}
attributes in the subset (\textbf{DG4}). Hovering on an attribute's name shows its description in a tooltip. Clicking the bi-colored glyph opens the \textbf{Attribute Detail View}.

\paragraphHeadingSpace\noindent\textbf{(B) Attribute Detail View} is an overlay \edit{showing} details \edit{of} the~\add{attribute} quality and usage, like LinkedIn's~\cite{linkedin} profile completeness (\textbf{DG3}). Like the bi-colored glyph, the left column shows data quality dimensions and the right shows usage dimensions along with the scores visualized on 5-point icon-array rating scales, e.g., ``placecontext.geo.city''~\includegraphics[height=1.2\fontcharht\font`\B]{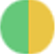} has a 100\% completeness score 
(\includegraphics[height=1.2\fontcharht\font`\B]{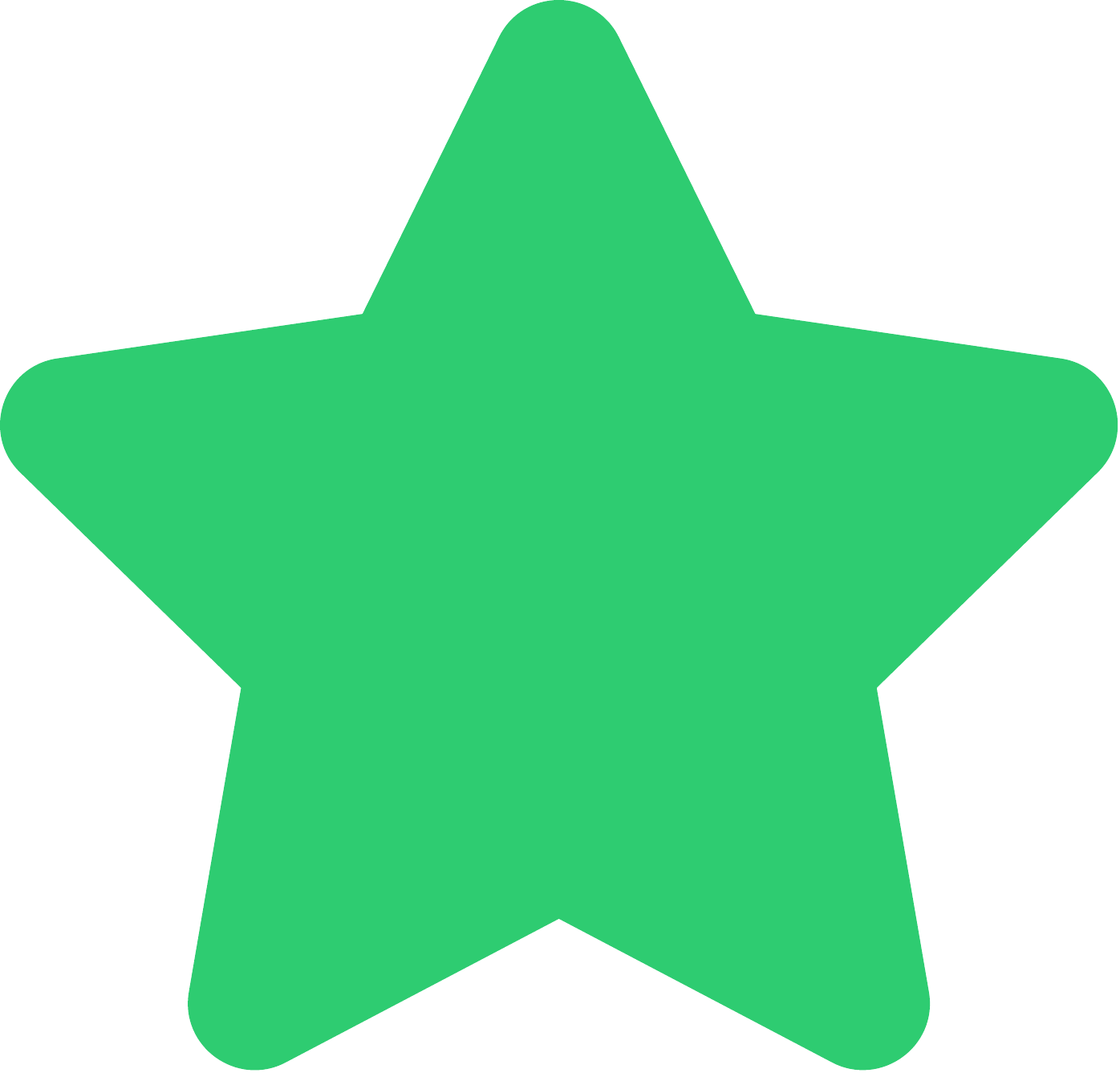}\includegraphics[height=1.2\fontcharht\font`\B]{figures/star-green-eps-converted-to.pdf}\includegraphics[height=1.2\fontcharht\font`\B]{figures/star-green-eps-converted-to.pdf}\includegraphics[height=1.2\fontcharht\font`\B]{figures/star-green-eps-converted-to.pdf}\includegraphics[height=1.2\fontcharht\font`\B]{figures/star-green-eps-converted-to.pdf}) 
and an 87\% overall usage score 
(\includegraphics[height=1.2\fontcharht\font`\B]{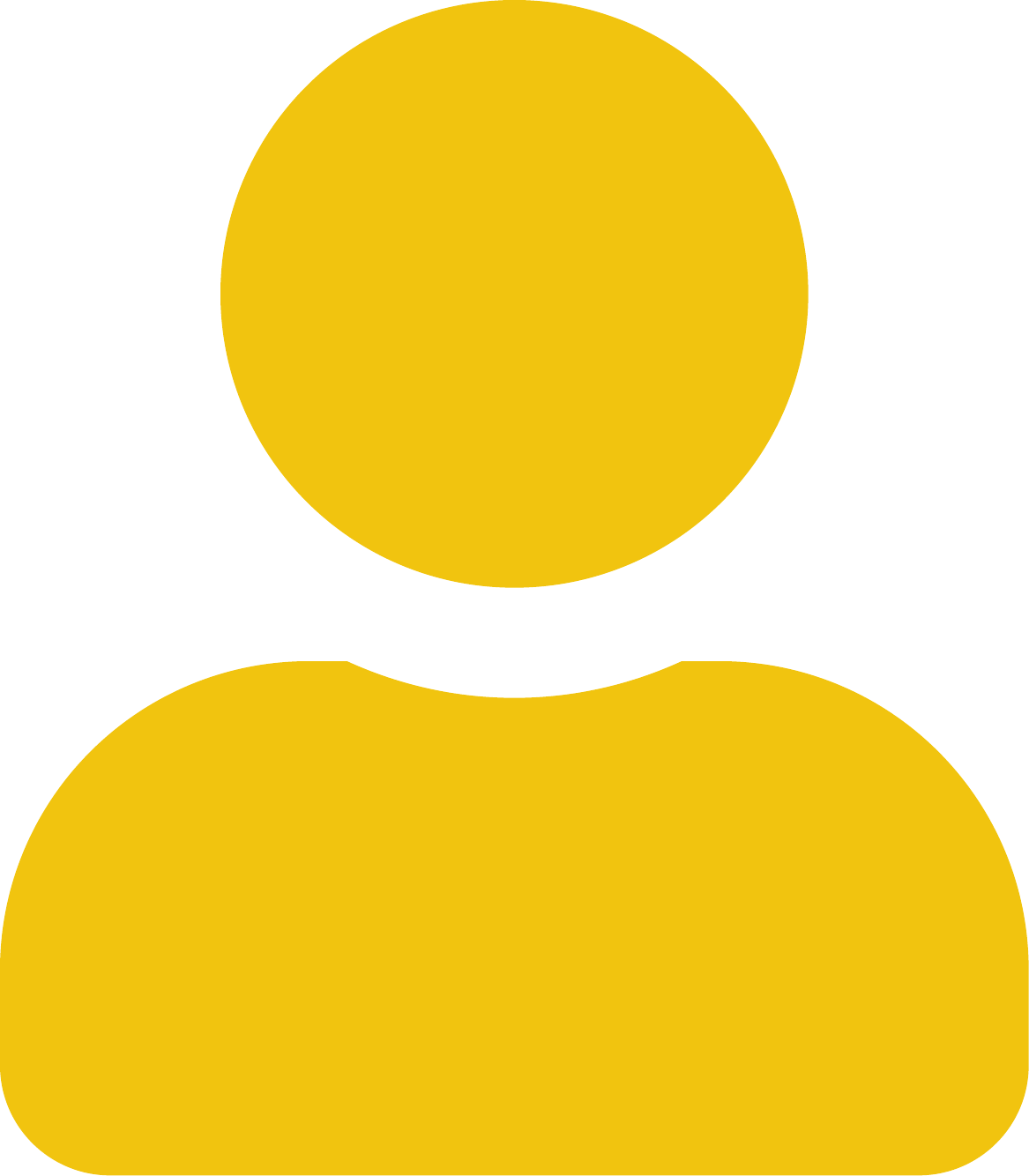}~\includegraphics[height=1.2\fontcharht\font`\B]{figures/user-yellow-eps-converted-to.pdf}~\includegraphics[height=1.2\fontcharht\font`\B]{figures/user-yellow-eps-converted-to.pdf}~\includegraphics[height=1.2\fontcharht\font`\B]{figures/user-yellow-eps-converted-to.pdf}~\includegraphics[height=1.2\fontcharht\font`\B]{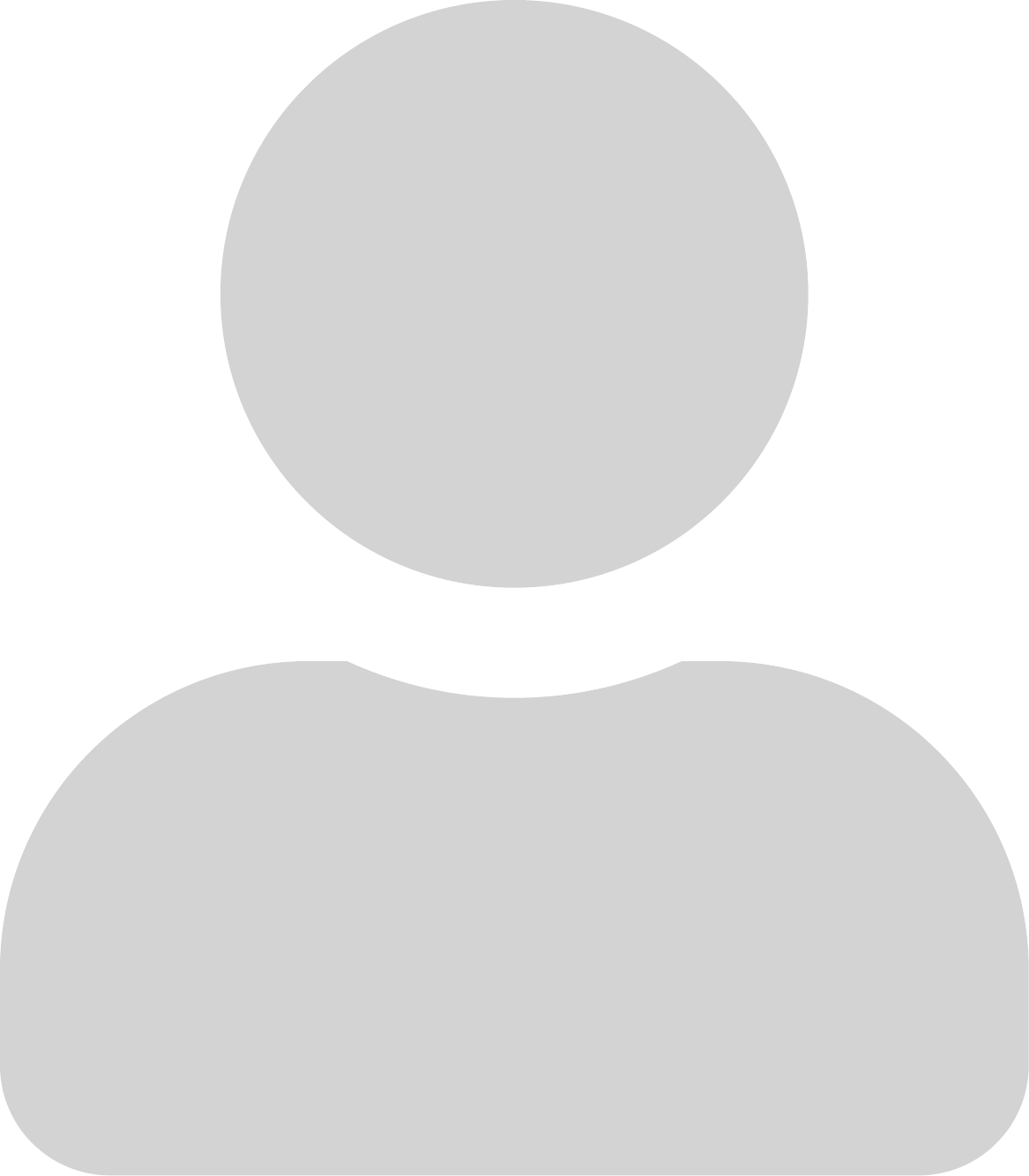}). 
Hovering the info icon \includegraphics[height=1.2\fontcharht\font`\B]{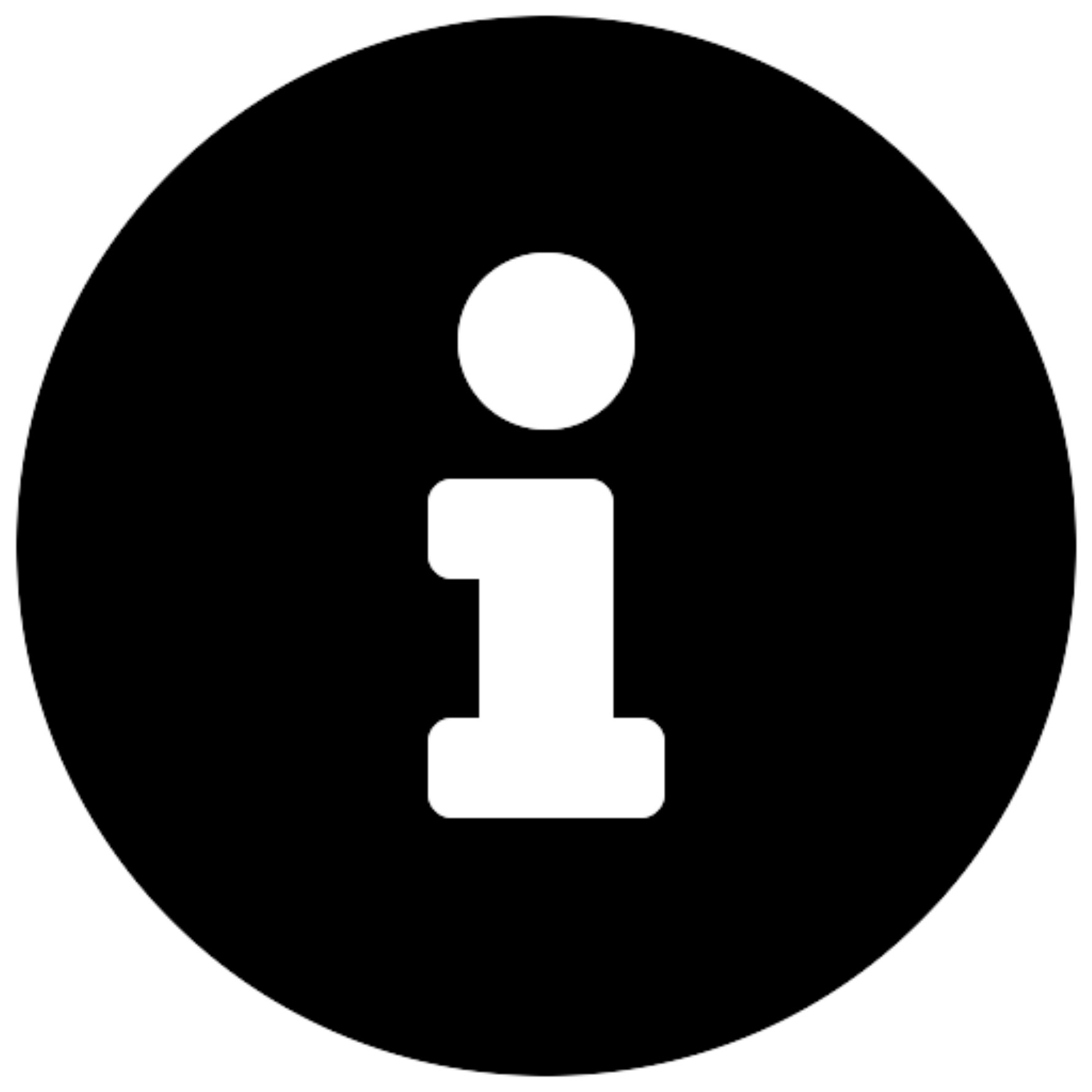}\ shows the dimension's definition (e.g., \emph{``Completeness is the percentage (\%) of non-missing values in the attribute''}) and any preconfigured rules for the calculation (e.g., \emph{``sales.purchases.price is considered correct if it is $\ge$ 0''}) to help educate the user (\textbf{DG2}).

\paragraphHeadingSpace\noindent\textbf{(C) Data View} shows the entire dataset in an interactive table. The first row shows a summary view of attribute characteristics such as cardinality (number of unique values), missing values, and distribution plots (area charts for \add{numerical}~\includegraphics[height=1.2\fontcharht\font`\B]{figures/hashtag-eps-converted-to.pdf}, bar charts for \add{categorical}~\includegraphics[height=1.2\fontcharht\font`\B]{figures/font-eps-converted-to.pdf}
attributes that show the underlying data distribution in black and the filtered data distribution in blue) (\textbf{DG3}). 
Table cells that have missing or incorrect values (e.g., \emph{``sales.purchases.price''}=``NaN'') are highlighted in red with details shown on hover (\textbf{DG3}). Standard operations such as search, pagination, and sorting are integrated within the table controls. Users can also sort by quality and usage dimensions at the record level (\textbf{DG4}). In Figure~\ref{fig:step-1}, the records are sorted by \emph{completeness} (the ``Sort Values'' dropdown in the \textbf{Data View}) and the columns are sorted by \emph{correctness} (the ``Sort'' \nobreak dropdown in the \textbf{Attribute View}), both in the ascending order 
\includegraphics[height=1.2\fontcharht\font`\B]{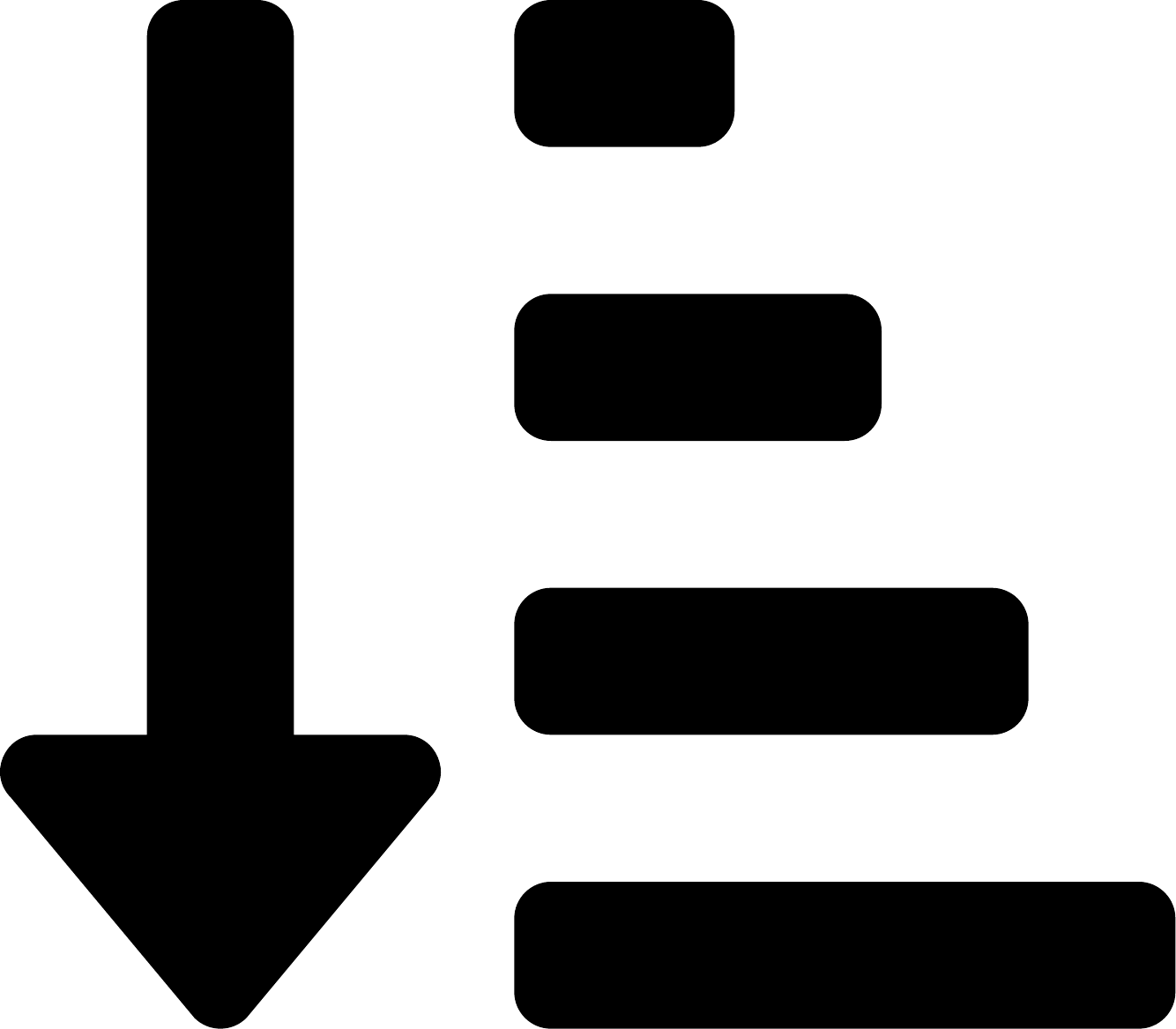}.

\paragraphHeadingSpace\noindent\textbf{(D) Attribute Filter View} enables users to filter the dataset by applying filters for each attribute by dragging them (from the \textbf{Attribute View} or the \textbf{Data View}) into this view's drop-zone (\textbf{DG4}). Multi-select dropdowns for categorical~\includegraphics[height=1.2\fontcharht\font`\B]{figures/font-eps-converted-to.pdf} and range-sliders for numerical~\includegraphics[height=1.2\fontcharht\font`\B]{figures/hashtag-eps-converted-to.pdf} attributes along with visual scents~\add{(embedded visualizations that provide information scent cues for navigating information spaces~\cite{willett2007scented})} for the distribution of attribute values in the original dataset (in black) and after applying filters (in blue) help the user determine appropriate filter criteria (\textbf{DG3}). Unlike selection of attributes, where one must explicitly check checkboxes to add to the subset, \app automatically selects all remaining records after filtering into the subset.

\paragraphHeadingSpace\noindent\textbf{(E) Quality Filters View} enables users to filter the dataset by quality dimensions at~\add{both} an attribute \edit{and} a record level (\textbf{DG4}). For example, applying the attribute-level completeness filter $\in$~[60,~100] removes all data attributes (columns) that have a completeness score outside the range. Similarly, a record-level completeness filter $\in$~[50,~75.61] filters out all records (rows) outside that range.

\paragraphHeadingSpace\noindent\textbf{(F) Usage Filters View}, like the \textbf{Quality Filters View}, enables users to filter the dataset based on usage dimensions (\textbf{DG4}), For example, applying the attribute-level \emph{in-subsets} usage filter $\in$~[30,~100] removes all attributes that were selected by less than 30\% of users.

\paragraphHeadingSpace\noindent\textbf{(G) Minimap View} provides a novel, visual overview of the proportion of attributes and records originally in the dataset (gray), currently visible after applying filters (blue), and selected in the dataset subset (green) (\textbf{DG4}). We disabled the green (selected) state by default as our pilot users found it to be overwhelming (Section~\ref{sec:alternatives}). The width and height of the rectangular area encode the number of attributes and records, respectively. This view is discretized into small rectangles proportional to the dataset size.

\subsection{Step 2: Review Selected Subset}
This \emph{review} step consists of the \textbf{(H) Attribute View} and \textbf{(I) Data View} with \emph{just} the 
\includegraphics[height=1.2\fontcharht\font`\B]{figures/check-square-green-eps-converted-to.pdf}
selected attributes and records (Figure~\ref{fig:step-2-3}). 
Viewing all selected attributes stacked together enables users to inspect the relative distributions of high, medium, and low quality and usage scores; this~\add{view} also makes it easy to inspect the distribution of the red highlights (missing or incorrect values) in the selected table cells; both of these~\add{tasks} would be difficult in \textbf{Step 1} in the presence of deselected attributes.
This step makes users pause and reflect on their subset selection performance before moving onto building a dashboard (\textbf{DG1}).

\subsection{Step 3: Create Dashboard}
After reviewing the selected subset, this step helps users create and save univariate and bivariate visualizations, collectively forming a dashboard (Figure~\ref{fig:step-2-3}) (\textbf{DG1}). \edit{This step} consists of:

\paragraphHeadingSpace\noindent\textbf{(J) Attribute View} is the same as the \textbf{Attribute View} in \textbf{Step 2}.

\paragraphHeadingSpace\noindent\textbf{(K) Encodings View} allows users to create visualizations by specifying a chart type (bar chart, scatter plot, line chart), dragging attributes onto visual encodings (X, Y), and determining aggregations (sum, mean, max, min) wherever applicable (\textbf{DG4}).

\paragraphHeadingSpace\noindent\textbf{(L) Visualization Canvas} renders the visualization based on the specifications configured in the \textbf{Encodings View}. Users can save a visualization by giving it a title and clicking the save icon 
\includegraphics[height=1.2\fontcharht\font`\B]{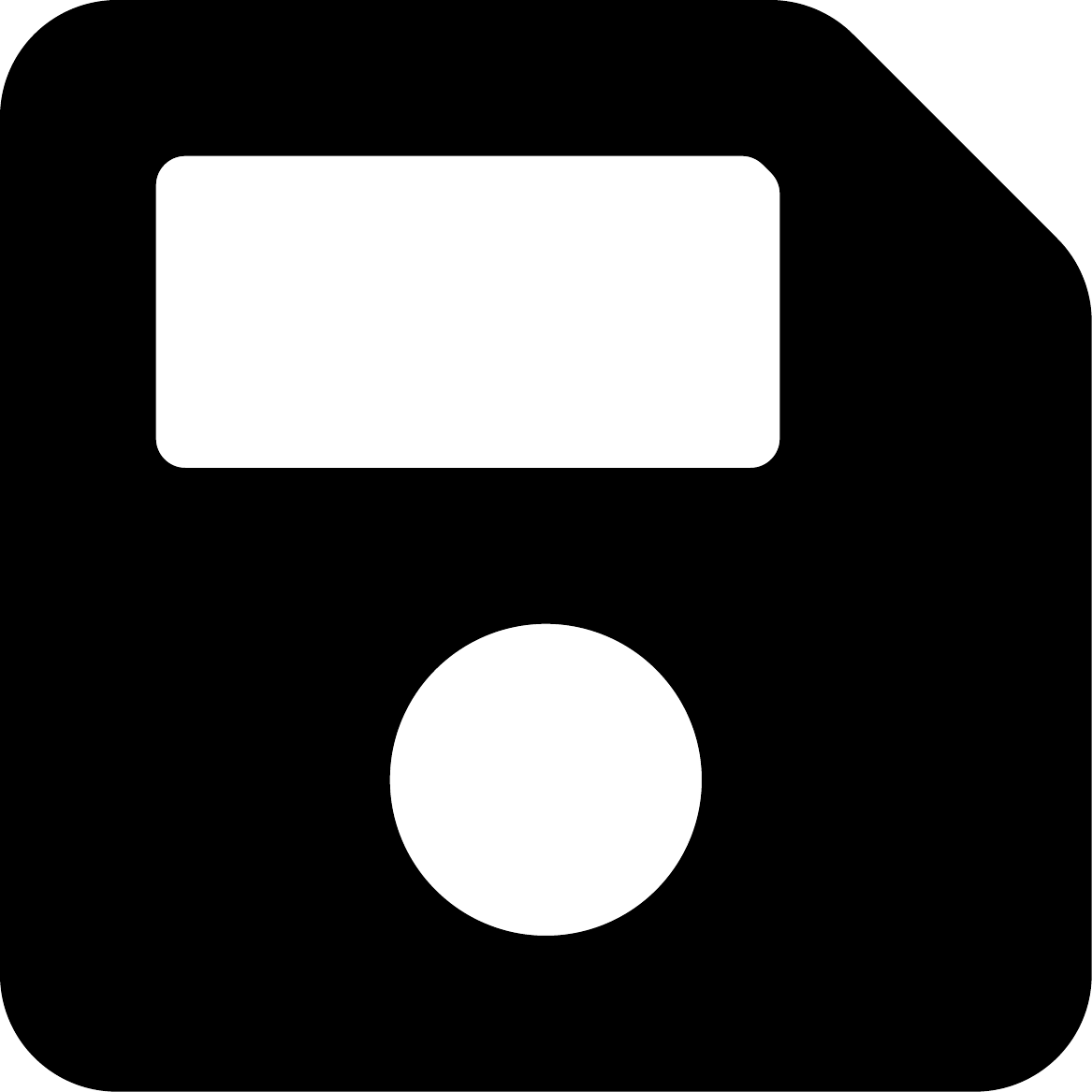}.

\paragraphHeadingSpace\noindent\textbf{(M) Saved Visualizations View} shows the list of all visualizations saved from the \textbf{Visualization Canvas}.
\edit{This view} also allows users to delete 
\includegraphics[height=1.2\fontcharht\font`\B]{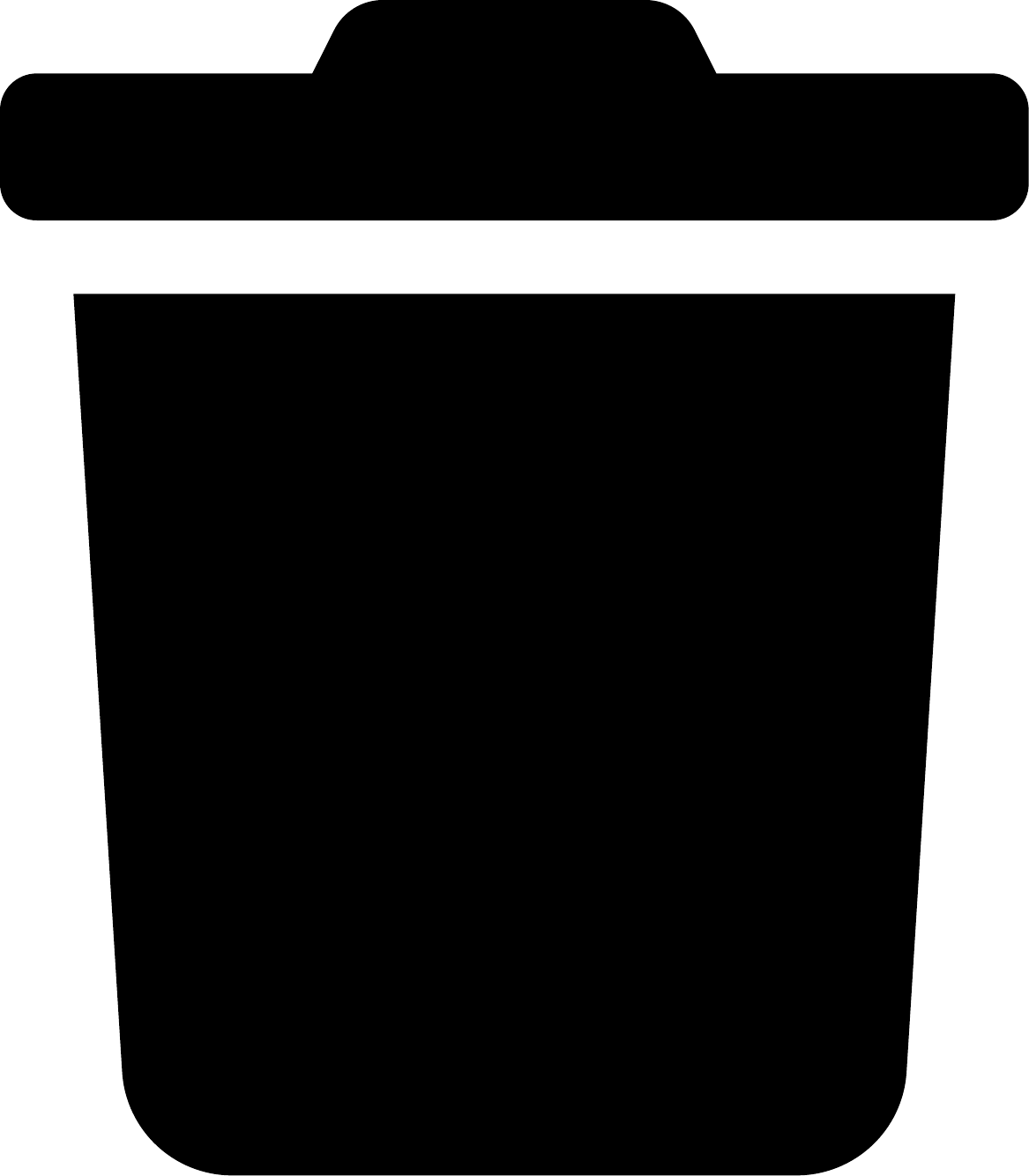}
one or all saved visualizations as needed (\textbf{DG4}).

\subsection{Implementation}
We developed the \app frontend in Angular~\cite{angular}, which interfaces with a Python~\cite{python} server in real-time over the HTTP REST~\cite{richards2006representational} and websocket~\cite{fette2011websocket} protocols. The datasets, user interaction logs (collected from the frontend), and auxiliary information were all stored in PostgreSQL, and queried later using SQL (\textbf{DG6}).

\subsection{Design Alternatives}
\label{sec:alternatives}
Before finalizing the design of \app, we presented an initial version of the interface to \emph{four} pilot users to assess the feasibility of certain designs as well as the fidelity of the evaluation task planned for the user study (Section~\ref{sec:study}). Some of our design considerations that did not make it to the current version are described next. 

Before fixating on the bi-colored glyphs next to the attribute names, we experimented with other visual variables such as size (e.g., a larger circle means higher score) and shape (e.g., quality is square and usage is a circle). 
We did not choose these \edit{alternatives} in order to satisfy \textbf{DG5} (configure \app to support one, none, or both of quality and usage information); the bi-colored glyphs were more aesthetic as they retained a consistent circular shape while using different colors to describe different dimensions across configurations. Next, we picked a discrete three-class (high, medium, low) scale over a continuous scale to help users perceptually distinguish between (and form groups of) attributes by color hue instead of the less effective saturation~\cite{koffka2013principles}. For the five-class rating scales in the \emph{Attribute Details View}, we considered a progress bar-like continuous widget that encodes the size (length), but eventually chose discrete icon arrays as they are easy to read~\cite{galesic2009using}. 
For selecting records into the subset, we considered if they should, like attributes, be selected manually through checkboxes; however, this one-by-one selection was deemed tedious and was hence discarded. Finally, to facilitate data preparation along with analysis, we had several workflow-related considerations, e.g., How many steps should we have? Should they be linear? Is the review step necessary?
Our pilot users helped us finalize the flexible, linear, three-step workflow.

\subsection{\app Example Scenarios}



To illustrate how \app can help users prepare relevant subsets from large, unfamiliar datasets, we developed two usage scenarios about two hypothetical users - Sunny (data engineer) and Kiran (data analyst); these scenarios were developed in collaboration with the domain experts to ensure domain relevance (Section~\ref{sec:interview}).

\subsection{Case 1: Expert User, Improved Performance}\hfill

\paragraphHeadingSpace\noindent Sunny, an experienced data engineer, often prepares data subsets for analysts who then\cut{perform different tasks with them, e.g.,} prepare business reports. They open \app, upload a recent batch of customer transactions data for an e-commerce app, and begin analysis. Given their domain expertise, they quickly lookup known attributes via the search field and select \emph{five} attributes for their subset: ``sales.product.name''~\includegraphics[height=1.2\fontcharht\font`\B]{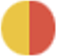}, ``sales.purchase.price''~\includegraphics[height=1.2\fontcharht\font`\B]{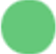} (in USD), ``timestamp''~\includegraphics[height=1.2\fontcharht\font`\B]{figures/glyph-l-green-r-red.png} (of purchase), ``placecontext.geo.countrycode''~\includegraphics[height=1.2\fontcharht\font`\B]{figures/glyph-l-green-r-red.png} (e.g., `IN' for India), and ``environment.operatingsystem''~\includegraphics[height=1.2\fontcharht\font`\B]{figures/glyph-l-yellow-r-red.png} (e.g., `iOS'). 

They switch to \textbf{\edit{Step 2: Review Selected Subset}} where they observe several cells in the data table (which now only shows the five selected attributes) with a red background. \edit{In particular, the} ``placecontext.geo.countrycode''~\includegraphics[height=1.2\fontcharht\font`\B]{figures/glyph-l-green-r-red.png} column is highlighting cells with the value ``\emph{AA}'' (\includegraphics[height=1.2\fontcharht\font`\B]{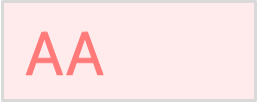}) and the ``environment.operatingsystem''~\includegraphics[height=1.2\fontcharht\font`\B]{figures/glyph-l-yellow-r-red.png} column is highlighting cells with blank (missing) values (\includegraphics[height=1.2\fontcharht\font`\B]{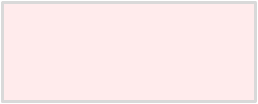}). Realizing no country has ``\emph{AA}'' as their code (as per \app's \emph{correctness} constraint and\cut{also} from their own knowledge) and that a majority (706 out of 1000)~\add{of} values for operating system are missing, they go back to \edit{\textbf{Step 1: Review Raw Data}} to make amends. 

They drag the ``placecontext.geo.countrycode''~\includegraphics[height=1.2\fontcharht\font`\B]{figures/glyph-l-green-r-red.png} attribute from the \textbf{Attribute View} into the \emph{Filter Panel} to remove all records with ``\emph{AA}'' values (\includegraphics[height=1.2\fontcharht\font`\B]{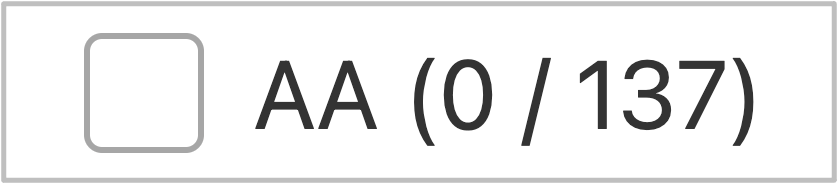}) and separately alert the data collection team about this issue. To absolutely ensure that their data are correct across all attributes, they apply a record-level ``Correctness'' filter (\includegraphics[height=1.2\fontcharht\font`\B]{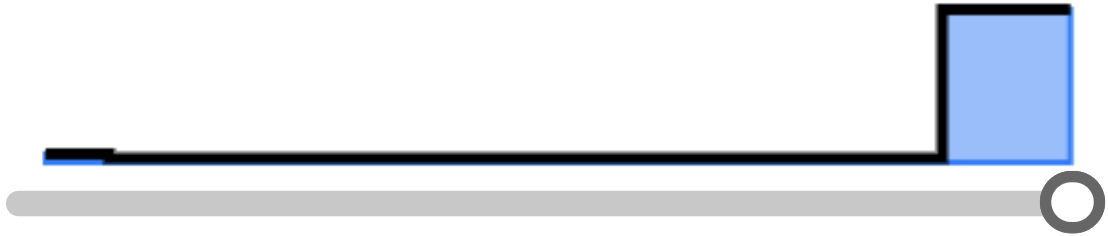}) to only keep 100\% correct records. Finally, they deselect ``environment.operatingsystem''~\includegraphics[height=1.2\fontcharht\font`\B]{figures/glyph-l-yellow-r-red.png} from the subset and instead select another attribute ``environment.browserdetails.useragent''~\includegraphics[height=1.2\fontcharht\font`\B]{figures/glyph-l-green-r-red.png} that has similar information, e.g., \emph{`Mozilla/5.0 (iPhone; CPU OS 12\_0 like Mac OS X; en\_US)'} and although it has not been used often before (right half is red), it is of high overall quality (left half is green). In this way, throughout their working session, \app helped Sunny become aware of issues with their data, guiding them to prepare a more complete and correct subset.

\subsection{Case 2: New User, Effective Onboarding} \hfill

\paragraphHeadingSpace\noindent 
Kiran recently joined a data analytics company and is tasked with becoming familiar with a client's data for designing future dashboards.
They upload a \cut{recent} client dataset of e-commerce transactions into \app and start analyzing. The dataset is large and unfamiliar\cut{to them}. They start inspecting the attribute names and descriptions from the \textbf{Attribute View} and the corresponding values and distribution plots in the \textbf{Data View} (\includegraphics[height=1.3\fontcharht\font`\B]{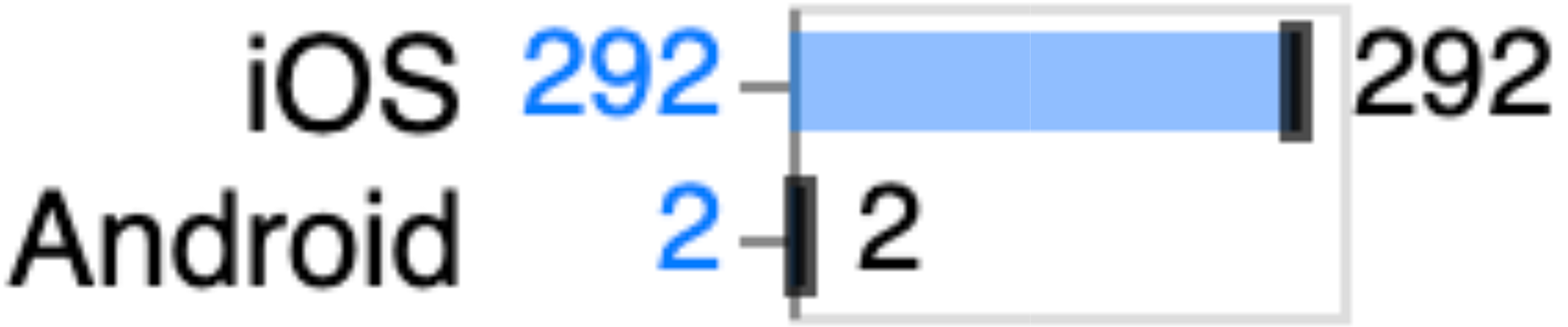}). Overwhelmed by the sheer size of the data and wanting to speed up their onboarding, they modify their strategy to only target \emph{important} attributes. 

They try to reduce the attribute search space by applying attribute-level filters in the \textbf{Quality Filters View} and \textbf{Usage Filters View} as proxies for \emph{importance}. Specifically, they inspect the distribution\add{s}\cut{scents} over the respective range sliders and filter out attributes with an \emph{overall} quality score~$<$~75~(\includegraphics[height=1.2\fontcharht\font`\B]{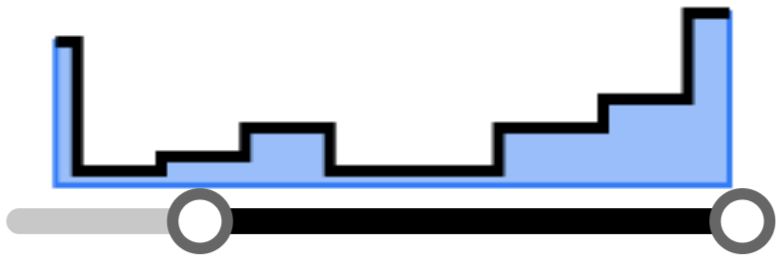}) and an \emph{overall} usage score~$<$~25~(\includegraphics[height=1.2\fontcharht\font`\B]{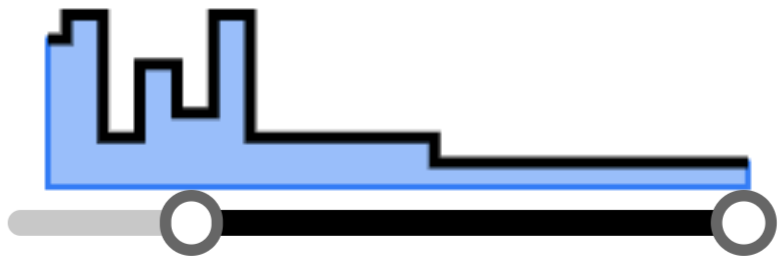}), reducing the number of attributes to a manageable 17. Finally, they sort these attributes by \emph{overall} quality score in the descending order (\includegraphics[height=1.2\fontcharht\font`\B]{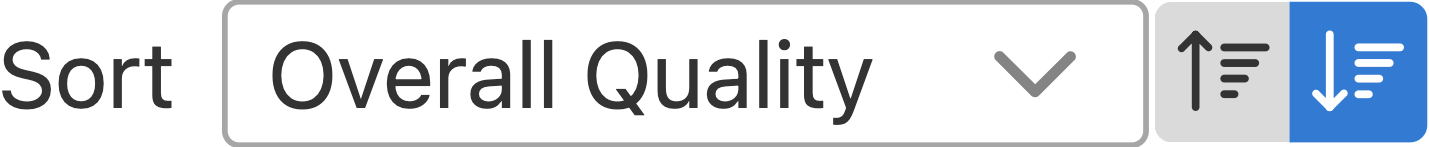}) and start inspecting their name, description, and \includegraphics[height=1\fontcharht\font`\B]{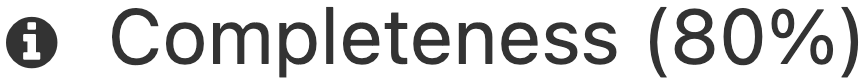} and \includegraphics[height=1\fontcharht\font`\B]{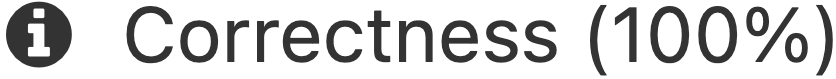} scores in the \textbf{Attribute Detail View} (via the bi-colored circular glyphs \includegraphics[height=1.2\fontcharht\font`\B]{figures/glyph-l-yellow-r-red.png}). In this way, \app helped Kiran get onboarded~\add{to a new, unfamiliar dataset} quickly and effectively.




\section{Evaluation}
\label{sec:study}



We conducted a user study to evaluate and understand how the quality and usage information in \app guides users in preparing effective data subsets for subsequent analysis. 

\paragraphHeadingSpace\bpstart{Task:} 
We designed a task involving subset selection and visual analysis wherein participants are expected to:
\begin{quote}
    \emph{``Explore a dataset of online customer behavior on an e-commerce website, prepare an effective subset\footnote{Note that a data subset comprises attributes and records less than or equal to those in the original presented dataset.} to determine meaningful drivers of \$ (dollar) sales revenue for the company, and create a dashboard of at least three visualizations to convey their findings.''}
\end{quote}

\paragraphHeadingSpace\bpstart{Participants:} We recruited 36 participants consisting of professionals and researchers from industry and academia: \emph{students}~(23), \emph{business consultants}~(2), \emph{senior data analysts}~(2), \emph{assistant professor}, \emph{associate product manager}, \emph{data science manager}, \emph{postdoctoral scholar}, \emph{program manager}, \emph{quality assurance engineer}, \emph{scientist (clinical trials)}, \emph{software developer}, and \emph{UX designer}. \edit{Participants} were pursuing or had received \emph{bachelors}~(3), \emph{masters}~(14), or \emph{doctoral}~(19) degrees in \emph{computer science}~(21), \emph{human-centered computing}~(4), \emph{human-computer interaction}~(2), \emph{business administration}~(3), \emph{pharmaceutical sciences}, \emph{economics}, \emph{electronics engineering}, \emph{systems engineering}, \emph{data science}, or \emph{information studies}. Demographically, they were in the \emph{18-24}~(13), \emph{25-34}~(19), \emph{35-44}~(3), or \emph{preferred not to say}~(1) age groups (in years) and of \emph{female}~(16), \emph{male}~(19), \emph{other}~(0), or \emph{preferred not to say}~(1) genders. They self-reported their experience performing any kind of data analysis using visual analysis tools (e.g., Excel, Tableau) or programming as either \emph{everyday or part of the job}~(10), \emph{often}~(13), \emph{occasionally}~(13), \emph{rarely}~(0), or \emph{never}~(0).

\paragraphHeadingSpace\bpstart{Dataset:} For the purpose of a thorough evaluation of all \app capabilities and to ensure completion of the task within the stipulated study duration, we used a random sample of 1000 records and 42 attributes (columns) from an open-source digital marketing dataset~\cite{digitalMarketingDataset} and infused certain quality issues pertaining to \textit{correctness} and \textit{objectivity} (by setting appropriate constraints). We marked quality and usage (and overall) scores such that $\ge$90 is marked as \emph{high} 
\includegraphics[height=1.2\fontcharht\font`\B]{figures/circle-green-eps-converted-to.pdf},  
$\ge$67 but $<$90 as \emph{medium} 
\includegraphics[height=1.2\fontcharht\font`\B]{figures/circle-yellow-eps-converted-to.pdf},  
and the rest as \emph{low} \includegraphics[height=1.2\fontcharht\font`\B]{figures/circle-red-eps-converted-to.pdf}.
We fixed these thresholds to realize a reasonable distribution of attributes and records across the three (high, medium, low) categories, so that participants are neither demoralized (all scores are low) nor overconfident (all scores are high).

\paragraphHeadingSpace\bpstart{System Configurations as User Study Conditions:} To achieve \textbf{DG5}, we designed \app to support four configurations: (1)~neither quality nor usage, (2)~only quality, (3)~only usage, and (4)~both quality and usage. Of these four configurations, we did not explicitly evaluate the (3) only usage configuration because our expert interviews highlighted addressing data quality concerns as most important and that usage information alone must never power ``data-driven'' analysis and decision-making, \edit{at least not} without more important aspects such as quality. Hence, we utilized the other three \app configurations as standalone study conditions in a between-subjects evaluation, described next.

\paragraphHeadingSpace\noindent{\textbf{\textcolor{BB}{[B] Baseline}}}: With this configuration, we aim to understand user strategies \emph{without} quality and usage information, also simulating what many current systems do (e.g., Tableau~\cite{tableau}). Specifically, the bi-colored glyphs next to the attribute name, filter and sort options, and visual scents (in the table) for usage and quality are all hidden.

\paragraphHeadingSpace\noindent\textbf{\textcolor{QQ}{[Q] Quality}}: With this \edit{configuration}, we aim to understand how users utilize only quality information to perform the study task, also simulating what many current systems do (e.g., Profiler~\cite{kandel2012profiler}, Trifacta~\cite{trifactawrangler}). This condition would also enable us to compare against the following \emph{D} configuration (that has both quality and usage information). Specifically, only single-colored circular glyphs next to the attribute name, sort and filter options, and visual scents (in the data table) that are relevant to quality are visible and enabled. 

\paragraphHeadingSpace\noindent\textcolor{DD}{\textbf{[D] \app}}: This all encompassing configuration shows both data quality and usage information in the interface. Specifically, all features described in Section~\ref{subsection:user-interface} are enabled. Usage information for the \emph{D} condition were computed by processing the interaction logs of the participants in the \emph{B} and \emph{Q} conditions (24 participants). We computed each attribute's \emph{in-subsets} score as the percentage of participants who selected that attribute to be in their subsets, \emph{in-filters} score as the percentage of participants who filtered by that attribute, \emph{in-visualizations} score as the percentage of participants who assigned that attribute to a visual encoding, and an \emph{overall} score as the maximum of the three aforementioned scores. Similarly, for each record, we computed the \emph{in-subsets} score (also the \emph{overall} score in this case) by computing the percentage of participants who selected that record (automatically as a result of applied filters) to be in their subsets. To disregard temporary, unplanned, and accidental selections during analysis, we compute this information only based on the final state of the interface at the end of the task (selected subset, applied filters, saved visualizations).


\paragraphHeadingSpace\bpstart{Study Session:} We~\cut{randomly}assigned participants to one of the three study conditions (\edit{\emph{B},} \emph{Q}, \emph{D}) while trying to balance for their backgrounds, demographics and visual data analysis literacies. Each study session lasted between 60 and 90 minutes, with \emph{D} taking longer than \emph{Q} than \emph{B} due to differences in participants' training and practice times. We compensated each participant with a \$15 gift card for their time. We conducted the study remotely using Microsoft Teams~\cite{teams}\edit{; the} experimenter provided participants access to the study environment by sharing their (experimenter's) computer screen and granting input control to the participant. After providing consent, participants saw a video tutorial (\emph{B}:5, \emph{Q}:7,~\emph{D}:10 minutes long) that demonstrated the features of \app\footnote{\add{\emph{D} participants saw both quality and usage; \emph{Q} only saw quality; and \emph{C} saw neither; hence the difference in the duration of the respective video tutorials.}}. \edit{Participants} then performed a practice task on a \emph{dataset of houses} (adapted from~\cite{de2011ames}) to get acquainted with the UI before \edit{starting} the actual task.

The actual task \edit{lasted} a maximum duration of 30 minutes. Participants were not required to think aloud during the task to simulate a realistic work setting (although some participants felt comfortable doing so). During the task, participants' interactions with the system (e.g., the filters they applied, the data subsets they selected) were logged. The study ended with participants completing a questionnaire to rate the usefulness of \app's features and a semi-structured debriefing interview for 10 minutes in which participants reflected on their overall experience, provided feedback, and answered \edit{other} questions. \edit{At} the end of the debriefing interview, the experimenter also demonstrated the \emph{D} configuration to both \emph{B} and \emph{Q} participants to get their initial reactions and elicit feedback on how the new set of aids would have hypothetically helped them accomplish their task differently. Each debriefing interview was screen- and audio-recorded for subsequent qualitative analysis.

\subsection{Hypotheses}

We structure our study analysis according to the hypotheses below, predetermined before the study based on our expectations from the intended purpose of the tool, former perception studies, feedback from pilot studies, and our own instincts. $>$ implies \emph{more or greater than}; $<$ implies \emph{less or smaller than}. 

\paragraphHeadingSpace\begin{enumerate}
    
    \item[\textbf{H1}]   {\textcolor{BB}{B (Baseline)}} $>$ {\textcolor{QQ}{Q (Quality)}} $>$ {\textcolor{DD}{D (\app)}} in terms of the number of attributes and records in the selected subsets.

    \item[\textbf{H2}] \textcolor{BB}{B} $>$ \textcolor{QQ}{Q} $>$ \textcolor{DD}{D} in terms of the proportion of attributes and records with \emph{low} quality and usage in the selected subsets.
    
    \item[\textbf{H3}] \textcolor{BB}{B} $<$ \textcolor{QQ}{Q} $<$ \textcolor{DD}{D} in terms of the proportion of attributes and records with \emph{high} quality and usage in the selected subsets.
    
    \item[\textbf{H4}] \textcolor{BB}{B} $<$ \textcolor{QQ}{Q} $<$ \textcolor{DD}{D} in terms of success and confidence after the task.
    
    \item[\textbf{H5}] \textcolor{BB}{B} $<$ \textcolor{QQ}{Q} $<$ \textcolor{DD}{D} in terms of amount of effort, temporal demand, mental demand, and frustration while doing the task.
    
    
    
    
    \item[\textbf{H6}] Participants will find quality information to have greater utility than usage information while doing the task.
    


\end{enumerate}


\begin{figure*}[ht]
    \centering
    \includegraphics[width=\textwidth]{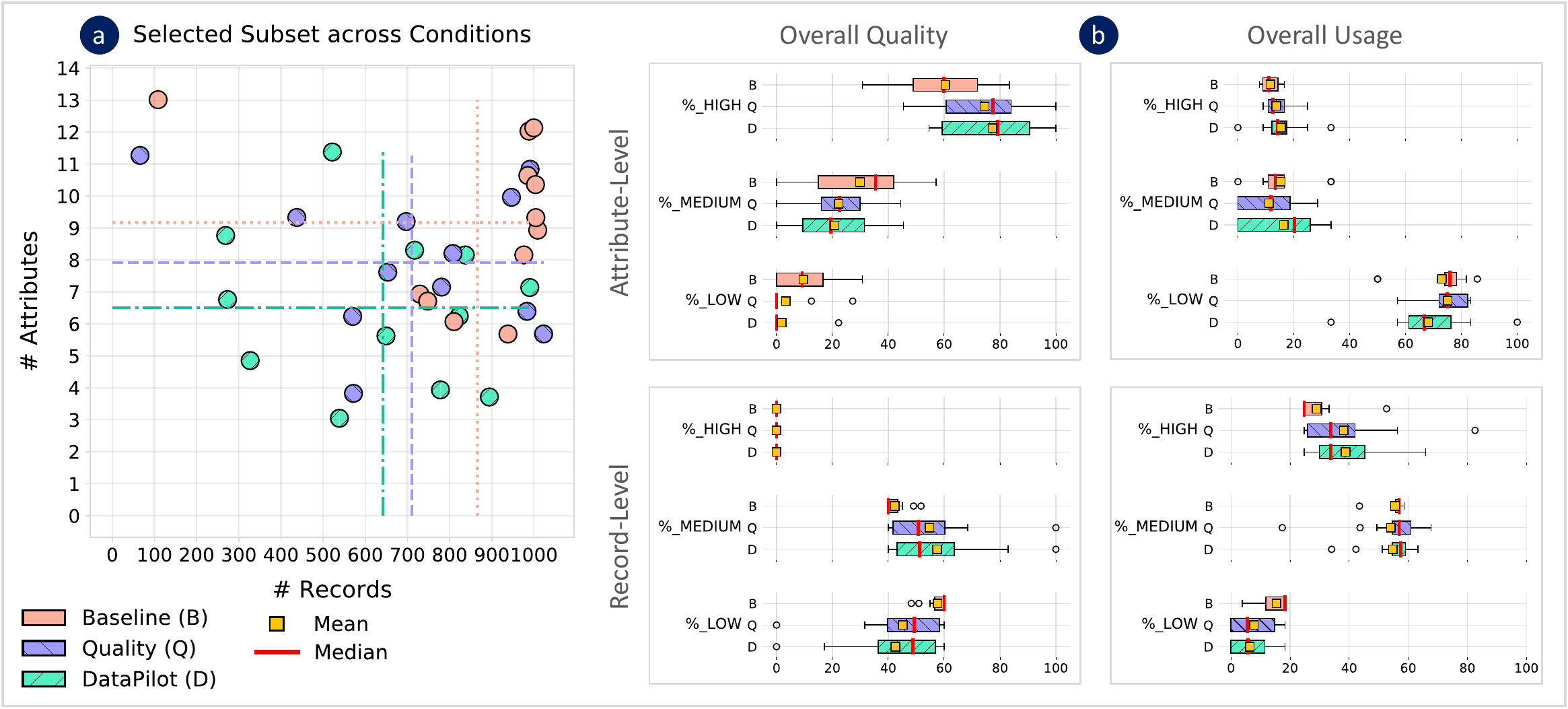}
    \caption{(a) Number of attributes and records in the participants' selected subsets and (b) attribute-level and record-level distributions of high, medium, low overall scores for both quality and usage across the three study conditions (\textcolor{BB}{B}, \textcolor{QQ}{Q}, \textcolor{DD}{D}).}
    \Description{Quantitative findings related to the data subsets selected by the user study participants. Reading from left to right: (a) a scatterplot is shown with the number of selected attributes on the Y axis, the number of selected records on the X axis, and each point corresponding to a participant colored by the study condition assigned to them; and (b) a 2x2 grid of views with Overall Quality and Overall Usage along the horizontal axis and Attribute-Level and Record-Level along the vertical axis; each of the four intersecting cells consists of grouped box plots that show the percentage distributions of high, medium, low overall scores for attributes and records (vertical axis) across both quality and usage information (horizontal axis).}
    \label{fig:attribute-record-subset-count}
\end{figure*}

\subsection{Results}
Below, we present findings from the user study and discuss them in the context of qualitative participant feedback. \textcolor{BB}{\pc{1,...,12}}, \textcolor{QQ}{\pq{1,...,12}}, \textcolor{DD}{\pe{1,...,12}} refer to the 36 participants in the \textcolor{BB}{Baseline~(B)}, \textcolor{QQ}{Quality~(Q)}, and \textcolor{DD}{\app~(D)} conditions, respectively. Participant quotes spoken during the debriefing interview and responses written in the questionnaires were both coded and categorized using affinity diagramming~\cite{gaffney1999affinity}, an inductive thematic analysis~\cite{boyatzis1998transforming} technique. One experimenter came up with an initial set of categories that were then refined during iterations with three other experimenters until a consensus was reached; the final codebook consisted of 6 high-level categories with 43 detailed, low-level codes.
Relevant study material consisting of the~\add{users' interaction logs,} questionnaires, interview transcripts, assigned qualitative codes,~\add{data analysis scripts,} and relevant figures with RainCloudPlots~\cite{allen2019raincloud} (instead of box plots) are made available in the supplemental material.

\subsubsection{Feedback~\add{on \app's Quality and Usage Information}}\hfill

\paragraphHeadingSpace\bpstart{\app, the system.} Overall, participants found \app to be useful, reporting above average system usability (SUS~\cite{brooke1996sus}) scores across the three conditions as \{\textcolor{BB}{B: 80.21}, \textcolor{QQ}{Q: 74.17}, \textcolor{DD}{D: 71.67}\}.
\pe{4} commented \edit{that} \emph{``Providing detailed auxiliary information such as the quality and usage of each data attribute is very important and missing in current tools like Tableau and PowerBI.''} \pq{8} also explained why quality and usage information are important noting, \emph{``80-90\% of true data analysis, data science, machine learning is [the data preparation] step. These [quality and usage] measurements that you're creating to allow users to start [working on their tasks] and make them explore some of the unintended consequences is very powerful. It has ample opportunity for future discovery to continuously make this a better product, so very very fascinating stuff.''}

\paragraphHeadingSpace\bpstart{Quality information.} Participants had overall positive feedback for the quality information. \pq{10} commented \edit{that} \emph{``There are invisible problems with your data and you don't necessarily find out until you start playing around with the visualizations. [Furthermore,] in aggregate visualizations, you either have limited or no ability to identify quality problems so I appreciate that \app is just very explicit about these quality issues.''} \pq{7} noted that \emph{``It is important for systems to provide such out-of-the-box insights so that users like me who don't write code don't completely ignore these aspects and can rely on the green attributes and just get started with analysis.''} 
\pq{8} saw \emph{``a lot of value to enable users to more quickly filter [attributes and records] through the signal of these measurements of quality as opposed to learning [them] on their own.''} However, \edit{\pq{8}} also expressed caution about \emph{``confounding factors, especially missing data, because many times data is not missing at random it is actually missing and telling a story,''}
\add{suggesting quality information can provide a good starting place but additional analysis by users may still be required.}

\paragraphHeadingSpace\bpstart{Usage information.} There was mixed feedback regarding the usage information. Participants with positive feedback suggested using usage information to perform fast and efficient analysis (\pq{6}), to seek validation \emph{``by performing little investigations''} (\pq{1}), \emph{``to check if they have a similar opinion as others''} (\pe{4}), \emph{``to identify new things where other people are not looking''} (\pe{3}), to seek guidance from predecessors (e.g., \pq{2,11}), to avoid repeating past mistakes (\pe{3}), and to choose between conflicting choices (e.g., \emph{``for some attributes it's not easy to decide...but usage can help choose''} - \pe{8}). Participants with mixed and negative feedback said they would not care (\pe{3}) or rely on what other people did as they do not know anything about \edit{the other users} and would have to assume they did a great job with their analysis (\pq{1}, \pe{10}). Participants also raised concerns around bias and following the crowd as \emph{``one might miss out on an uncommon attribute that is also useful''} (\pc{7}).




\subsubsection{Comparing Prepared Subsets}

\begin{table}[t]
    \centering
    \caption{Statistics associated with the prepared dataset subsets in terms of their \textbf{``Size''} and distribution of \emph{high} (``\%~H''), \emph{medium} (``\%~M''), \emph{low} (``\%~L'') values for attribute- (\textbf{``A''}) and record-level (\textbf{``R''}) quality and usage scores across the three study conditions (\textcolor{BB}{\textbf{B}}, \textcolor{QQ}{\textbf{Q}}, \textcolor{DD}{\textbf{D}}). The bolded and highlighted values in each row support our hypothesis, specifically H1, H2, H3, e.g., \textbf{\textcolor{DD}{6.5 (D)}} has the smallest $\mu$ of number (\textbf{``Size''}) of attributes (\textbf{``A''}) selected in the subset, supporting H1. 
    No record (``R'') had a high (``\% H'') overall quality score because the chosen dataset was sparse.
    In addition, medium (``\% M'') values were not part of our hypotheses; thus, the table cells corresponding to these values are neither highlighted nor formatted.}
    \label{tab:prepared-subsets}
    \vspace{-2mm}


\begin{tabular}{|l|l|rr|rr|rr|}
\hline
\multicolumn{2}{|c|}{\multirow{2}{*}{\textbf{}}} &
  \multicolumn{2}{c|}{\textbf{\textcolor{BB}{Baseline (B)}}} &
  \multicolumn{2}{c|}{\textbf{\textcolor{QQ}{Quality (Q)}}} &
  \multicolumn{2}{c|}{\textbf{\textcolor{DD}{DataPilot (D)}}} \\ \cline{3-8} 
\multicolumn{2}{|c|}{} &
  \multicolumn{1}{c|}{\textbf{$\mu$}} &
  \multicolumn{1}{c|}{\textbf{$\sigma$}} &
  \multicolumn{1}{c|}{\textbf{$\mu$}} &
  \multicolumn{1}{c|}{\textbf{$\sigma$}} &
  \multicolumn{1}{c|}{\textbf{$\mu$}} &
  \multicolumn{1}{c|}{\textbf{$\sigma$}} \\ \hline
  \multicolumn{8}{|c|}{\textbf{Size of Prepared (Selected Subsets)}} \\ \hline
\multicolumn{1}{|c|}{\multirow{2}{*}{Size}} &
  A &
  \multicolumn{1}{r|}{9.17} &
  2.44 &
  \multicolumn{1}{r|}{7.92} &
  2.19 &
  \multicolumn{1}{r|}{\textbf{\cellcolor{cellhighlight}{6.5}}} &
  2.32 \\ \cline{2-8} 
  \hhline{~|-}
\multicolumn{1}{|c|}{}             & R & \multicolumn{1}{r|}{866.17} & 253.08 & \multicolumn{1}{r|}{710.83} & 282.85 & \multicolumn{1}{r|}{\textbf{\cellcolor{cellhighlight}{642.17}}} & 249.18 \\ \hline
\multicolumn{8}{|c|}{\textbf{Distribution of Overall Quality Scores}} \\ \hline
\multirow{2}{*}{\% H}   & A & \multicolumn{1}{r|}{60.45}  & 16.63  & \multicolumn{1}{r|}{74.47}  & 18     & \multicolumn{1}{r|}{\textbf{\cellcolor{cellhighlight}{77.32}}}  & 17.73  \\ \cline{2-8} 
                                   & R & \multicolumn{1}{r|}{0}      & 0      & \multicolumn{1}{r|}{0}      & 0      & \multicolumn{1}{r|}{0}      & 0      \\ \hline
\multirow{2}{*}{\% M} & A & \multicolumn{1}{r|}{29.90}  & 20.43  & \multicolumn{1}{r|}{22.22}  & 13.94  & \multicolumn{1}{r|}{20.83}  & 16.36  \\ \cline{2-8} 
                                   & R & \multicolumn{1}{r|}{42.36}  & 4.12   & \multicolumn{1}{r|}{54.78}  & 17.09  & \multicolumn{1}{r|}{57.45}  & 18.7   \\ \hline
\multirow{2}{*}{\% L}    & A & \multicolumn{1}{r|}{9.65}   & 10.32  & \multicolumn{1}{r|}{3.31}   & 8.36   & \multicolumn{1}{r|}{\textbf{\cellcolor{cellhighlight}{1.85}}}   & 6.42   \\ \cline{2-8} 
  \hhline{~|-}
                                   & R & \multicolumn{1}{r|}{57.64}  & 4.12   & \multicolumn{1}{r|}{45.22}  & 17.09  & \multicolumn{1}{r|}{\textbf{\cellcolor{cellhighlight}{42.55}}}  & 18.70  \\ \hline
\multicolumn{8}{|c|}{\textbf{Distribution of Overall Usage Scores}} \\ \hline
\multirow{2}{*}{\% H}   & A & \multicolumn{1}{r|}{11.67}  & 33.20  & \multicolumn{1}{r|}{13.72}  & 4.54   & \multicolumn{1}{r|}{\textbf{\cellcolor{cellhighlight}{15.45}}}  & 8.27   \\ \cline{2-8} 
  \hhline{~|-}
                                   & R & \multicolumn{1}{r|}{29.03}  & 8.14   & \multicolumn{1}{r|}{38.16}  & 16.89  & 
                                   \multicolumn{1}{r|}{\textbf{\cellcolor{cellhighlight}{38.78}}}  & 12.97  \\ \hline
\multirow{2}{*}{\% M} & A & \multicolumn{1}{r|}{15.29}  & 9.53   & \multicolumn{1}{r|}{11.18}  & 9.81   & \multicolumn{1}{r|}{16.46}  & 12.95  \\ \cline{2-8} 
                                   & R & \multicolumn{1}{r|}{55.54}  & 3.90   & \multicolumn{1}{r|}{54.11}  & 13.16  & \multicolumn{1}{r|}{54.82}  & 8.59   \\ \hline
\multirow{2}{*}{\% L}    & A & \multicolumn{1}{r|}{73.04}  & 11.37  & \multicolumn{1}{r|}{75.09}  & 7.87   & \multicolumn{1}{r|}{\textbf{\cellcolor{cellhighlight}{68.08}}}  & 16.45  \\ \cline{2-8} 
  \hhline{~|-}
                                   & R & \multicolumn{1}{r|}{57.64}  & 4.12  & \multicolumn{1}{r|}{45.22}  & 17.09  & \multicolumn{1}{r|}{\textbf{\cellcolor{cellhighlight}{42.55}}}  & 18.70  \\ \hline
\end{tabular}


\vspace{4mm}
\end{table}

\add{Table~\ref{tab:prepared-subsets}} \edit{and Figure~\ref{fig:attribute-record-subset-count} show the sizes of subsets (total number of attributes out of 42 and records out of 1000) selected by the participants (Figure~\ref{fig:attribute-record-subset-count}a) and the distribution of \emph{high}, \emph{medium}, \emph{low} values of attribute- and record-level quality and usage information (Figure~\ref{fig:attribute-record-subset-count}b).} Validating \textbf{H1}, \emph{D}\textcolor{DD}{\cut{($\mu_A$=6.5, $\sigma_A$=2.32; $\mu_R$=642.17, $\sigma_R$=249.18)}} chose the \edit{fewest~}attributes and records followed by \emph{Q}\textcolor{QQ}{\cut{($\mu_A$=7.92, $\sigma_A$=2.19; $\mu_R$=710.83, $\sigma_R$=282.85)}} followed by \emph{B}\textcolor{BB}{\cut{($\mu_A$=9.17, $\sigma_A$=2.44; $\mu_R$=866.17, $\sigma_R$=253.08)}}\cut{, where $\mu_{A, R}$, $\sigma_{A, R}$ = mean and standard deviation of number of attributes (A) and records (R), respectively}. 

Furthermore, \emph{D} \textcolor{DD}{\cut{($\mu_A$=77.32, $\sigma_A$=17.73)}} chose a higher percentage of \emph{high overall quality} attributes than \emph{Q}\cut{}\textcolor{QQ}{\cut{($\mu_A$=74.47, $\sigma_A$=18)}} than \emph{B}\cut{}\textcolor{BB}{\cut{($\mu_A$=60.45, $\sigma_A$=16.63)}}. Because the dataset was sparse (a majority of values in each record were empty), no record had a \emph{high overall quality} score, hence the corresponding $\mu_{R}$, $\sigma_{R}$ values for \textcolor{BB}{B}, \textcolor{QQ}{Q}, \textcolor{DD}{D} were all 0. \emph{D}\cut{}\textcolor{DD}{\cut{($\mu_A$=15.45, $\sigma_A$=8.27; $\mu_R$=38.78, $\sigma_R$=12.97)}} also chose a higher percentage of \emph{high overall usage} attributes and records than \emph{Q}\cut{}\textcolor{QQ}{\cut{($\mu_A$=13.72, $\sigma_A$=4.54; $\mu_R$=38.16, $\sigma_R$=16.89)}} than \emph{B}\cut{}\textcolor{BB}{\cut{($\mu_A$=11.67, $\sigma_A$=33.20; $\mu_R$=29.03, $\sigma_R$=8.14)}}. These results validate \textbf{H3}.
 
Similarly, \emph{D}\cut{}\textcolor{DD}{\cut{($\mu_A$=1.85, $\sigma_A$=6.42; $\mu_R$=42.55, $\sigma_R$=18.70)}}\cut{also} chose a lower percentage of \emph{low overall quality} attributes and records than \emph{Q}\cut{}\textcolor{QQ}{\cut{($\mu_A$=3.31, $\sigma_A$=8.36; $\mu_R$=45.22, $\sigma_R$=17.09)}} than \emph{B}\cut{}\textcolor{BB}{\cut{($\mu_A$=9.65, $\sigma_A$=10.32; $\mu_R$=57.64, $\sigma_R$=4.12)}}. Furthermore, \emph{D}\cut{}\textcolor{DD}{\cut{($\mu_A$=68.08, $\sigma_A$=16.45; $\mu_R$=42.55, $\sigma_R$=18.70)}}\cut{also} chose a lower percentage of \emph{low overall usage} attributes and records than \emph{Q}\cut{}\textcolor{QQ}{\cut{($\mu_A$=75.09, $\sigma_A$=7.87; $\mu_R$=45.22, $\sigma_R$=17.09)}} and \emph{B}\cut{}\textcolor{BB}{\cut{($\mu_A$=73.04, $\sigma_A$=11.37; $\mu_R$=57.64, $\sigma_R$=4.12)}}, validating \textbf{H2}. These findings suggest that quality and usage information nudged users to prepare smaller\edit{,} more effective subsets.

\subsubsection{Task Fidelity Scores}

\begin{figure}[ht]
    \centering
    \includegraphics[width=0.85\columnwidth]{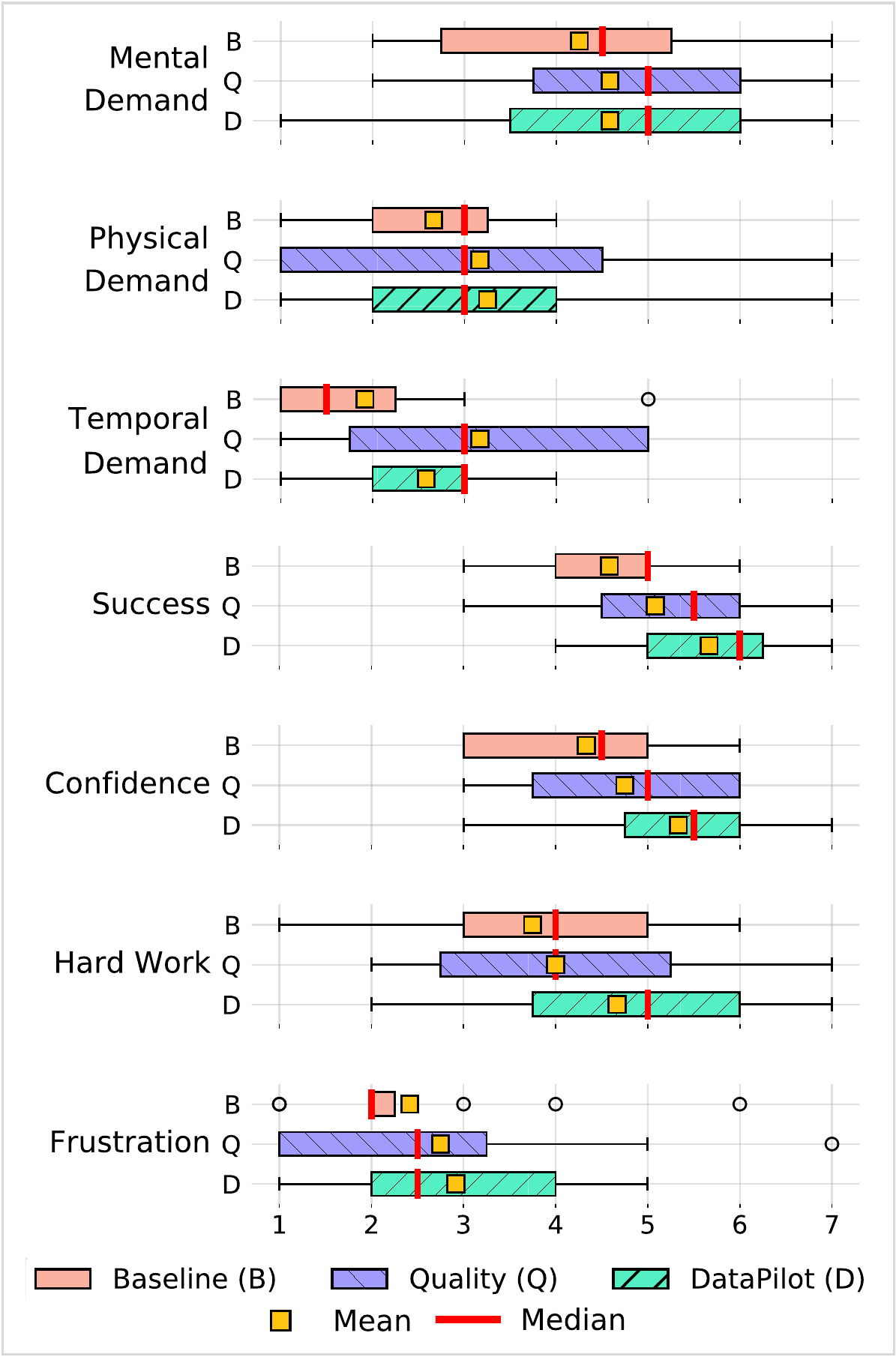}
    \caption{Assessment of the fidelity of the study task as reported by participants on a seven-point \emph{Disagree}~(1) to \emph{Agree}~(7) scale. \emph{D} participants reported higher or comparable mental demand, hard work, and frustration but greater success and confidence at the end of the task than \emph{Q} than \emph{B}.}
    \Description{Assessment of the fidelity of the study task as reported by user study participants on a seven-point Disagree to Agree scale. Seven metrics (Mental Demand, Physical Demand, Temporal Demand, Success, Confidence, Hard Work, and Frustration) are shown along with the distributions of corresponding participant-reported scores as a box plot; there is one box-plot for each study condition and it is also colored based on it. A key takeaway was that participants in the DataPilot (D) condition reported higher or comparable mental demand, hard work, and frustration but greater success and confidence at the end of the task than Quality (Q) and Baseline (B) conditions.}
    \label{fig:task-fidelity}
\end{figure}

Figure~\ref{fig:task-fidelity} shows participant feedback~\edit{on} the fidelity of the task on a seven-point \emph{Disagree}~(1) to \emph{Agree}~(7) scale.
\emph{D} reported higher or comparable mental demand (\textcolor{DD}{$M_D$=5}; \textcolor{QQ}{$M_Q$=5}; \textcolor{BB}{$M_B$=4.5}; M=median), hard work (\textcolor{DD}{$M_D$=5}; \textcolor{QQ}{$M_Q$=4}; \textcolor{BB}{$M_B$=4}), and frustration (\textcolor{DD}{$M_D$=2.5}; \textcolor{QQ}{$M_Q$=2.5}; \textcolor{BB}{$M_B$=2}) than \emph{Q} than \emph{B}, finding some evidence in support of \textbf{H5}. We attribute this result to the increased complexity due to additional user interface elements in \emph{D}, that may have affected users' cognitive load. However, \emph{D} reported greater success (\textcolor{DD}{$M_D$=6}; \textcolor{QQ}{$M_Q$=5.5}; \textcolor{BB}{$M_B$=5}) and confidence (\textcolor{DD}{$M_D$=5.5}; \textcolor{QQ}{$M_Q$=5}; \textcolor{BB}{$M_B$=4.5}) in the end, validating \textbf{H4} and suggesting that the auxiliary information helped participants perform the task more effectively. 




\subsubsection{Importance of General, Quality, and Usage Information}

\begin{figure*}[ht]
    \centering
    \includegraphics[width=\textwidth]{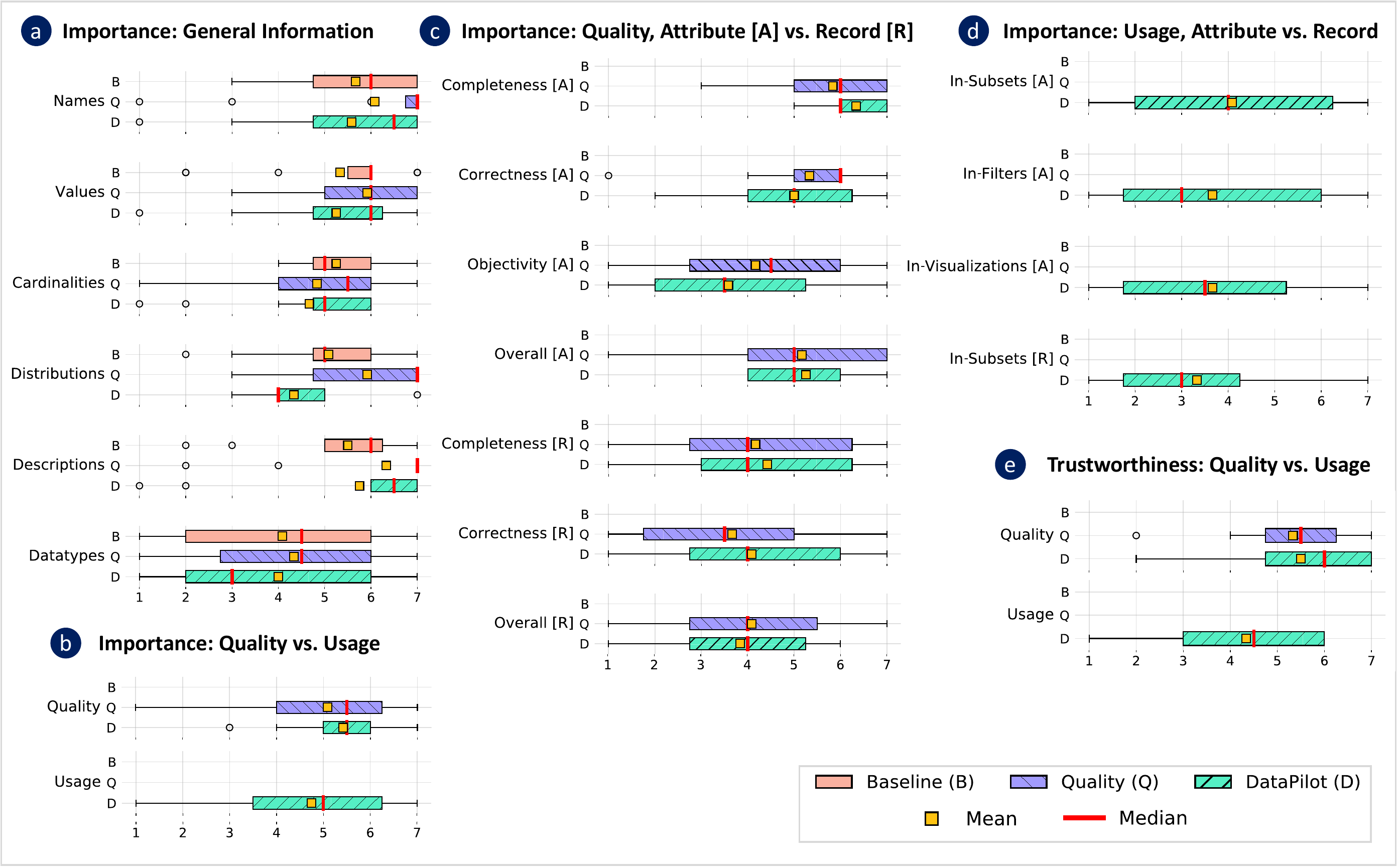}
    \caption{Importance and trustworthiness scores of general, quality and usage information for attributes and records across the three study conditions. There are no box plots for some study conditions, e.g., \textcolor{BB}{Baseline (B)} in (b)-(e), as they were not applicable.}
    \Description{Importance and trustworthiness scores of general, quality and usage information for attributes and records across the three study conditions. As in Figure 4, there are multiple metrics that are grouped into five high-level themes (Importance of General Information; Importance of Quality versus Usage; Importance of Quality across Attributes and Records; Importance of Usage across Attributes and Records; and Trustworthiness of Quality versus Usage). Each of these themes comprises multiple metrics, e.g., Importance of General Information consists of six metrics (Name; Values; Cardinalities; Distributions; Descriptions; and Datatypes). Each of these metrics is complemented with box-plots showing the distribution of corresponding metric scores as reported by user study participants on a seven-point Disagree to Agree scale, grouped and colored by the study condition assigned to them. For some metrics there is no box plot as it was not applicable. A key takeaway was that participants found Quality information more useful than Usage information.}
    \label{fig:importance-scores}
\end{figure*}

We asked participants about the importance of different kinds of general, quality, and usage information in the interface on a \emph{Not at all important}~(1) to \emph{Very important}~(7) scale. Except attribute \emph{datatypes}, other general information such as attribute \emph{names}, \emph{values}, \emph{distributions}, \emph{cardinalities}, and \emph{descriptions} were mostly useful (Figure~\ref{fig:importance-scores}a).

Figures~\ref{fig:importance-scores}b,~\ref{fig:importance-scores}c show that overall, both \emph{Q} and \emph{D} participants found quality information to be useful (\textcolor{DD}{$M_D$=5}; \textcolor{QQ}{$M_Q$=5}; M=median). At the attribute-level, \emph{completeness} (\textcolor{DD}{$M_D$=6}; \textcolor{QQ}{$M_Q$=6}) was more important than \emph{correctness} (\textcolor{DD}{$M_D$=5}; \textcolor{QQ}{$M_Q$=6}) and \emph{overall} (\textcolor{DD}{$M_D$=5}; \textcolor{QQ}{$M_Q$=5}), while \emph{objectivity} (\textcolor{DD}{$M_D$=3.5}; \textcolor{QQ}{$M_Q$=4.5}) received mixed scores. Many participants \edit{felt} completeness \edit{was} the most important\cut{dimension} (\pq{5}, \pe{3,6,9,10}) because \emph{``[they were] not the one who set the rules for correctness and objectivity''} (\pe{6}). 
Scores were mixed for the~record-level dimensions: \emph{overall} (\textcolor{DD}{$M_D$=4}; \textcolor{QQ}{$M_Q$=4}), \emph{correctness} (\textcolor{DD}{$M_D$=4}; \textcolor{QQ}{$M_Q$=3.5}), and \emph{completeness} (\textcolor{DD}{$M_D$=4}; \textcolor{QQ}{$M_Q$=4}). \pq{4} tried to make their subset as authentic as possible with mostly complete records but \pq{7} did not as they felt it would be counterproductive after applying attribute-level filters. \emph{B} participants, when~\cut{they were}presented with quality information during the debriefing\cut{interview}, stated that they either assumed there were no missing values (\pc{2,10}), forgot to look for them and vowed to be more alert next time~(\pc{9}), or thought of but ignored them~(\pc{4,7}). 

Figures~\ref{fig:importance-scores}b,~\ref{fig:importance-scores}d show that overall, \emph{D} participants had mixed \nobreak feedback about the usage information (\textcolor{DD}{$M_D$=5}; M=median). \pe{2,7,8} found them useful, \pe{1} not so much, and \pe{3,4,5,10} raised concerns \edit{about} bias and loss of originality, suggesting \edit{usage} be provided with care in specific situations. At the attribute-level, \emph{overall} (\textcolor{DD}{$M_D$=5}) was more important than \emph{in-subsets} (\textcolor{DD}{$M_D$=4}), \emph{in-visualizations} (\textcolor{DD}{$M_D$=3.5}), and \emph{in-filters} (\textcolor{DD}{$M_D$=3}). Most participants also stated \emph{overall} to be the most important dimension except \pe{6} who \emph{``went for the highest [usage] in filters.''} Participants found the record-level dimensions less useful (\emph{in-subsets}: \textcolor{DD}{$M_D$=3}). \emph{Q} and \emph{B} participants, when they were presented simulated usage information during the debriefing interview reflected that usage can \emph{``give [them] more confidence in selecting attributes''} (\pq{4}), help verify their work (\pq{1}), and be guided by others' work (\pc{8}, \pq{2}). 
Overall, participants found quality to be more important than usage, as noted by \pe{4}, \emph{``Data quality is way more important in our daily life and only if there are several people working on the same dataset or tool, then data usage may be helpful''} and \pq{12}, \emph{``If an attribute is of high quality but low usage, I would still pick that attribute.''} Collectively, these results validate \textbf{H6}.

\section{Discussion}


\subsection{Participant strategies to select subsets.}

\bpstart{Only quality.} Ten \emph{Q} and two \emph{D} participants relied only on quality: \pq{4} discarded incomplete records by applying a \emph{completeness} filter, \pe{1,5} filtered out attributes based on \emph{completeness}, and \pq{3} looked for high \includegraphics[height=1.2\fontcharht\font`\B]{figures/circle-green-eps-converted-to.pdf} \emph{overall} quality attributes \edit{via} the colored glyphs.

\paragraphHeadingSpace\bpstart{Only usage.} No \emph{D} participant relied only on usage, vindicating our domain experts' judgment that quality is still the most critical information during data-driven preparation and analysis.

\paragraphHeadingSpace\bpstart{Both quality and usage.} Seven out of twelve \emph{D} participants used both quality and usage. For example, \pe{9} applied quality filters and then focused on the bi-colored glyphs to avoid the low \includegraphics[height=1.2\fontcharht\font`\B]{figures/circle-red-eps-converted-to.pdf} usage attributes. \pe{8} sorted attributes by \emph{overall} usage scores before applying quality filters, \pe{11} inspected the \emph{in-subsets} usage dimension after applying quality filters, and \pe{4,6} used quality to make initial selections and then usage to verify and validate.

\paragraphHeadingSpace\bpstart{Neither quality nor usage.} All \emph{B} (as they did not see any auxiliary information), two \emph{Q} (\pq{1,2}), and three \emph{D} participants (\pe{2,3,10}) primarily relied on general attribute information (e.g., attribute names and descriptions) and correlation and trend analysis (e.g., by creating visualizations) to select their subsets.

\paragraphHeadingSpace\bpstart{Other non data-driven strategies.} Participants also \edit{relied} on their preconceptions (\pq{3}, \pe{4}), common sense (\pe{1}), intuition (\pe{2,3,5,7}), and trial and error practices (\pe{3,6}) as secondary strategies, highlighting the role of human-intelligence in data-driven analysis. Modeling auxiliary information such as quality, usage can minimize uncertainties and inconsistencies associated with such strategies.

\subsection{Reflections on the three-step workflow.}
We designed \app to facilitate a three-step workflow: (1)~\textbf{\edit{Review Raw Data}}, (2)~\textbf{Review \add{Selected} Subset}, and (3)~\textbf{Create Dashboard}, that forces the user to first select attributes and records of interest before creating visualizations. This approach deviates from~\cut{common}\edit{many} visual data analysis workflows wherein either there are no steps and no means to (de)select attributes (e.g., Voyager~\cite{wongsuphasawat2015voyager}, Lumos~\cite{narechania2021lumos}) or all attributes are selected by default and users can only hide irrelevant ones (e.g., Tableau~\cite{tableau}). 
\add{In Power BI~\cite{powerbi}, users are first presented with a \emph{separate} ``Query Editor'' to transform data before analysis; however, because data preparation is an iterative process, users can utilize the ``Transform Data'' feature to open the ``Query Editor'' window at any time during analysis. Tableau Prep~\cite{tableauprep} on the other hand, is a \emph{separate} tool that provides data preparation affordances before use in Tableau~\cite{tableau}; Tableau Prep does, however, have the ``Open sample in Tableau Desktop'' feature for users to test how a sample of the data currently under preparation would appear during the eventual analysis in Tableau.}

Regarding \app's \add{flexible, three-step} workflow, participants found it useful as \emph{``it made [them] think about what is important, whereas in Tableau, one imports the dataset and then immediately goes on to the chart making step, dragging and dropping attributes hoping to find something interesting''} (\pc{1}). \pe{11} commented, \emph{``I think \textbf{Step 1} is the most important step for me in creation of the dataset. I know that charts are very important but they are appropriately put at the third step otherwise it would get overwhelming while having all the attributes.''} For some participants, the workflow \edit{helped them} focus on individual aspects of data (\pe{10}), \edit{was time saving because they could quickly identify if the attributes that sounded important and interesting were not worth looking at (\pq{10}), and prevented junk data from reaching the chart creation step (\pe{7})}. \pc{3} used the \emph{\edit{Review Raw Data}} step as more exploratory and found it convenient \emph{``to move back and forth between the steps to remove certain attributes that [they] don't need''} and~\cut{also}liked the \emph{Review \add{Selected} Subset} step as \emph{``they get to see just their smaller, cleaner subset of data.''} However some others requested support for \emph{``creating charts using all attributes''}~(\pq{10}) so that \emph{``[they] don't have to tab back and forth''}~(\pe{10}).

\subsection{\edit{Trust, bias, convergence, ethics concerns.}}
We asked \emph{Q} and \emph{D} participants to rate the trustworthiness of the auxiliary information they interacted with. As shown in Figure~\ref{fig:importance-scores}e, both \emph{Q} and \emph{D} participants found quality (\textcolor{QQ}{$M_Q$=5.5}; \textcolor{DD}{$M_D$=6}) to be more trustworthy than usage\cut{information} (\textcolor{DD}{$M_D$=4.5}). 
Whereas \pe{12} simply \emph{``trusted the overall [quality] score,''} some others exhibited hesitation in trusting the quality scores, referencing the preconfigured constraints for the \emph{correctness} and \emph{objectivity} dimensions and the lack of clarity around how these were defined; some participants stated that \emph{``[they] don't trust [their] manager or the settings they've made''}~(\pe{3}) as \emph{``they may not be doing it in a reliable or an unbiased way''~(\pq{5}).}

\cut{Some}Participants similarly expressed a lack of trust \edit{for} usage, particularly~\add{about} the behavior or decisions of other people since they do not know them (\pe{10}), their experience (\pq{9}), their expertise (\pq{1,9}, \pe{1}), \edit{or} their tasks (\pq{5}, \pe{4,10,12}). \pe{10} \edit{noted}, \emph{``I'll trust [usage information] if I know the ten people working with me and everyone is doing this [same] task.''} 
\pe{1} commented, \emph{``I don't want to depend on previous people's understanding of the system.''} 
\pe{12} wondered if the people who created these data had different objectives, hence \emph{``[they] might be using data to get totally different insights as compared to another team, so [they weren't] comfortable trusting~usage.''} 

In addition to discussing trust, participants reported usage insights as a potential source of bias. \pq{7} commented \edit{that} \emph{``[usage information] would definitely bias the perspective of people who are doing data exploration themselves...it is not necessarily expert exploration but leaning towards exploitation.''} \pq{5} \edit{explained that} \emph{``If I have to see how other people are using the dataset, then that would bias me.''} \pq{7} referenced convergence theory~\cite{manstead1995blackwell}, \emph{``If I select an attribute and notice that not many people have selected it in the past, then I will be compelled to deselect it. On the other hand, if previous users did not select an otherwise relevant and important attribute, then it will likely stay lowly consumed due to the convergence effect.''}

Privacy was another concern reported by both participants and our domain experts. Applicability and availability aside, participants questioned if it is ethical to extract usage data and share it publicly, even if it is anonymous and aggregated. They also raised concerns around users unwilling to share their usage history\cut{with others}. In fact, Thom-Santelli et al.~\cite{thom2010you} found that tension may arise between users (experts and novice contributors) especially when they perceive a threat (e.g., when a workplace bonus or promotion is at stake).

With the above considerations, we believe providing quality and usage information during subset selection and analysis was an effective way to alleviate many of our domain experts' concerns; however, providing additional context and advanced configuration capabilities is the next important step, especially for power users.


\section{Lessons Learned}


\bpstart{Give importance to data preparation, not just visual data analysis.}
\cut{Bad data is costly~\cite{badDataCostsMillions} and m}Motivated by the ``garbage in, garbage out'' principle, applications need more data work~\cite{sambasivan2021everyone}, as was also echoed by our users. Achieving this balance in data analysis tools is desirable and raises the need for functionality to inspect and interact with auxiliary information such as \app's quality and usage. Furthermore, the workflow change to prioritize and integrate data preparation (e.g., by first selecting relevant attributes and records) into current visual data analysis tool workflows should be considered.

\paragraphHeadingSpace\bpstart{Present quality information for accurate and objective analysis.}
Evidenced by positive user feedback in terms of both importance and trust, quality\cut{information} can nudge users to pause and reflect upon the state of their data and take suitable corrective actions (e.g., clean the data) before performing analysis\edit{, or not use the data at all}.

\paragraphHeadingSpace\bpstart{Present usage information, albeit with care and caution.}
Users had mixed feedback about usage in terms of both importance and trust. While usage information can help nudge users to draw inspiration from previous judgments, it can also be counterproductive, leading to the propagation of negative practices (e.g., biased analytic behaviors) or hampering creativity and originality (e.g., preventing fresh, new ideas to flourish) within the organization. One way to achieve a good balance is to present usage information on demand, e.g., \emph{``when I get to know the dataset, I want to hide the second half [usage] of the [bi-colored glyph]''} (\pq{5}).


\paragraphHeadingSpace\bpstart{Different tasks call for flexible information.}
Data preparation and subsequent analysis are contextualized by specific task requirements and user preferences. Tools must provide the desired flexibility to, e.g., assign different weights to constituent quality and usage dimensions or determine different aggregation functions (max, mean) for calculating the \emph{overall} scores, override preset constraints to assess \emph{correctness} and \emph{objectivity}, and modify the thresholds to determine the \emph{high}, \emph{medium}, \emph{low} score cutoffs.

\paragraphHeadingSpace\bpstart{Additional degrees of guidance towards quality and usage characteristics may be pursued.}
Participants found that the visual interactive affordances for quality and usage information were useful for ``orienting''~\cite{ceneda2016characterizing} them with the dataset. Future tools could also explore higher guidance degrees (e.g., ``directing'' or ``prescribing''~\cite{ceneda2016characterizing}) to more actively steer users rather than just passively increasing awareness hoping they react in good conscience.

\paragraphHeadingSpace\bpstart{Organizations should start building and utilizing collective intelligence.}
Organizations should capture usage logs across their databases and applications to model usage information \edit{to increase general user awareness within and across teams.}\cut{and enable different user endpoints, e.g., a \emph{``Data in Use''} tab that shows the quality as well as popularity (usage) of attributes and records across multiple applications, increasing general user awareness within and across teams.}
Moreover, as new data is regularly ingested (and old data archived), persistence and subsequent monitoring of the auxiliary information can help detect shifting trends, flag anomalous events, and generally track data provenance, ensuring accurate and efficient data management.

\section{Limitations and Future Work}
We noted five key limitations related to our studies and tool.

\paragraphHeadingSpace\bpstart{\add{Study limitations.}}
One, during the design study, we structured our interviews\cut{with domain experts} in a group setting; while these interviewees had a strong working relationship, this mode of discussion may result in complexities around gender and organizational hierarchies and must be accounted for (e.g., through 1-1 interviews). Two, during the user study, we made a fair assumption that our participants were unfamiliar with the dataset and hence exhibited similar expertise, supporting internal validity; however, this assumption may not hold true for real-world cases from an external validity standpoint~\cite{schmuckler2001ecological}. Future work may incorporate weighting mechanisms to more accurately approximate usage based on recency of use (e.g., give more importance to recent data), user expertise (favor experts), or the criticality of the application that utilized the data.
Three, because our participants were not domain experts, we did not have experts assess the selected subsets or final dashboards; future user studies with domain experts should further evaluate the quality of these results. 
Four, we focused on the particular task of exporting visualizations for a dashboard, which may have impacted how the attributes and records were chosen; future work should consider developing additional tools to study downstream analytics tasks other than subset selection such as ranking and clustering. 
Five, although data quality dimensions are not easily transferable across domains~\cite{leonelli2019data}, study participants suggested utilizing \app for searching datasets on Kaggle~\cite{kaggle2022misc} (\pq{10}), generating tutorials for software use (\pq{2}, \pe{5}), and preparing fair and accurate datasets for machine learning (\pe{4}), which are left for future work.

\paragraphHeadingSpace\bpstart{\add{Tool limitations.}}
One, \app currently supports quality information for tabular datasets; future work may explore other structured (e.g., relational databases) and unstructured (e.g., text, documents) datasets. 
Two, there are also other data-dependent (e.g., consistent representation, ease of manipulation, and timeliness~\cite{pipino2002data}) and process-dependent (e.g., data collection~\cite{friedman2010data}) aspects of quality, and similarly, other aspects of usage beyond a subset selection and dashboard building task (e.g., co-usage frequencies of multiple attributes in a visualization, frequency of visualization interactions such as zooming and panning~\cite{brehmer2013multi}) that may be operationalized in the future. 
Three, \app's dashboard view currently supports creation of disconnected visualizations; future work may explore the effects of interactive affordances such as brushing and linking.
Four, to ensure scalability, \app computes quality and usage scores using SQL queries (objectivity is computed using both SQL and Python); some of these new dimensions, however, may be difficult to operationalize using SQL and hence challenging to scale. 
Five, the completeness, correctness, and objectivity quality constraints are currently hard-coded in the \app source code in a SQL-like syntax. Future work can provide interactive affordances for the user to configure these constraints and also clean the data (e.g., handle missing values) directly via the user interface.


\section{Conclusion}


\app is a visual data preparation and analysis tool that models two kinds of auxiliary information, \emph{quality}~\cut{ -- ``the validity and appropriateness of data required to perform certain analytical tasks''} and \emph{usage},~\cut{ -- ``the historical utilization characteristics of data across multiple users''} to assist users in analyzing a large and unfamiliar tabular dataset, selecting a relevant subset, and building a visualization dashboard. 
\app is an outcome of a design study with 14 data workers over a period of two months who communicated the importance of data quality and also suggested surfacing data usage characteristics to guide users during data preparation. 
A user study with 36 participants suggested that quality and usage information together help users select smaller, effective data subsets with greater success and confidence. 
\cut{To balance exploration versus exploitation, our participants sounded caution about users relying excessively on usage information.}
We posit that through quality \emph{and} usage information, organizations can build collective intelligence, increasing transparency and accuracy to foster closer collaboration and cooperation among teams. 
\cut{Challenging convention, our findings also call for visual data analysis tools to prioritize data preparation affordances.}


\begin{acks}
We thank our data worker interviewees, study participants, members of the Georgia Tech Visualization Lab, and anonymous reviewers for providing feedback at different stages of this work.
\end{acks}

\balance
\bibliographystyle{ACM-Reference-Format}
\bibliography{paper}


\begin{thebibliography}{129}


\ifx \showCODEN    \undefined \def \showCODEN     #1{\unskip}     \fi
\ifx \showDOI      \undefined \def \showDOI       #1{#1}\fi
\ifx \showISBNx    \undefined \def \showISBNx     #1{\unskip}     \fi
\ifx \showISBNxiii \undefined \def \showISBNxiii  #1{\unskip}     \fi
\ifx \showISSN     \undefined \def \showISSN      #1{\unskip}     \fi
\ifx \showLCCN     \undefined \def \showLCCN      #1{\unskip}     \fi
\ifx \shownote     \undefined \def \shownote      #1{#1}          \fi
\ifx \showarticletitle \undefined \def \showarticletitle #1{#1}   \fi
\ifx \showURL      \undefined \def \showURL       {\relax}        \fi
\providecommand\bibfield[2]{#2}
\providecommand\bibinfo[2]{#2}
\providecommand\natexlab[1]{#1}
\providecommand\showeprint[2][]{arXiv:#2}

\bibitem[por(2012)]%
        {portal2012ogdindia}
 \bibinfo{year}{2012}\natexlab{}.
\newblock \bibinfo{title}{Open Government Data (OGD) Platform India}.
\newblock
\newblock
\urldef\tempurl%
\url{https://data.gov.in/}
\showURL{%
Retrieved 21-Nov-2022 from \tempurl}


\bibitem[por(2021)]%
        {portal2021cern}
 \bibinfo{year}{2021}\natexlab{}.
\newblock \bibinfo{title}{CERN Open Data Portal}.
\newblock
\newblock
\urldef\tempurl%
\url{https://opendata.cern.ch}
\showURL{%
Retrieved 21-Nov-2022 from \tempurl}


\bibitem[Abedjan et~al\mbox{.}(2016)]%
        {abedjan2016detecting}
\bibfield{author}{\bibinfo{person}{Ziawasch Abedjan}, \bibinfo{person}{Xu Chu},
  \bibinfo{person}{Dong Deng}, \bibinfo{person}{Raul~Castro Fernandez},
  \bibinfo{person}{Ihab~F. Ilyas}, \bibinfo{person}{Mourad Ouzzani},
  \bibinfo{person}{Paolo Papotti}, \bibinfo{person}{Michael Stonebraker}, {and}
  \bibinfo{person}{Nan Tang}.} \bibinfo{year}{2016}\natexlab{}.
\newblock \showarticletitle{Detecting Data Errors: Where Are We and What Needs
  to Be Done?}
\newblock \bibinfo{journal}{\emph{Proc. VLDB Endow.}} \bibinfo{volume}{9},
  \bibinfo{number}{12} (\bibinfo{date}{aug} \bibinfo{year}{2016}),
  \bibinfo{pages}{993–1004}.
\newblock
\showISSN{2150-8097}
\urldef\tempurl%
\url{https://doi.org/10.14778/2994509.2994518}
\showDOI{\tempurl}


\bibitem[Alexander et~al\mbox{.}(2009)]%
        {alexander2009revisiting}
\bibfield{author}{\bibinfo{person}{Jason Alexander}, \bibinfo{person}{Andy
  Cockburn}, \bibinfo{person}{Stephen Fitchett}, \bibinfo{person}{Carl Gutwin},
  {and} \bibinfo{person}{Saul Greenberg}.} \bibinfo{year}{2009}\natexlab{}.
\newblock \showarticletitle{Revisiting read wear: analysis, design, and
  evaluation of a footprints scrollbar}. In
  \bibinfo{booktitle}{\emph{Proceedings of the SIGCHI Conference on Human
  Factors in Computing Systems}}. \bibinfo{pages}{1665--1674}.
\newblock


\bibitem[Allen et~al\mbox{.}(2019)]%
        {allen2019raincloud}
\bibfield{author}{\bibinfo{person}{Micah Allen}, \bibinfo{person}{Davide
  Poggiali}, \bibinfo{person}{Kirstie Whitaker}, \bibinfo{person}{Tom~Rhys
  Marshall}, {and} \bibinfo{person}{Rogier~A Kievit}.}
  \bibinfo{year}{2019}\natexlab{}.
\newblock \showarticletitle{Raincloud plots: a multi-platform tool for robust
  data visualization}.
\newblock \bibinfo{journal}{\emph{Wellcome open research}}  \bibinfo{volume}{4}
  (\bibinfo{year}{2019}).
\newblock


\bibitem[Almahmoud et~al\mbox{.}(2021)]%
        {almahmoud2021teams}
\bibfield{author}{\bibinfo{person}{Jumana Almahmoud}, \bibinfo{person}{Robert
  DeLine}, {and} \bibinfo{person}{Steven~M Drucker}.}
  \bibinfo{year}{2021}\natexlab{}.
\newblock \showarticletitle{How Teams Communicate about the Quality of ML
  Models: A Case Study at an International Technology Company}.
\newblock \bibinfo{journal}{\emph{Proceedings of the ACM on Human-Computer
  Interaction}} \bibinfo{volume}{5}, \bibinfo{number}{GROUP}
  (\bibinfo{year}{2021}), \bibinfo{pages}{1--24}.
\newblock


\bibitem[Amazon(2022)]%
        {amazondeequ}
\bibfield{author}{\bibinfo{person}{Amazon}.} \bibinfo{year}{2022}\natexlab{}.
\newblock \bibinfo{title}{Deequ}.
\newblock
\newblock
\urldef\tempurl%
\url{https://aws.amazon.com/blogs/big-data/test-data-quality-at-scale-with-deequ/}
\showURL{%
Retrieved July 29, 2022 from \tempurl}


\bibitem[Arbesser et~al\mbox{.}(2017)]%
        {arbesser2016visplause}
\bibfield{author}{\bibinfo{person}{Clemens Arbesser}, \bibinfo{person}{Florian
  Spechtenhauser}, \bibinfo{person}{Thomas Mühlbacher}, {and}
  \bibinfo{person}{Harald Piringer}.} \bibinfo{year}{2017}\natexlab{}.
\newblock \showarticletitle{Visplause: Visual Data Quality Assessment of Many
  Time Series Using Plausibility Checks}.
\newblock \bibinfo{journal}{\emph{IEEE Transactions on Visualization and
  Computer Graphics}} \bibinfo{volume}{23}, \bibinfo{number}{1}
  (\bibinfo{year}{2017}), \bibinfo{pages}{641--650}.
\newblock
\urldef\tempurl%
\url{https://doi.org/10.1109/TVCG.2016.2598592}
\showDOI{\tempurl}


\bibitem[Arnold et~al\mbox{.}(2019)]%
        {arnold2019factsheets}
\bibfield{author}{\bibinfo{person}{Matthew Arnold}, \bibinfo{person}{Rachel~KE
  Bellamy}, \bibinfo{person}{Michael Hind}, \bibinfo{person}{Stephanie Houde},
  \bibinfo{person}{Sameep Mehta}, \bibinfo{person}{Aleksandra Mojsilovi{\'c}},
  \bibinfo{person}{Ravi Nair}, \bibinfo{person}{K~Natesan Ramamurthy},
  \bibinfo{person}{Alexandra Olteanu}, \bibinfo{person}{David Piorkowski},
  {et~al\mbox{.}}} \bibinfo{year}{2019}\natexlab{}.
\newblock \showarticletitle{FactSheets: Increasing trust in AI services through
  supplier's declarations of conformity}.
\newblock \bibinfo{journal}{\emph{IBM Journal of Research and Development}}
  \bibinfo{volume}{63}, \bibinfo{number}{4/5} (\bibinfo{year}{2019}),
  \bibinfo{pages}{6--1}.
\newblock


\bibitem[Arroyo et~al\mbox{.}(2006)]%
        {arroyo2006usability}
\bibfield{author}{\bibinfo{person}{Ernesto Arroyo}, \bibinfo{person}{Ted
  Selker}, {and} \bibinfo{person}{Willy Wei}.} \bibinfo{year}{2006}\natexlab{}.
\newblock \showarticletitle{Usability tool for analysis of web designs using
  mouse tracks}. In \bibinfo{booktitle}{\emph{CHI'06 extended abstracts on
  Human factors in computing systems}}. \bibinfo{pages}{484--489}.
\newblock


\bibitem[Bhardwaj et~al\mbox{.}(2014)]%
        {bhardwaj2014datahub}
\bibfield{author}{\bibinfo{person}{Anant Bhardwaj}, \bibinfo{person}{Souvik
  Bhattacherjee}, \bibinfo{person}{Amit Chavan}, \bibinfo{person}{Amol
  Deshpande}, \bibinfo{person}{Aaron~J Elmore}, \bibinfo{person}{Samuel
  Madden}, {and} \bibinfo{person}{Aditya~G Parameswaran}.}
  \bibinfo{year}{2014}\natexlab{}.
\newblock \showarticletitle{Datahub: Collaborative data science \& dataset
  version management at scale}.
\newblock \bibinfo{journal}{\emph{arXiv preprint arXiv:1409.0798}}
  (\bibinfo{year}{2014}).
\newblock


\bibitem[Borland et~al\mbox{.}(2019)]%
        {borland2019selection}
\bibfield{author}{\bibinfo{person}{David Borland}, \bibinfo{person}{Wenyuan
  Wang}, \bibinfo{person}{Jonathan Zhang}, \bibinfo{person}{Joshua Shrestha},
  {and} \bibinfo{person}{David Gotz}.} \bibinfo{year}{2019}\natexlab{}.
\newblock \showarticletitle{Selection bias tracking and detailed subset
  comparison for high-dimensional data}.
\newblock \bibinfo{journal}{\emph{IEEE Transactions on Visualization and
  Computer Graphics}} \bibinfo{volume}{26}, \bibinfo{number}{1}
  (\bibinfo{year}{2019}), \bibinfo{pages}{429--439}.
\newblock


\bibitem[Boyatzis(1998)]%
        {boyatzis1998transforming}
\bibfield{author}{\bibinfo{person}{Richard~E Boyatzis}.}
  \bibinfo{year}{1998}\natexlab{}.
\newblock \bibinfo{booktitle}{\emph{Transforming qualitative information:
  Thematic analysis and code development}}.
\newblock \bibinfo{publisher}{sage}.
\newblock


\bibitem[Brehmer and Munzner(2013)]%
        {brehmer2013multi}
\bibfield{author}{\bibinfo{person}{Matthew Brehmer} {and}
  \bibinfo{person}{Tamara Munzner}.} \bibinfo{year}{2013}\natexlab{}.
\newblock \showarticletitle{A multi-level typology of abstract visualization
  tasks}.
\newblock \bibinfo{journal}{\emph{IEEE transactions on visualization and
  computer graphics}} \bibinfo{volume}{19}, \bibinfo{number}{12}
  (\bibinfo{year}{2013}), \bibinfo{pages}{2376--2385}.
\newblock


\bibitem[Brooke et~al\mbox{.}(1996)]%
        {brooke1996sus}
\bibfield{author}{\bibinfo{person}{John Brooke} {et~al\mbox{.}}}
  \bibinfo{year}{1996}\natexlab{}.
\newblock \showarticletitle{SUS-A quick and dirty usability scale}.
\newblock \bibinfo{journal}{\emph{Usability evaluation in industry}}
  \bibinfo{volume}{189}, \bibinfo{number}{194} (\bibinfo{year}{1996}),
  \bibinfo{pages}{4--7}.
\newblock


\bibitem[Ceneda et~al\mbox{.}(2017)]%
        {ceneda2016characterizing}
\bibfield{author}{\bibinfo{person}{Davide Ceneda}, \bibinfo{person}{Theresia
  Gschwandtner}, \bibinfo{person}{Thorsten May}, \bibinfo{person}{Silvia
  Miksch}, \bibinfo{person}{Hans-Jörg Schulz}, \bibinfo{person}{Marc Streit},
  {and} \bibinfo{person}{Christian Tominski}.} \bibinfo{year}{2017}\natexlab{}.
\newblock \showarticletitle{Characterizing Guidance in Visual Analytics}.
\newblock \bibinfo{journal}{\emph{IEEE Transactions on Visualization and
  Computer Graphics}} (\bibinfo{year}{2017}).
\newblock
\urldef\tempurl%
\url{https://doi.org/10.1109/TVCG.2016.2598468}
\showDOI{\tempurl}


\bibitem[Chang et~al\mbox{.}(2019)]%
        {chang2019nist}
\bibfield{author}{\bibinfo{person}{Wo~L Chang}, \bibinfo{person}{Arnab Roy},
  \bibinfo{person}{Mark Underwood}, {et~al\mbox{.}}}
  \bibinfo{year}{2019}\natexlab{}.
\newblock \showarticletitle{NIST Big Data Interoperability Framework: Volume 4,
  Security and Privacy}.
\newblock  (\bibinfo{year}{2019}).
\newblock


\bibitem[Chu and Ilyas(2016)]%
        {chu2016qualitative}
\bibfield{author}{\bibinfo{person}{Xu Chu} {and} \bibinfo{person}{Ihab~F.
  Ilyas}.} \bibinfo{year}{2016}\natexlab{}.
\newblock \showarticletitle{Qualitative Data Cleaning}.
\newblock \bibinfo{journal}{\emph{Proc. VLDB Endow.}} \bibinfo{volume}{9},
  \bibinfo{number}{13} (\bibinfo{date}{sep} \bibinfo{year}{2016}),
  \bibinfo{pages}{1605–1608}.
\newblock
\showISSN{2150-8097}
\urldef\tempurl%
\url{https://doi.org/10.14778/3007263.3007320}
\showDOI{\tempurl}


\bibitem[Chu et~al\mbox{.}(2016)]%
        {chu2016data}
\bibfield{author}{\bibinfo{person}{Xu Chu}, \bibinfo{person}{Ihab~F. Ilyas},
  \bibinfo{person}{Sanjay Krishnan}, {and} \bibinfo{person}{Jiannan Wang}.}
  \bibinfo{year}{2016}\natexlab{}.
\newblock \showarticletitle{Data Cleaning: Overview and Emerging Challenges}.
  In \bibinfo{booktitle}{\emph{Proceedings of the 2016 International Conference
  on Management of Data}} (San Francisco, California, USA)
  \emph{(\bibinfo{series}{SIGMOD '16})}. \bibinfo{publisher}{Association for
  Computing Machinery}, \bibinfo{address}{New York, NY, USA},
  \bibinfo{pages}{2201–2206}.
\newblock
\showISBNx{9781450335317}
\urldef\tempurl%
\url{https://doi.org/10.1145/2882903.2912574}
\showDOI{\tempurl}


\bibitem[Chu et~al\mbox{.}(2013)]%
        {chu2013holistic}
\bibfield{author}{\bibinfo{person}{Xu Chu}, \bibinfo{person}{Ihab~F Ilyas},
  {and} \bibinfo{person}{Paolo Papotti}.} \bibinfo{year}{2013}\natexlab{}.
\newblock \showarticletitle{Holistic data cleaning: Putting violations into
  context}. In \bibinfo{booktitle}{\emph{2013 IEEE 29th International
  Conference on Data Engineering (ICDE)}}. IEEE, \bibinfo{pages}{458--469}.
\newblock


\bibitem[Council et~al\mbox{.}(2000)]%
        {national2000people}
\bibfield{author}{\bibinfo{person}{National~Research Council} {et~al\mbox{.}}}
  \bibinfo{year}{2000}\natexlab{}.
\newblock \bibinfo{booktitle}{\emph{How people learn: Brain, mind, experience,
  and school: Expanded edition}}.
\newblock \bibinfo{publisher}{National Academies Press}.
\newblock


\bibitem[Dastin(2018)]%
        {amazonHiringAISexist}
\bibfield{author}{\bibinfo{person}{Jeffrey Dastin}.}
  \bibinfo{year}{2018}\natexlab{}.
\newblock \bibinfo{title}{Amazon scraps secret AI recruiting tool that showed
  bias against women}.
\newblock
\newblock
\urldef\tempurl%
\url{https://www.reuters.com/article/us-amazon-com-jobs-automation-insight/amazon-scraps-secret-ai-recruiting-tool-that-showed-bias-against-women-idUSKCN1MK08G}
\showURL{%
\tempurl}


\bibitem[De~Cock(2011)]%
        {de2011ames}
\bibfield{author}{\bibinfo{person}{Dean De~Cock}.}
  \bibinfo{year}{2011}\natexlab{}.
\newblock \showarticletitle{Ames, Iowa: Alternative to the Boston housing data
  as an end of semester regression project}.
\newblock \bibinfo{journal}{\emph{Journal of Statistics Education}}
  \bibinfo{volume}{19}, \bibinfo{number}{3} (\bibinfo{year}{2011}).
\newblock
\urldef\tempurl%
\url{https://doi.org/10.1080/10691898.2011.11889627}
\showDOI{\tempurl}


\bibitem[Deng et~al\mbox{.}(2017)]%
        {deng2017data}
\bibfield{author}{\bibinfo{person}{Dong Deng}, \bibinfo{person}{Raul~Castro
  Fernandez}, \bibinfo{person}{Ziawasch Abedjan}, \bibinfo{person}{Sibo Wang},
  \bibinfo{person}{Michael Stonebraker}, \bibinfo{person}{Ahmed~K Elmagarmid},
  \bibinfo{person}{Ihab~F Ilyas}, \bibinfo{person}{Samuel Madden},
  \bibinfo{person}{Mourad Ouzzani}, {and} \bibinfo{person}{Nan Tang}.}
  \bibinfo{year}{2017}\natexlab{}.
\newblock \showarticletitle{The Data Civilizer System.}. In
  \bibinfo{booktitle}{\emph{Cidr}}.
\newblock


\bibitem[Dremio(2022)]%
        {dremio}
\bibfield{author}{\bibinfo{person}{Dremio}.} \bibinfo{year}{2022}\natexlab{}.
\newblock \bibinfo{title}{Dremio}.
\newblock
\newblock
\urldef\tempurl%
\url{https://www.dremio.com/}
\showURL{%
Retrieved July 29, 2022 from \tempurl}


\bibitem[Drosos et~al\mbox{.}(2020a)]%
        {drosos2020wrex}
\bibfield{author}{\bibinfo{person}{Ian Drosos}, \bibinfo{person}{Titus Barik},
  \bibinfo{person}{Philip~J. Guo}, \bibinfo{person}{Robert DeLine}, {and}
  \bibinfo{person}{Sumit Gulwani}.} \bibinfo{year}{2020}\natexlab{a}.
\newblock \showarticletitle{Wrex: A Unified Programming-by-Example Interaction
  for Synthesizing Readable Code for Data Scientists}. In
  \bibinfo{booktitle}{\emph{Proceedings of the 2020 CHI Conference on Human
  Factors in Computing Systems}} (Honolulu, HI, USA)
  \emph{(\bibinfo{series}{CHI '20})}. \bibinfo{publisher}{Association for
  Computing Machinery}, \bibinfo{address}{New York, NY, USA},
  \bibinfo{pages}{1–12}.
\newblock
\showISBNx{9781450367080}
\urldef\tempurl%
\url{https://doi.org/10.1145/3313831.3376442}
\showDOI{\tempurl}


\bibitem[Drosos et~al\mbox{.}(2020b)]%
        {wrex2020drosos}
\bibfield{author}{\bibinfo{person}{Ian Drosos}, \bibinfo{person}{Titus Barik},
  \bibinfo{person}{Philip~J. Guo}, \bibinfo{person}{Robert DeLine}, {and}
  \bibinfo{person}{Sumit Gulwani}.} \bibinfo{year}{2020}\natexlab{b}.
\newblock \showarticletitle{Wrex: A Unified Programming-by-Example Interaction
  for Synthesizing Readable Code for Data Scientists}
  \emph{(\bibinfo{series}{CHI '20})}. \bibinfo{publisher}{Association for
  Computing Machinery}, \bibinfo{address}{New York, NY, USA},
  \bibinfo{pages}{1–12}.
\newblock
\showISBNx{9781450367080}
\urldef\tempurl%
\url{https://doi.org/10.1145/3313831.3376442}
\showDOI{\tempurl}


\bibitem[Dumais et~al\mbox{.}(2014)]%
        {dumais2014understanding}
\bibfield{author}{\bibinfo{person}{Susan Dumais}, \bibinfo{person}{Robin
  Jeffries}, \bibinfo{person}{Daniel~M Russell}, \bibinfo{person}{Diane Tang},
  {and} \bibinfo{person}{Jaime Teevan}.} \bibinfo{year}{2014}\natexlab{}.
\newblock \showarticletitle{Understanding user behavior through log data and
  analysis}.
\newblock In \bibinfo{booktitle}{\emph{Ways of Knowing in HCI}}.
  \bibinfo{publisher}{Springer}, \bibinfo{pages}{349--372}.
\newblock


\bibitem[Ehsan et~al\mbox{.}(2021)]%
        {ehsan2021expanding}
\bibfield{author}{\bibinfo{person}{Upol Ehsan}, \bibinfo{person}{Q~Vera Liao},
  \bibinfo{person}{Michael Muller}, \bibinfo{person}{Mark~O Riedl}, {and}
  \bibinfo{person}{Justin~D Weisz}.} \bibinfo{year}{2021}\natexlab{}.
\newblock \showarticletitle{Expanding explainability: towards social
  transparency in AI systems}. In \bibinfo{booktitle}{\emph{Proceedings of the
  2021 CHI Conference on Human Factors in Computing Systems}}.
\newblock


\bibitem[Elmqvist et~al\mbox{.}(2011)]%
        {elmqvist2011fluid}
\bibfield{author}{\bibinfo{person}{Niklas Elmqvist},
  \bibinfo{person}{Andrew~Vande Moere}, \bibinfo{person}{Hans-Christian
  Jetter}, \bibinfo{person}{Daniel Cernea}, \bibinfo{person}{Harald Reiterer},
  {and} \bibinfo{person}{TJ Jankun-Kelly}.} \bibinfo{year}{2011}\natexlab{}.
\newblock \showarticletitle{Fluid interaction for information visualization}.
\newblock \bibinfo{journal}{\emph{Information Visualization}}
  \bibinfo{volume}{10}, \bibinfo{number}{4} (\bibinfo{year}{2011}),
  \bibinfo{pages}{327--340}.
\newblock


\bibitem[Erickson and Kellogg(2000)]%
        {socialtranslucence2000erickson}
\bibfield{author}{\bibinfo{person}{Thomas Erickson} {and}
  \bibinfo{person}{Wendy~A. Kellogg}.} \bibinfo{year}{2000}\natexlab{}.
\newblock \showarticletitle{Social Translucence: An Approach to Designing
  Systems That Support Social Processes}.
\newblock \bibinfo{journal}{\emph{ACM Trans. Comput.-Hum. Interact.}}
  \bibinfo{volume}{7}, \bibinfo{number}{1} (\bibinfo{date}{mar}
  \bibinfo{year}{2000}).
\newblock
\showISSN{1073-0516}
\urldef\tempurl%
\url{https://doi.org/10.1145/344949.345004}
\showDOI{\tempurl}


\bibitem[Farid et~al\mbox{.}(2016)]%
        {farid2016clams}
\bibfield{author}{\bibinfo{person}{Mina Farid}, \bibinfo{person}{Alexandra
  Roatis}, \bibinfo{person}{Ihab~F. Ilyas}, \bibinfo{person}{Hella-Franziska
  Hoffmann}, {and} \bibinfo{person}{Xu Chu}.} \bibinfo{year}{2016}\natexlab{}.
\newblock \showarticletitle{CLAMS: Bringing Quality to Data Lakes}
  \emph{(\bibinfo{series}{SIGMOD '16})}. \bibinfo{publisher}{Association for
  Computing Machinery}, \bibinfo{address}{New York, NY, USA},
  \bibinfo{pages}{2089–2092}.
\newblock
\showISBNx{9781450335317}
\urldef\tempurl%
\url{https://doi.org/10.1145/2882903.2899391}
\showDOI{\tempurl}


\bibitem[Fernandez et~al\mbox{.}(2018)]%
        {fernandez2018aurum}
\bibfield{author}{\bibinfo{person}{Raul~Castro Fernandez},
  \bibinfo{person}{Ziawasch Abedjan}, \bibinfo{person}{Famien Koko},
  \bibinfo{person}{Gina Yuan}, \bibinfo{person}{Samuel Madden}, {and}
  \bibinfo{person}{Michael Stonebraker}.} \bibinfo{year}{2018}\natexlab{}.
\newblock \showarticletitle{Aurum: A data discovery system}. In
  \bibinfo{booktitle}{\emph{2018 IEEE 34th International Conference on Data
  Engineering (ICDE)}}. IEEE.
\newblock


\bibitem[Fette and Melnikov(2011)]%
        {fette2011websocket}
\bibfield{author}{\bibinfo{person}{Ian Fette} {and} \bibinfo{person}{Alexey
  Melnikov}.} \bibinfo{year}{2011}\natexlab{}.
\newblock \bibinfo{booktitle}{\emph{The websocket protocol}}.
\newblock \bibinfo{type}{{T}echnical {R}eport}.
\newblock


\bibitem[Fleckenstein et~al\mbox{.}(2018)]%
        {fleckenstein2018modern}
\bibfield{author}{\bibinfo{person}{Mike Fleckenstein},
  \bibinfo{person}{Lorraine Fellows}, {and} \bibinfo{person}{Krista Ferrante}.}
  \bibinfo{year}{2018}\natexlab{}.
\newblock \showarticletitle{Data Quality}.
\newblock In \bibinfo{booktitle}{\emph{Modern data strategy}}.
  \bibinfo{publisher}{Springer}.
\newblock


\bibitem[Fodor(2002)]%
        {fodor2002survey}
\bibfield{author}{\bibinfo{person}{Imola~K Fodor}.}
  \bibinfo{year}{2002}\natexlab{}.
\newblock \bibinfo{booktitle}{\emph{A survey of dimension reduction
  techniques}}.
\newblock \bibinfo{type}{{T}echnical {R}eport}. \bibinfo{institution}{Lawrence
  Livermore National Lab., CA (US)}.
\newblock


\bibitem[Foundation(2022)]%
        {python}
\bibfield{author}{\bibinfo{person}{Python~Software Foundation}.}
  \bibinfo{year}{2022}\natexlab{}.
\newblock \bibinfo{title}{Python}.
\newblock
\newblock
\urldef\tempurl%
\url{https://www.python.org/}
\showURL{%
Retrieved July 29, 2022 from \tempurl}


\bibitem[Friedman et~al\mbox{.}(2010)]%
        {friedman2010data}
\bibfield{author}{\bibinfo{person}{Lawrence~M Friedman},
  \bibinfo{person}{Curt~D Furberg}, {and} \bibinfo{person}{David~L DeMets}.}
  \bibinfo{year}{2010}\natexlab{}.
\newblock \showarticletitle{Data collection and quality control}.
\newblock In \bibinfo{booktitle}{\emph{Fundamentals of clinical trials}}.
  \bibinfo{publisher}{Springer}.
\newblock


\bibitem[F{\"u}rber(2016)]%
        {furber2016semantic}
\bibfield{author}{\bibinfo{person}{Christian F{\"u}rber}.}
  \bibinfo{year}{2016}\natexlab{}.
\newblock \showarticletitle{Semantic Technologies}.
\newblock In \bibinfo{booktitle}{\emph{Data Quality Management with Semantic
  Technologies}}. \bibinfo{publisher}{Springer}, \bibinfo{pages}{56--68}.
\newblock


\bibitem[Gaffney(1999)]%
        {gaffney1999affinity}
\bibfield{author}{\bibinfo{person}{Gerry Gaffney}.}
  \bibinfo{year}{1999}\natexlab{}.
\newblock \showarticletitle{Affinity diagramming}.
\newblock \bibinfo{journal}{\emph{Retrieved January}}  \bibinfo{volume}{3}
  (\bibinfo{year}{1999}), \bibinfo{pages}{2013}.
\newblock


\bibitem[Galesic et~al\mbox{.}(2009)]%
        {galesic2009using}
\bibfield{author}{\bibinfo{person}{Mirta Galesic}, \bibinfo{person}{Rocio
  Garcia-Retamero}, {and} \bibinfo{person}{Gerd Gigerenzer}.}
  \bibinfo{year}{2009}\natexlab{}.
\newblock \showarticletitle{Using icon arrays to communicate medical risks:
  overcoming low numeracy.}
\newblock \bibinfo{journal}{\emph{Health Psychology}} \bibinfo{volume}{28},
  \bibinfo{number}{2} (\bibinfo{year}{2009}), \bibinfo{pages}{210}.
\newblock
\urldef\tempurl%
\url{https://doi.org/10.1037/a0014474}
\showDOI{\tempurl}


\bibitem[Gitelman(2013)]%
        {gitelman2013raw}
\bibfield{author}{\bibinfo{person}{Lisa Gitelman}.}
  \bibinfo{year}{2013}\natexlab{}.
\newblock \bibinfo{booktitle}{\emph{Raw data is an oxymoron}}.
\newblock \bibinfo{publisher}{MIT press}.
\newblock


\bibitem[Google(2022)]%
        {angular}
\bibfield{author}{\bibinfo{person}{Google}.} \bibinfo{year}{2022}\natexlab{}.
\newblock \bibinfo{title}{Angular}.
\newblock
\newblock
\urldef\tempurl%
\url{https://angular.io/}
\showURL{%
Retrieved July 29, 2022 from \tempurl}


\bibitem[Gotz et~al\mbox{.}(2016)]%
        {gotz2016adaptive}
\bibfield{author}{\bibinfo{person}{David Gotz}, \bibinfo{person}{Shun Sun},
  {and} \bibinfo{person}{Nan Cao}.} \bibinfo{year}{2016}\natexlab{}.
\newblock \showarticletitle{Adaptive contextualization: Combating bias during
  high-dimensional visualization and data selection}. In
  \bibinfo{booktitle}{\emph{Proceedings of the 21st International Conference on
  Intelligent User Interfaces}}. \bibinfo{pages}{85--95}.
\newblock


\bibitem[Gutwin(2002)]%
        {gutwin2002traces}
\bibfield{author}{\bibinfo{person}{Carl Gutwin}.}
  \bibinfo{year}{2002}\natexlab{}.
\newblock \showarticletitle{Traces: Visualizing the immediate past to support
  group interaction}. In \bibinfo{booktitle}{\emph{Graphics interface}}.
  Citeseer, \bibinfo{pages}{43--50}.
\newblock


\bibitem[Halevy et~al\mbox{.}(2009)]%
        {halevy2009unreasonable}
\bibfield{author}{\bibinfo{person}{Alon Halevy}, \bibinfo{person}{Peter
  Norvig}, {and} \bibinfo{person}{Fernando Pereira}.}
  \bibinfo{year}{2009}\natexlab{}.
\newblock \showarticletitle{The unreasonable effectiveness of data}.
\newblock \bibinfo{journal}{\emph{IEEE Intelligent Systems}}
  \bibinfo{volume}{24}, \bibinfo{number}{2} (\bibinfo{year}{2009}),
  \bibinfo{pages}{8--12}.
\newblock


\bibitem[Ham(2013)]%
        {ham2013openrefine}
\bibfield{author}{\bibinfo{person}{Kelli Ham}.}
  \bibinfo{year}{2013}\natexlab{}.
\newblock \showarticletitle{OpenRefine (version 2.5). http://openrefine. org.
  Free, open-source tool for cleaning and transforming data}.
\newblock \bibinfo{journal}{\emph{Journal of the Medical Library Association:
  JMLA}} \bibinfo{volume}{101}, \bibinfo{number}{3} (\bibinfo{year}{2013}),
  \bibinfo{pages}{233}.
\newblock
\urldef\tempurl%
\url{https://doi.org/10.3163/1536-5050.101.3.020}
\showDOI{\tempurl}


\bibitem[Haug et~al\mbox{.}(2011)]%
        {haug2011costs}
\bibfield{author}{\bibinfo{person}{Anders Haug}, \bibinfo{person}{Frederik
  Zachariassen}, {and} \bibinfo{person}{Dennis Van~Liempd}.}
  \bibinfo{year}{2011}\natexlab{}.
\newblock \showarticletitle{The costs of poor data quality}.
\newblock \bibinfo{journal}{\emph{Journal of Industrial Engineering and
  Management (JIEM)}} \bibinfo{volume}{4}, \bibinfo{number}{2}
  (\bibinfo{year}{2011}), \bibinfo{pages}{168--193}.
\newblock


\bibitem[Hazelwood et~al\mbox{.}(2018)]%
        {hazelwood2018applied}
\bibfield{author}{\bibinfo{person}{Kim Hazelwood}, \bibinfo{person}{Sarah
  Bird}, \bibinfo{person}{David Brooks}, \bibinfo{person}{Soumith Chintala},
  \bibinfo{person}{Utku Diril}, \bibinfo{person}{Dmytro Dzhulgakov},
  \bibinfo{person}{Mohamed Fawzy}, \bibinfo{person}{Bill Jia},
  \bibinfo{person}{Yangqing Jia}, \bibinfo{person}{Aditya Kalro},
  {et~al\mbox{.}}} \bibinfo{year}{2018}\natexlab{}.
\newblock \showarticletitle{Applied machine learning at facebook: A datacenter
  infrastructure perspective}. In \bibinfo{booktitle}{\emph{2018 IEEE
  International Symposium on High Performance Computer Architecture (HPCA)}}.
  IEEE, \bibinfo{pages}{620--629}.
\newblock


\bibitem[Heer et~al\mbox{.}(2007)]%
        {heer2007voyagers}
\bibfield{author}{\bibinfo{person}{Jeffrey Heer}, \bibinfo{person}{Fernanda~B
  Vi{\'e}gas}, {and} \bibinfo{person}{Martin Wattenberg}.}
  \bibinfo{year}{2007}\natexlab{}.
\newblock \showarticletitle{Voyagers and voyeurs: supporting asynchronous
  collaborative information visualization}. In
  \bibinfo{booktitle}{\emph{Proceedings of the SIGCHI conference on Human
  factors in computing systems}}. \bibinfo{pages}{1029--1038}.
\newblock


\bibitem[Herzog et~al\mbox{.}(2007)]%
        {herzog2007data}
\bibfield{author}{\bibinfo{person}{Thomas~N Herzog}, \bibinfo{person}{Fritz~J
  Scheuren}, {and} \bibinfo{person}{William~E Winkler}.}
  \bibinfo{year}{2007}\natexlab{}.
\newblock \showarticletitle{What is Data Quality and Why Should We Care?}
\newblock In \bibinfo{booktitle}{\emph{Data quality and record linkage
  techniques}}. \bibinfo{publisher}{Springer}, \bibinfo{pages}{7--15}.
\newblock


\bibitem[Hill et~al\mbox{.}(1992)]%
        {hill1992edit}
\bibfield{author}{\bibinfo{person}{William~C Hill}, \bibinfo{person}{James~D
  Hollan}, \bibinfo{person}{Dave Wroblewski}, {and} \bibinfo{person}{Tim
  McCandless}.} \bibinfo{year}{1992}\natexlab{}.
\newblock \showarticletitle{Edit wear and read wear}. In
  \bibinfo{booktitle}{\emph{Proceedings of the SIGCHI conference on Human
  factors in computing systems}}. \bibinfo{pages}{3--9}.
\newblock


\bibitem[Holland et~al\mbox{.}(2020)]%
        {holland2020dataset}
\bibfield{author}{\bibinfo{person}{Sarah Holland}, \bibinfo{person}{Ahmed
  Hosny}, \bibinfo{person}{Sarah Newman}, \bibinfo{person}{Joshua Joseph},
  {and} \bibinfo{person}{Kasia Chmielinski}.} \bibinfo{year}{2020}\natexlab{}.
\newblock \showarticletitle{The dataset nutrition label}.
\newblock \bibinfo{journal}{\emph{Data Protection and Privacy, Volume 12: Data
  Protection and Democracy}}  \bibinfo{volume}{12} (\bibinfo{year}{2020}),
  \bibinfo{pages}{1}.
\newblock


\bibitem[Howe et~al\mbox{.}(2011)]%
        {howe2011automatic}
\bibfield{author}{\bibinfo{person}{Bill Howe}, \bibinfo{person}{Garret Cole},
  \bibinfo{person}{Nodira Khoussainova}, {and} \bibinfo{person}{Leilani
  Battle}.} \bibinfo{year}{2011}\natexlab{}.
\newblock \showarticletitle{Automatic Example Queries for Ad Hoc Databases}. In
  \bibinfo{booktitle}{\emph{Proceedings of the 2011 ACM SIGMOD International
  Conference on Management of Data}} (Athens, Greece)
  \emph{(\bibinfo{series}{SIGMOD '11})}. \bibinfo{publisher}{Association for
  Computing Machinery}, \bibinfo{address}{New York, NY, USA}.
\newblock
\showISBNx{9781450306614}
\urldef\tempurl%
\url{https://doi.org/10.1145/1989323.1989487}
\showDOI{\tempurl}


\bibitem[Isenberg et~al\mbox{.}(2011)]%
        {isenberg2011collaborative}
\bibfield{author}{\bibinfo{person}{Petra Isenberg}, \bibinfo{person}{Niklas
  Elmqvist}, \bibinfo{person}{Jean Scholtz}, \bibinfo{person}{Daniel Cernea},
  \bibinfo{person}{Kwan-Liu Ma}, {and} \bibinfo{person}{Hans Hagen}.}
  \bibinfo{year}{2011}\natexlab{}.
\newblock \showarticletitle{Collaborative visualization: Definition,
  challenges, and research agenda}.
\newblock \bibinfo{journal}{\emph{Information Visualization}}
  \bibinfo{volume}{10}, \bibinfo{number}{4} (\bibinfo{year}{2011}),
  \bibinfo{pages}{310--326}.
\newblock


\bibitem[John et~al\mbox{.}(1994)]%
        {john1994irrelevant}
\bibfield{author}{\bibinfo{person}{George~H. John}, \bibinfo{person}{Ron
  Kohavi}, {and} \bibinfo{person}{Karl Pfleger}.}
  \bibinfo{year}{1994}\natexlab{}.
\newblock \showarticletitle{Irrelevant Features and the Subset Selection
  Problem}.
\newblock In \bibinfo{booktitle}{\emph{Machine Learning Proceedings 1994}},
  \bibfield{editor}{\bibinfo{person}{William~W. Cohen} {and}
  \bibinfo{person}{Haym Hirsh}} (Eds.). \bibinfo{publisher}{Morgan Kaufmann},
  \bibinfo{address}{San Francisco (CA)}, \bibinfo{pages}{121--129}.
\newblock
\showISBNx{978-1-55860-335-6}
\urldef\tempurl%
\url{https://doi.org/10.1016/B978-1-55860-335-6.50023-4}
\showDOI{\tempurl}


\bibitem[Jović et~al\mbox{.}(2015)]%
        {featureselection2015jovic}
\bibfield{author}{\bibinfo{person}{A. Jović}, \bibinfo{person}{K. Brkić},
  {and} \bibinfo{person}{N. Bogunović}.} \bibinfo{year}{2015}\natexlab{}.
\newblock \showarticletitle{A review of feature selection methods with
  applications}. In \bibinfo{booktitle}{\emph{2015 38th International
  Convention on Information and Communication Technology, Electronics and
  Microelectronics (MIPRO)}}. \bibinfo{pages}{1200--1205}.
\newblock
\urldef\tempurl%
\url{https://doi.org/10.1109/MIPRO.2015.7160458}
\showDOI{\tempurl}


\bibitem[Kaggle(2019)]%
        {kaggle2019survey}
\bibfield{author}{\bibinfo{person}{Kaggle}.} \bibinfo{year}{2019}\natexlab{}.
\newblock \bibinfo{title}{Kaggle ML \& DS Survey}.
\newblock
\newblock
\urldef\tempurl%
\url{https://www.kaggle.com/c/kaggle-survey-2019.}
\showURL{%
Retrieved 02/25/2022 from \tempurl}


\bibitem[Kaggle(2022)]%
        {kaggle2022misc}
\bibfield{author}{\bibinfo{person}{Kaggle}.} \bibinfo{year}{2022}\natexlab{}.
\newblock \bibinfo{title}{Kaggle}.
\newblock
\newblock
\urldef\tempurl%
\url{https://www.kaggle.com/}
\showURL{%
Retrieved July 29, 2022 from \tempurl}


\bibitem[Kandel et~al\mbox{.}(2011)]%
        {kandel2011wrangler}
\bibfield{author}{\bibinfo{person}{Sean Kandel}, \bibinfo{person}{Andreas
  Paepcke}, \bibinfo{person}{Joseph Hellerstein}, {and}
  \bibinfo{person}{Jeffrey Heer}.} \bibinfo{year}{2011}\natexlab{}.
\newblock \showarticletitle{Wrangler: Interactive visual specification of data
  transformation scripts}. In \bibinfo{booktitle}{\emph{Proceedings of the
  SIGCHI Conference on Human Factors in Computing Systems}}.
  \bibinfo{pages}{3363--3372}.
\newblock


\bibitem[Kandel et~al\mbox{.}(2012a)]%
        {kandel2012enterprise}
\bibfield{author}{\bibinfo{person}{Sean Kandel}, \bibinfo{person}{Andreas
  Paepcke}, \bibinfo{person}{Joseph~M Hellerstein}, {and}
  \bibinfo{person}{Jeffrey Heer}.} \bibinfo{year}{2012}\natexlab{a}.
\newblock \showarticletitle{Enterprise data analysis and visualization: An
  interview study}.
\newblock \bibinfo{journal}{\emph{IEEE transactions on visualization and
  computer graphics}} \bibinfo{volume}{18}, \bibinfo{number}{12}
  (\bibinfo{year}{2012}), \bibinfo{pages}{2917--2926}.
\newblock


\bibitem[Kandel et~al\mbox{.}(2012b)]%
        {kandel2012profiler}
\bibfield{author}{\bibinfo{person}{Sean Kandel}, \bibinfo{person}{Ravi Parikh},
  \bibinfo{person}{Andreas Paepcke}, \bibinfo{person}{Joseph~M. Hellerstein},
  {and} \bibinfo{person}{Jeffrey Heer}.} \bibinfo{year}{2012}\natexlab{b}.
\newblock \showarticletitle{Profiler: Integrated Statistical Analysis and
  Visualization for Data Quality Assessment}. In
  \bibinfo{booktitle}{\emph{Proceedings of the International Working Conference
  on Advanced Visual Interfaces}} (Capri Island, Italy)
  \emph{(\bibinfo{series}{AVI '12})}. \bibinfo{publisher}{Association for
  Computing Machinery}, \bibinfo{address}{New York, NY, USA},
  \bibinfo{pages}{547–554}.
\newblock
\showISBNx{9781450312875}
\urldef\tempurl%
\url{https://doi.org/10.1145/2254556.2254659}
\showDOI{\tempurl}


\bibitem[Kim et~al\mbox{.}(2008)]%
        {kim2008provenance}
\bibfield{author}{\bibinfo{person}{Jihie Kim}, \bibinfo{person}{Ewa Deelman},
  \bibinfo{person}{Yolanda Gil}, \bibinfo{person}{Gaurang Mehta}, {and}
  \bibinfo{person}{Varun Ratnakar}.} \bibinfo{year}{2008}\natexlab{}.
\newblock \showarticletitle{Provenance trails in the wings/pegasus system}.
\newblock \bibinfo{journal}{\emph{Concurrency and Computation: Practice and
  Experience}} \bibinfo{volume}{20}, \bibinfo{number}{5}
  (\bibinfo{year}{2008}), \bibinfo{pages}{587--597}.
\newblock


\bibitem[Kim et~al\mbox{.}(2017)]%
        {kim2017data}
\bibfield{author}{\bibinfo{person}{Miryung Kim}, \bibinfo{person}{Thomas
  Zimmermann}, \bibinfo{person}{Robert DeLine}, {and} \bibinfo{person}{Andrew
  Begel}.} \bibinfo{year}{2017}\natexlab{}.
\newblock \showarticletitle{Data scientists in software teams: State of the art
  and challenges}.
\newblock \bibinfo{journal}{\emph{IEEE Transactions on Software Engineering}}
  \bibinfo{volume}{44}, \bibinfo{number}{11} (\bibinfo{year}{2017}),
  \bibinfo{pages}{1024--1038}.
\newblock


\bibitem[Koesten et~al\mbox{.}(2019)]%
        {koesten2019collaborative}
\bibfield{author}{\bibinfo{person}{Laura Koesten}, \bibinfo{person}{Emilia
  Kacprzak}, \bibinfo{person}{Jeni Tennison}, {and} \bibinfo{person}{Elena
  Simperl}.} \bibinfo{year}{2019}\natexlab{}.
\newblock \showarticletitle{Collaborative practices with structured data: Do
  tools support what users need?}. In \bibinfo{booktitle}{\emph{Proceedings of
  the 2019 CHI Conference on Human Factors in Computing Systems}}.
  \bibinfo{pages}{1--14}.
\newblock


\bibitem[Koesten et~al\mbox{.}(2020)]%
        {koesten2020dataset}
\bibfield{author}{\bibinfo{person}{Laura Koesten}, \bibinfo{person}{Pavlos
  Vougiouklis}, \bibinfo{person}{Elena Simperl}, {and} \bibinfo{person}{Paul
  Groth}.} \bibinfo{year}{2020}\natexlab{}.
\newblock \showarticletitle{Dataset reuse: Toward translating principles to
  practice}.
\newblock \bibinfo{journal}{\emph{Patterns}} \bibinfo{volume}{1},
  \bibinfo{number}{8} (\bibinfo{year}{2020}), \bibinfo{pages}{100136}.
\newblock


\bibitem[Koffka(2013)]%
        {koffka2013principles}
\bibfield{author}{\bibinfo{person}{Kurt Koffka}.}
  \bibinfo{year}{2013}\natexlab{}.
\newblock \bibinfo{booktitle}{\emph{Principles of Gestalt psychology}}.
\newblock \bibinfo{publisher}{Routledge}.
\newblock


\bibitem[Lancar(2022)]%
        {digitalMarketingDataset}
\bibfield{author}{\bibinfo{person}{Jonathan Lancar}.}
  \bibinfo{year}{2022}\natexlab{}.
\newblock \bibinfo{title}{Luma}.
\newblock
  \bibinfo{howpublished}{\url{https://github.com/adobe/experience-platform-dsw-reference/blob/master/datasets/luma/luma\_post\_extended.csv}}.
\newblock


\bibitem[Laranjeiro et~al\mbox{.}(2015)]%
        {laranjeiro2015survey}
\bibfield{author}{\bibinfo{person}{Nuno Laranjeiro}, \bibinfo{person}{Seyma~Nur
  Soydemir}, {and} \bibinfo{person}{Jorge Bernardino}.}
  \bibinfo{year}{2015}\natexlab{}.
\newblock \showarticletitle{A survey on data quality: classifying poor data}.
  In \bibinfo{booktitle}{\emph{2015 IEEE 21st Pacific rim international
  symposium on dependable computing (PRDC)}}. IEEE, \bibinfo{pages}{179--188}.
\newblock


\bibitem[Leonelli(2019)]%
        {leonelli2019data}
\bibfield{author}{\bibinfo{person}{Sabina Leonelli}.}
  \bibinfo{year}{2019}\natexlab{}.
\newblock \showarticletitle{Data governance is key to interpretation:
  Reconceptualizing data in data science}.
\newblock \bibinfo{journal}{\emph{Harvard Data Science Review}}
  \bibinfo{volume}{1}, \bibinfo{number}{1} (\bibinfo{year}{2019}),
  \bibinfo{pages}{10--1162}.
\newblock


\bibitem[Liu et~al\mbox{.}(2020)]%
        {liu2020data}
\bibfield{author}{\bibinfo{person}{Zipeng Liu}, \bibinfo{person}{Zhicheng Liu},
  {and} \bibinfo{person}{Tamara Munzner}.} \bibinfo{year}{2020}\natexlab{}.
\newblock \showarticletitle{Data-driven Multi-level Segmentation of Image
  Editing Logs}. In \bibinfo{booktitle}{\emph{Proceedings of the 2020 CHI
  Conference on Human Factors in Computing Systems}}. \bibinfo{pages}{1--12}.
\newblock


\bibitem[Luo et~al\mbox{.}(2020a)]%
        {luo2020interactive}
\bibfield{author}{\bibinfo{person}{Yuyu Luo}, \bibinfo{person}{Chengliang
  Chai}, \bibinfo{person}{Xuedi Qin}, \bibinfo{person}{Nan Tang}, {and}
  \bibinfo{person}{Guoliang Li}.} \bibinfo{year}{2020}\natexlab{a}.
\newblock \showarticletitle{Interactive cleaning for progressive visualization
  through composite questions}. In \bibinfo{booktitle}{\emph{2020 IEEE 36th
  International Conference on Data Engineering (ICDE)}}. IEEE.
\newblock


\bibitem[Luo et~al\mbox{.}(2020b)]%
        {luo2020visclean}
\bibfield{author}{\bibinfo{person}{Yuyu Luo}, \bibinfo{person}{Chengliang
  Chai}, \bibinfo{person}{Xuedi Qin}, \bibinfo{person}{Nan Tang}, {and}
  \bibinfo{person}{Guoliang Li}.} \bibinfo{year}{2020}\natexlab{b}.
\newblock \showarticletitle{VisClean: Interactive Cleaning for Progressive
  Visualization}.
\newblock \bibinfo{journal}{\emph{Proc. VLDB Endow.}} (\bibinfo{date}{aug}
  \bibinfo{year}{2020}).
\newblock
\showISSN{2150-8097}
\urldef\tempurl%
\url{https://doi.org/10.14778/3415478.3415484}
\showDOI{\tempurl}


\bibitem[Mahanti(2019)]%
        {mahanti2019data}
\bibfield{author}{\bibinfo{person}{Rupa Mahanti}.}
  \bibinfo{year}{2019}\natexlab{}.
\newblock \showarticletitle{Data, Data Quality, and Cost of Poor Data Quality}.
\newblock In \bibinfo{booktitle}{\emph{Data quality: dimensions, measurement,
  strategy, management, and governance}}. \bibinfo{publisher}{Quality Press}.
\newblock


\bibitem[Manstead et~al\mbox{.}(1995)]%
        {manstead1995blackwell}
\bibfield{author}{\bibinfo{person}{Antony~SR Manstead},
  \bibinfo{person}{Miles~Ed Hewstone}, \bibinfo{person}{Susan~T Fiske},
  \bibinfo{person}{Michael~A Hogg}, \bibinfo{person}{Harry~T Reis}, {and}
  \bibinfo{person}{G{\"u}n~R Semin}.} \bibinfo{year}{1995}\natexlab{}.
\newblock \bibinfo{booktitle}{\emph{The Blackwell encyclopedia of social
  psychology.}}
\newblock \bibinfo{publisher}{Blackwell Reference/Blackwell Publishers}.
\newblock


\bibitem[Mayfield et~al\mbox{.}(2010)]%
        {mayfield2010eracer}
\bibfield{author}{\bibinfo{person}{Chris Mayfield}, \bibinfo{person}{Jennifer
  Neville}, {and} \bibinfo{person}{Sunil Prabhakar}.}
  \bibinfo{year}{2010}\natexlab{}.
\newblock \showarticletitle{ERACER: a database approach for statistical
  inference and data cleaning}. In \bibinfo{booktitle}{\emph{Proceedings of the
  2010 ACM SIGMOD International Conference on Management of data}}.
  \bibinfo{pages}{75--86}.
\newblock


\bibitem[Medi{\'c} et~al\mbox{.}(2016)]%
        {medic2016new}
\bibfield{author}{\bibinfo{person}{Sr{\dj}an Medi{\'c}},
  \bibinfo{person}{Biljana Karlovi{\'c}}, {and} \bibinfo{person}{Zrinko
  Cindri{\'c}}.} \bibinfo{year}{2016}\natexlab{}.
\newblock \showarticletitle{New standard ISO 9001: 2015 and its effect on
  organisations}.
\newblock \bibinfo{journal}{\emph{Interdisciplinary Description of Complex
  Systems: INDECS}} \bibinfo{volume}{14}, \bibinfo{number}{2}
  (\bibinfo{year}{2016}), \bibinfo{pages}{188--193}.
\newblock


\bibitem[Mehrabi et~al\mbox{.}(2021)]%
        {mehrabi2021survey}
\bibfield{author}{\bibinfo{person}{Ninareh Mehrabi}, \bibinfo{person}{Fred
  Morstatter}, \bibinfo{person}{Nripsuta Saxena}, \bibinfo{person}{Kristina
  Lerman}, {and} \bibinfo{person}{Aram Galstyan}.}
  \bibinfo{year}{2021}\natexlab{}.
\newblock \showarticletitle{A survey on bias and fairness in machine learning}.
\newblock \bibinfo{journal}{\emph{ACM Computing Surveys (CSUR)}}
  \bibinfo{volume}{54}, \bibinfo{number}{6} (\bibinfo{year}{2021}),
  \bibinfo{pages}{1--35}.
\newblock


\bibitem[Metaplane(2022)]%
        {metaplane}
\bibfield{author}{\bibinfo{person}{Metaplane}.}
  \bibinfo{year}{2022}\natexlab{}.
\newblock \bibinfo{title}{Metaplane}.
\newblock
\newblock
\urldef\tempurl%
\url{https://www.metaplane.dev/}
\showURL{%
Retrieved July 29, 2022 from \tempurl}


\bibitem[Microsoft(2022a)]%
        {linkedin}
\bibfield{author}{\bibinfo{person}{Microsoft}.}
  \bibinfo{year}{2022}\natexlab{a}.
\newblock \bibinfo{title}{LinkedIn}.
\newblock
\newblock
\urldef\tempurl%
\url{https://linkedin.com/}
\showURL{%
Retrieved July 29, 2022 from \tempurl}


\bibitem[Microsoft(2022b)]%
        {powerbi}
\bibfield{author}{\bibinfo{person}{Microsoft}.}
  \bibinfo{year}{2022}\natexlab{b}.
\newblock \bibinfo{title}{Power BI Desktop}.
\newblock
\newblock
\urldef\tempurl%
\url{https://powerbi.microsoft.com/en-us/}
\showURL{%
Retrieved May 25, 2022 from \tempurl}


\bibitem[Microsoft(2022c)]%
        {teams}
\bibfield{author}{\bibinfo{person}{Microsoft}.}
  \bibinfo{year}{2022}\natexlab{c}.
\newblock \bibinfo{title}{Teams}.
\newblock
\newblock
\urldef\tempurl%
\url{https://www.microsoft.com/en-us/microsoft-teams/group-chat-software}
\showURL{%
Retrieved July 29, 2022 from \tempurl}


\bibitem[Nagle et~al\mbox{.}(2017)]%
        {nagle2017only}
\bibfield{author}{\bibinfo{person}{Tadhg Nagle}, \bibinfo{person}{Thomas~C
  Redman}, {and} \bibinfo{person}{David Sammon}.}
  \bibinfo{year}{2017}\natexlab{}.
\newblock \showarticletitle{Only 3\% of companies’ data meets basic quality
  standards}.
\newblock \bibinfo{journal}{\emph{Harvard Business Review}}
  (\bibinfo{year}{2017}).
\newblock


\bibitem[Narechania et~al\mbox{.}(2021)]%
        {narechania2021lumos}
\bibfield{author}{\bibinfo{person}{Arpit Narechania}, \bibinfo{person}{Adam
  Coscia}, \bibinfo{person}{Emily Wall}, {and} \bibinfo{person}{Alex Endert}.}
  \bibinfo{year}{2021}\natexlab{}.
\newblock \showarticletitle{Lumos: Increasing awareness of analytic behavior
  during visual data analysis}.
\newblock \bibinfo{journal}{\emph{IEEE Transactions on Visualization and
  Computer Graphics}} (\bibinfo{year}{2021}).
\newblock


\bibitem[Nielsen and Pernice(2010)]%
        {nielsen2010eyetracking}
\bibfield{author}{\bibinfo{person}{Jakob Nielsen} {and} \bibinfo{person}{Kara
  Pernice}.} \bibinfo{year}{2010}\natexlab{}.
\newblock \bibinfo{booktitle}{\emph{Eyetracking web usability}}.
\newblock \bibinfo{publisher}{New Riders}.
\newblock


\bibitem[North et~al\mbox{.}(2011)]%
        {north2011analytic}
\bibfield{author}{\bibinfo{person}{Chris North}, \bibinfo{person}{Remco Chang},
  \bibinfo{person}{Alex Endert}, \bibinfo{person}{Wenwen Dou},
  \bibinfo{person}{Richard May}, \bibinfo{person}{Bill Pike}, {and}
  \bibinfo{person}{Glenn Fink}.} \bibinfo{year}{2011}\natexlab{}.
\newblock \showarticletitle{Analytic provenance: process+ interaction+
  insight}.
\newblock In \bibinfo{booktitle}{\emph{CHI'11 Extended Abstracts on Human
  Factors in Computing Systems}}. \bibinfo{pages}{33--36}.
\newblock


\bibitem[Otto et~al\mbox{.}(2009)]%
        {otto2009identification}
\bibfield{author}{\bibinfo{person}{Boris Otto}, \bibinfo{person}{Kai~M
  H{\"u}ner}, {and} \bibinfo{person}{Hubert {\"O}sterle}.}
  \bibinfo{year}{2009}\natexlab{}.
\newblock \showarticletitle{Identification of Business Oriented Data Quality
  Metrics.}. In \bibinfo{booktitle}{\emph{ICIQ}}. \bibinfo{pages}{122--134}.
\newblock


\bibitem[Peng et~al\mbox{.}(2021)]%
        {peng2021dataprep}
\bibfield{author}{\bibinfo{person}{Jinglin Peng}, \bibinfo{person}{Weiyuan Wu},
  \bibinfo{person}{Brandon Lockhart}, \bibinfo{person}{Song Bian},
  \bibinfo{person}{Jing~Nathan Yan}, \bibinfo{person}{Linghao Xu},
  \bibinfo{person}{Zhixuan Chi}, \bibinfo{person}{Jeffrey~M Rzeszotarski},
  {and} \bibinfo{person}{Jiannan Wang}.} \bibinfo{year}{2021}\natexlab{}.
\newblock \showarticletitle{DataPrep.EDA: Task-Centric Exploratory Data
  Analysis for Statistical Modeling in Python}. In
  \bibinfo{booktitle}{\emph{Proceedings of the 2021 International Conference on
  Management of Data}}. \bibinfo{pages}{2271--2280}.
\newblock


\bibitem[Pipino et~al\mbox{.}(2002)]%
        {pipino2002data}
\bibfield{author}{\bibinfo{person}{Leo~L. Pipino}, \bibinfo{person}{Yang~W.
  Lee}, {and} \bibinfo{person}{Richard~Y. Wang}.}
  \bibinfo{year}{2002}\natexlab{}.
\newblock \showarticletitle{Data Quality Assessment}.
\newblock \bibinfo{journal}{\emph{Commun. ACM}} \bibinfo{volume}{45},
  \bibinfo{number}{4} (\bibinfo{date}{apr} \bibinfo{year}{2002}),
  \bibinfo{pages}{211–218}.
\newblock
\showISSN{0001-0782}
\urldef\tempurl%
\url{https://doi.org/10.1145/505248.506010}
\showDOI{\tempurl}


\bibitem[Portal(2016)]%
        {portal2016eu}
\bibfield{author}{\bibinfo{person}{EU~Open~Data Portal}.}
  \bibinfo{year}{2016}\natexlab{}.
\newblock \showarticletitle{EU open data}.
\newblock  (\bibinfo{year}{2016}).
\newblock


\bibitem[Pudil and Novovi{\v{c}}ov{\'a}(1998)]%
        {pudil1998novel}
\bibfield{author}{\bibinfo{person}{Pavel Pudil} {and} \bibinfo{person}{Jana
  Novovi{\v{c}}ov{\'a}}.} \bibinfo{year}{1998}\natexlab{}.
\newblock \showarticletitle{Novel methods for feature subset selection with
  respect to problem knowledge}.
\newblock In \bibinfo{booktitle}{\emph{Feature extraction, construction and
  selection}}. \bibinfo{publisher}{Springer}, \bibinfo{pages}{101--116}.
\newblock


\bibitem[Pyle(1999)]%
        {pyle1999data}
\bibfield{author}{\bibinfo{person}{Dorian Pyle}.}
  \bibinfo{year}{1999}\natexlab{}.
\newblock \bibinfo{booktitle}{\emph{Data preparation for data mining}}.
\newblock \bibinfo{publisher}{morgan kaufmann}.
\newblock


\bibitem[Ragan et~al\mbox{.}(2015)]%
        {ragan2015characterizing}
\bibfield{author}{\bibinfo{person}{Eric~D Ragan}, \bibinfo{person}{Alex
  Endert}, \bibinfo{person}{Jibonananda Sanyal}, {and} \bibinfo{person}{Jian
  Chen}.} \bibinfo{year}{2015}\natexlab{}.
\newblock \showarticletitle{Characterizing provenance in visualization and data
  analysis: an organizational framework of provenance types and purposes}.
\newblock \bibinfo{journal}{\emph{IEEE transactions on visualization and
  computer graphics}} \bibinfo{volume}{22}, \bibinfo{number}{1}
  (\bibinfo{year}{2015}), \bibinfo{pages}{31--40}.
\newblock


\bibitem[Raman and Hellerstein(2001)]%
        {raman2001potter}
\bibfield{author}{\bibinfo{person}{Vijayshankar Raman} {and}
  \bibinfo{person}{Joseph~M Hellerstein}.} \bibinfo{year}{2001}\natexlab{}.
\newblock \showarticletitle{Potter's wheel: An interactive data cleaning
  system}. In \bibinfo{booktitle}{\emph{VLDB}}, Vol.~\bibinfo{volume}{1}.
  \bibinfo{pages}{381--390}.
\newblock


\bibitem[Redman(2016)]%
        {badDataCostsMillions}
\bibfield{author}{\bibinfo{person}{Thomas~C. Redman}.}
  \bibinfo{year}{2016}\natexlab{}.
\newblock \bibinfo{title}{Bad Data Costs the U.S. \$3 Trillion Per Year}.
\newblock
\newblock
\urldef\tempurl%
\url{https://hbr.org/2016/09/bad-data-costs-the-u-s-3-trillion-per-year}
\showURL{%
\tempurl}


\bibitem[Redman(2018)]%
        {redman2018if}
\bibfield{author}{\bibinfo{person}{Thomas~C Redman}.}
  \bibinfo{year}{2018}\natexlab{}.
\newblock \showarticletitle{If your data is bad, your machine learning tools
  are useless}.
\newblock \bibinfo{journal}{\emph{Harvard Business Review}}
  \bibinfo{volume}{2} (\bibinfo{year}{2018}).
\newblock


\bibitem[Richards(2006)]%
        {richards2006representational}
\bibfield{author}{\bibinfo{person}{Robert Richards}.}
  \bibinfo{year}{2006}\natexlab{}.
\newblock \showarticletitle{Representational state transfer (rest)}.
\newblock In \bibinfo{booktitle}{\emph{Pro PHP XML and web services}}.
  \bibinfo{publisher}{Springer}, \bibinfo{pages}{633--672}.
\newblock


\bibitem[Rogelberg et~al\mbox{.}(2012)]%
        {rogelberg2012wasted}
\bibfield{author}{\bibinfo{person}{Steven~G Rogelberg},
  \bibinfo{person}{Linda~Rhoades Shanock}, {and} \bibinfo{person}{Cliff~W
  Scott}.} \bibinfo{year}{2012}\natexlab{}.
\newblock \showarticletitle{Wasted time and money in meetings: Increasing
  return on investment}.
\newblock \bibinfo{journal}{\emph{Small Group Research}} \bibinfo{volume}{43},
  \bibinfo{number}{2} (\bibinfo{year}{2012}), \bibinfo{pages}{236--245}.
\newblock


\bibitem[Sambasivan et~al\mbox{.}(2021)]%
        {sambasivan2021everyone}
\bibfield{author}{\bibinfo{person}{Nithya Sambasivan}, \bibinfo{person}{Shivani
  Kapania}, \bibinfo{person}{Hannah Highfill}, \bibinfo{person}{Diana Akrong},
  \bibinfo{person}{Praveen Paritosh}, {and} \bibinfo{person}{Lora~M Aroyo}.}
  \bibinfo{year}{2021}\natexlab{}.
\newblock \showarticletitle{“Everyone wants to do the model work, not the
  data work”: Data Cascades in High-Stakes AI}. In
  \bibinfo{booktitle}{\emph{proceedings of the 2021 CHI Conference on Human
  Factors in Computing Systems}}. \bibinfo{pages}{1--15}.
\newblock


\bibitem[Schmuckler(2001)]%
        {schmuckler2001ecological}
\bibfield{author}{\bibinfo{person}{Mark~A Schmuckler}.}
  \bibinfo{year}{2001}\natexlab{}.
\newblock \showarticletitle{What is ecological validity? A dimensional
  analysis}.
\newblock \bibinfo{journal}{\emph{Infancy}} \bibinfo{volume}{2},
  \bibinfo{number}{4} (\bibinfo{year}{2001}), \bibinfo{pages}{419--436}.
\newblock


\bibitem[Schulz et~al\mbox{.}(2017)]%
        {schulz2017systematic}
\bibfield{author}{\bibinfo{person}{Hans-J{\"o}rg Schulz},
  \bibinfo{person}{Thomas Nocke}, \bibinfo{person}{Magnus Heitzler}, {and}
  \bibinfo{person}{Heidrun Schumann}.} \bibinfo{year}{2017}\natexlab{}.
\newblock \showarticletitle{A systematic view on data descriptors for the
  visual analysis of tabular data}.
\newblock \bibinfo{journal}{\emph{Information Visualization}}
  \bibinfo{volume}{16}, \bibinfo{number}{3} (\bibinfo{year}{2017}),
  \bibinfo{pages}{232--256}.
\newblock


\bibitem[Sedlmair et~al\mbox{.}(2012)]%
        {designstudy2012sedlmair}
\bibfield{author}{\bibinfo{person}{Michael Sedlmair}, \bibinfo{person}{Miriah
  Meyer}, {and} \bibinfo{person}{Tamara Munzner}.}
  \bibinfo{year}{2012}\natexlab{}.
\newblock \showarticletitle{Design Study Methodology: Reflections from the
  Trenches and the Stacks}.
\newblock \bibinfo{journal}{\emph{IEEE Transactions on Visualization and
  Computer Graphics}} \bibinfo{volume}{18}, \bibinfo{number}{12}
  (\bibinfo{year}{2012}), \bibinfo{pages}{2431--2440}.
\newblock
\urldef\tempurl%
\url{https://doi.org/10.1109/TVCG.2012.213}
\showDOI{\tempurl}


\bibitem[Sidi et~al\mbox{.}(2012)]%
        {sidi2012data}
\bibfield{author}{\bibinfo{person}{Fatimah Sidi},
  \bibinfo{person}{Payam~Hassany Shariat~Panahy},
  \bibinfo{person}{Lilly~Suriani Affendey}, \bibinfo{person}{Marzanah~A.
  Jabar}, \bibinfo{person}{Hamidah Ibrahim}, {and} \bibinfo{person}{Aida
  Mustapha}.} \bibinfo{year}{2012}\natexlab{}.
\newblock \showarticletitle{Data quality: A survey of data quality dimensions}.
  In \bibinfo{booktitle}{\emph{2012 International Conference on Information
  Retrieval \& Knowledge Management}}. \bibinfo{pages}{300--304}.
\newblock
\urldef\tempurl%
\url{https://doi.org/10.1109/InfRKM.2012.6204995}
\showDOI{\tempurl}


\bibitem[Song and Shepperd(2007)]%
        {song2007missing}
\bibfield{author}{\bibinfo{person}{Qinbao Song} {and} \bibinfo{person}{Martin
  Shepperd}.} \bibinfo{year}{2007}\natexlab{}.
\newblock \showarticletitle{Missing Data Imputation Techniques}.
\newblock \bibinfo{journal}{\emph{Int. J. Bus. Intell. Data Min.}}
  \bibinfo{volume}{2}, \bibinfo{number}{3} (\bibinfo{date}{oct}
  \bibinfo{year}{2007}), \bibinfo{pages}{261–291}.
\newblock
\showISSN{1743-8195}
\urldef\tempurl%
\url{https://doi.org/10.1504/IJBIDM.2007.015485}
\showDOI{\tempurl}


\bibitem[Stein and Morrison(2014)]%
        {stein2014pwcreport}
\bibfield{author}{\bibinfo{person}{Brian Stein} {and} \bibinfo{person}{Alan
  Morrison}.} \bibinfo{year}{2014}\natexlab{}.
\newblock \showarticletitle{Data lakes and the promise of unsiloed data}.
\newblock \bibinfo{journal}{\emph{PricewaterhouseCooper, Technology Forecast:
  Rethinking integration}} (\bibinfo{year}{2014}).
\newblock
\urldef\tempurl%
\url{https://www.pwc.com/us/en/technology-forecast/2014/cloud-computing/assets/pdf/pwc-technology-forecast-data-lakes.pdf}
\showURL{%
\tempurl}


\bibitem[Tableau(2022a)]%
        {tableau}
\bibfield{author}{\bibinfo{person}{Tableau}.} \bibinfo{year}{2022}\natexlab{a}.
\newblock \bibinfo{title}{Tableau}.
\newblock
\newblock
\urldef\tempurl%
\url{https://www.tableau.com/}
\showURL{%
Retrieved July 29, 2022 from \tempurl}


\bibitem[Tableau(2022b)]%
        {tableauprep}
\bibfield{author}{\bibinfo{person}{Tableau}.} \bibinfo{year}{2022}\natexlab{b}.
\newblock \bibinfo{title}{Tableau Prep}.
\newblock
\newblock
\urldef\tempurl%
\url{https://www.tableau.com/products/prep}
\showURL{%
Retrieved May 25, 2022 from \tempurl}


\bibitem[Thom-Santelli et~al\mbox{.}(2010)]%
        {thom2010you}
\bibfield{author}{\bibinfo{person}{Jennifer Thom-Santelli},
  \bibinfo{person}{Dan Cosley}, {and} \bibinfo{person}{Geri Gay}.}
  \bibinfo{year}{2010}\natexlab{}.
\newblock \showarticletitle{What do you know? Experts, novices and
  territoriality in collaborative systems}. In
  \bibinfo{booktitle}{\emph{Proceedings of the SIGCHI Conference on Human
  Factors in Computing Systems}}. \bibinfo{pages}{1685--1694}.
\newblock


\bibitem[Thom-Santelli et~al\mbox{.}(2009)]%
        {thom2009s}
\bibfield{author}{\bibinfo{person}{Jennifer Thom-Santelli},
  \bibinfo{person}{Dan~R Cosley}, {and} \bibinfo{person}{Geri Gay}.}
  \bibinfo{year}{2009}\natexlab{}.
\newblock \showarticletitle{What's mine is mine: territoriality in
  collaborative authoring}. In \bibinfo{booktitle}{\emph{Proceedings of the
  SIGCHI Conference on Human Factors in Computing Systems}}.
  \bibinfo{pages}{1481--1484}.
\newblock


\bibitem[Trifacta(2022)]%
        {trifactawrangler}
\bibfield{author}{\bibinfo{person}{Trifacta}.} \bibinfo{year}{2022}\natexlab{}.
\newblock \bibinfo{title}{Wrangler}.
\newblock
\newblock
\urldef\tempurl%
\url{https://www.trifacta.com/}
\showURL{%
Retrieved July 29, 2022 from \tempurl}


\bibitem[Tynan and Drayton(1987)]%
        {tynan1987market}
\bibfield{author}{\bibinfo{person}{A~Caroline Tynan} {and}
  \bibinfo{person}{Jennifer Drayton}.} \bibinfo{year}{1987}\natexlab{}.
\newblock \showarticletitle{Market segmentation}.
\newblock \bibinfo{journal}{\emph{Journal of marketing management}}
  \bibinfo{volume}{2}, \bibinfo{number}{3} (\bibinfo{year}{1987}),
  \bibinfo{pages}{301--335}.
\newblock


\bibitem[Uber(2022)]%
        {uberdatabook}
\bibfield{author}{\bibinfo{person}{Uber}.} \bibinfo{year}{2022}\natexlab{}.
\newblock \bibinfo{title}{Uber Databook}.
\newblock
\newblock
\urldef\tempurl%
\url{https://eng.uber.com/databook/}
\showURL{%
Retrieved July 29, 2022 from \tempurl}


\bibitem[Ulrich et~al\mbox{.}(2022)]%
        {metadata2022ulrich}
\bibfield{author}{\bibinfo{person}{Hannes Ulrich}, \bibinfo{person}{Ann-Kristin
  Kock-Schoppenhauer}, \bibinfo{person}{Noemi Deppenwiese},
  \bibinfo{person}{Robert G{\"o}tt}, \bibinfo{person}{Jori Kern},
  \bibinfo{person}{Martin Lablans}, \bibinfo{person}{Raphael~W Majeed},
  \bibinfo{person}{Mark~R St{\"o}hr}, \bibinfo{person}{J{\"u}rgen Stausberg},
  \bibinfo{person}{Julian Varghese}, \bibinfo{person}{Martin Dugas}, {and}
  \bibinfo{person}{Josef Ingenerf}.} \bibinfo{year}{2022}\natexlab{}.
\newblock \showarticletitle{Understanding the Nature of Metadata: Systematic
  Review}.
\newblock \bibinfo{journal}{\emph{J Med Internet Res}} \bibinfo{volume}{24},
  \bibinfo{number}{1} (\bibinfo{date}{11 Jan} \bibinfo{year}{2022}),
  \bibinfo{pages}{e25440}.
\newblock
\showISSN{1438-8871}
\urldef\tempurl%
\url{https://doi.org/10.2196/25440}
\showDOI{\tempurl}


\bibitem[Umbrich et~al\mbox{.}(2015)]%
        {umbrich2015quality}
\bibfield{author}{\bibinfo{person}{J{\"u}rgen Umbrich},
  \bibinfo{person}{Sebastian Neumaier}, {and} \bibinfo{person}{Axel Polleres}.}
  \bibinfo{year}{2015}\natexlab{}.
\newblock \showarticletitle{Quality assessment and evolution of open data
  portals}. In \bibinfo{booktitle}{\emph{2015 3rd international conference on
  future internet of things and cloud}}. IEEE, \bibinfo{pages}{404--411}.
\newblock


\bibitem[Vaziri et~al\mbox{.}(2019)]%
        {vaziri2019measuring}
\bibfield{author}{\bibinfo{person}{Reza Vaziri}, \bibinfo{person}{Mehran
  Mohsenzadeh}, {and} \bibinfo{person}{Jafar Habibi}.}
  \bibinfo{year}{2019}\natexlab{}.
\newblock \showarticletitle{Measuring data quality with weighted metrics}.
\newblock \bibinfo{journal}{\emph{Total Quality Management \& Business
  Excellence}} \bibinfo{volume}{30}, \bibinfo{number}{5-6}
  (\bibinfo{year}{2019}), \bibinfo{pages}{708--720}.
\newblock


\bibitem[Viegas and Wattenberg(2006)]%
        {viegas2006communication}
\bibfield{author}{\bibinfo{person}{Fernanda~B Viegas} {and}
  \bibinfo{person}{Martin Wattenberg}.} \bibinfo{year}{2006}\natexlab{}.
\newblock \showarticletitle{Communication-minded visualization: A call to
  action}.
\newblock \bibinfo{journal}{\emph{IBM Systems Journal}} \bibinfo{volume}{45},
  \bibinfo{number}{4} (\bibinfo{year}{2006}), \bibinfo{pages}{801}.
\newblock


\bibitem[Viegas et~al\mbox{.}(2007)]%
        {manyeyes2007viegas}
\bibfield{author}{\bibinfo{person}{Fernanda~B. Viegas}, \bibinfo{person}{Martin
  Wattenberg}, \bibinfo{person}{Frank van Ham}, \bibinfo{person}{Jesse Kriss},
  {and} \bibinfo{person}{Matt McKeon}.} \bibinfo{year}{2007}\natexlab{}.
\newblock \showarticletitle{ManyEyes: A Site for Visualization at Internet
  Scale}.
\newblock  \bibinfo{volume}{13}, \bibinfo{number}{6} (\bibinfo{date}{nov}
  \bibinfo{year}{2007}), \bibinfo{pages}{1121–1128}.
\newblock
\showISSN{1077-2626}
\urldef\tempurl%
\url{https://doi.org/10.1109/TVCG.2007.70577}
\showDOI{\tempurl}


\bibitem[Wall et~al\mbox{.}(2017)]%
        {wall2017warning}
\bibfield{author}{\bibinfo{person}{Emily Wall}, \bibinfo{person}{Leslie~M
  Blaha}, \bibinfo{person}{Lyndsey Franklin}, {and} \bibinfo{person}{Alex
  Endert}.} \bibinfo{year}{2017}\natexlab{}.
\newblock \showarticletitle{Warning, bias may occur: A proposed approach to
  detecting cognitive bias in interactive visual analytics}. In
  \bibinfo{booktitle}{\emph{2017 ieee conference on visual analytics science
  and technology (vast)}}. IEEE, \bibinfo{pages}{104--115}.
\newblock


\bibitem[Wall et~al\mbox{.}(2021)]%
        {wall2021left}
\bibfield{author}{\bibinfo{person}{Emily Wall}, \bibinfo{person}{Arpit
  Narechania}, \bibinfo{person}{Adam Coscia}, \bibinfo{person}{Jamal Paden},
  {and} \bibinfo{person}{Alex Endert}.} \bibinfo{year}{2021}\natexlab{}.
\newblock \showarticletitle{Left, right, and gender: Exploring interaction
  traces to mitigate human biases}.
\newblock \bibinfo{journal}{\emph{IEEE Transactions on Visualization and
  Computer Graphics}} (\bibinfo{year}{2021}).
\newblock


\bibitem[Wang et~al\mbox{.}(2015)]%
        {wang2015docuviz}
\bibfield{author}{\bibinfo{person}{Dakuo Wang}, \bibinfo{person}{Judith~S
  Olson}, \bibinfo{person}{Jingwen Zhang}, \bibinfo{person}{Trung Nguyen},
  {and} \bibinfo{person}{Gary~M Olson}.} \bibinfo{year}{2015}\natexlab{}.
\newblock \showarticletitle{DocuViz: visualizing collaborative writing}. In
  \bibinfo{booktitle}{\emph{Proceedings of the 33rd Annual ACM conference on
  human factors in computing systems}}. \bibinfo{pages}{1865--1874}.
\newblock


\bibitem[Wedig and Madani(2006)]%
        {wedig2006large}
\bibfield{author}{\bibinfo{person}{Steve Wedig} {and} \bibinfo{person}{Omid
  Madani}.} \bibinfo{year}{2006}\natexlab{}.
\newblock \showarticletitle{A large-scale analysis of query logs for assessing
  personalization opportunities}. In \bibinfo{booktitle}{\emph{Proceedings of
  the 12th ACM SIGKDD international conference on Knowledge discovery and data
  mining}}. \bibinfo{pages}{742--747}.
\newblock


\bibitem[Willett et~al\mbox{.}(2007)]%
        {willett2007scented}
\bibfield{author}{\bibinfo{person}{Wesley Willett}, \bibinfo{person}{Jeffrey
  Heer}, {and} \bibinfo{person}{Maneesh Agrawala}.}
  \bibinfo{year}{2007}\natexlab{}.
\newblock \showarticletitle{Scented widgets: Improving navigation cues with
  embedded visualizations}.
\newblock \bibinfo{journal}{\emph{IEEE Transactions on Visualization and
  Computer Graphics}} \bibinfo{volume}{13}, \bibinfo{number}{6}
  (\bibinfo{year}{2007}), \bibinfo{pages}{1129--1136}.
\newblock


\bibitem[Willett et~al\mbox{.}(2011)]%
        {willett2011commentspace}
\bibfield{author}{\bibinfo{person}{Wesley Willett}, \bibinfo{person}{Jeffrey
  Heer}, \bibinfo{person}{Joseph Hellerstein}, {and} \bibinfo{person}{Maneesh
  Agrawala}.} \bibinfo{year}{2011}\natexlab{}.
\newblock \showarticletitle{CommentSpace: structured support for collaborative
  visual analysis}. In \bibinfo{booktitle}{\emph{Proceedings of the SIGCHI
  conference on Human Factors in Computing Systems}}.
\newblock


\bibitem[Wongsuphasawat et~al\mbox{.}(2015)]%
        {wongsuphasawat2015voyager}
\bibfield{author}{\bibinfo{person}{Kanit Wongsuphasawat},
  \bibinfo{person}{Dominik Moritz}, \bibinfo{person}{Anushka Anand},
  \bibinfo{person}{Jock Mackinlay}, \bibinfo{person}{Bill Howe}, {and}
  \bibinfo{person}{Jeffrey Heer}.} \bibinfo{year}{2015}\natexlab{}.
\newblock \showarticletitle{Voyager: Exploratory analysis via faceted browsing
  of visualization recommendations}.
\newblock \bibinfo{journal}{\emph{IEEE transactions on visualization and
  computer graphics}} \bibinfo{volume}{22}, \bibinfo{number}{1}
  (\bibinfo{year}{2015}), \bibinfo{pages}{649--658}.
\newblock


\bibitem[Yakout et~al\mbox{.}(2013)]%
        {yakout2013don}
\bibfield{author}{\bibinfo{person}{Mohamed Yakout}, \bibinfo{person}{Laure
  Berti-{\'E}quille}, {and} \bibinfo{person}{Ahmed~K Elmagarmid}.}
  \bibinfo{year}{2013}\natexlab{}.
\newblock \showarticletitle{Don't be scared: use scalable automatic repairing
  with maximal likelihood and bounded changes}. In
  \bibinfo{booktitle}{\emph{Proceedings of the 2013 ACM SIGMOD International
  Conference on Management of Data}}. \bibinfo{pages}{553--564}.
\newblock


\bibitem[Yan and He(2020)]%
        {yan2020auto}
\bibfield{author}{\bibinfo{person}{Cong Yan} {and} \bibinfo{person}{Yeye He}.}
  \bibinfo{year}{2020}\natexlab{}.
\newblock \showarticletitle{Auto-Suggest: Learning-to-Recommend Data
  Preparation Steps Using Data Science Notebooks}. In
  \bibinfo{booktitle}{\emph{Proceedings of the 2020 ACM SIGMOD International
  Conference on Management of Data}} (Portland, OR, USA)
  \emph{(\bibinfo{series}{SIGMOD '20})}. \bibinfo{publisher}{Association for
  Computing Machinery}, \bibinfo{address}{New York, NY, USA},
  \bibinfo{pages}{1539–1554}.
\newblock
\showISBNx{9781450367356}
\urldef\tempurl%
\url{https://doi.org/10.1145/3318464.3389738}
\showDOI{\tempurl}


\bibitem[Zafar et~al\mbox{.}(2017)]%
        {zafar2017trustworthy}
\bibfield{author}{\bibinfo{person}{Faheem Zafar}, \bibinfo{person}{Abid Khan},
  \bibinfo{person}{Saba Suhail}, \bibinfo{person}{Idrees Ahmed},
  \bibinfo{person}{Khizar Hameed}, \bibinfo{person}{Hayat~Mohammad Khan},
  \bibinfo{person}{Farhana Jabeen}, {and} \bibinfo{person}{Adeel Anjum}.}
  \bibinfo{year}{2017}\natexlab{}.
\newblock \showarticletitle{Trustworthy data: A survey, taxonomy and future
  trends of secure provenance schemes}.
\newblock \bibinfo{journal}{\emph{Journal of network and computer
  applications}}  \bibinfo{volume}{94} (\bibinfo{year}{2017}),
  \bibinfo{pages}{50--68}.
\newblock


\bibitem[Zhang et~al\mbox{.}(2020)]%
        {zhang2020data}
\bibfield{author}{\bibinfo{person}{Amy~X Zhang}, \bibinfo{person}{Michael
  Muller}, {and} \bibinfo{person}{Dakuo Wang}.}
  \bibinfo{year}{2020}\natexlab{}.
\newblock \showarticletitle{How do data science workers collaborate? roles,
  workflows, and tools}.
\newblock \bibinfo{journal}{\emph{Proceedings of the ACM on Human-Computer
  Interaction}} \bibinfo{volume}{4}, \bibinfo{number}{CSCW1}
  (\bibinfo{year}{2020}), \bibinfo{pages}{1--23}.
\newblock


\bibitem[Zhang et~al\mbox{.}(2003)]%
        {dataprep2003zhang}
\bibfield{author}{\bibinfo{person}{Shichao Zhang}, \bibinfo{person}{Chengqi
  Zhang}, {and} \bibinfo{person}{Qiang Yang}.} \bibinfo{year}{2003}\natexlab{}.
\newblock \showarticletitle{Data preparation for data mining}.
\newblock \bibinfo{journal}{\emph{Applied Artificial Intelligence}}
  \bibinfo{volume}{17}, \bibinfo{number}{5-6} (\bibinfo{year}{2003}),
  \bibinfo{pages}{375--381}.
\newblock
\urldef\tempurl%
\url{https://doi.org/10.1080/713827180}
\showDOI{\tempurl}
\showeprint{https://doi.org/10.1080/713827180}


\end{thebibliography}

\end{document}